\documentclass[a4paper,10pt]{report}
\usepackage{a4wide,makeidx}                             
\usepackage[english]{babel}                             
\usepackage{amsmath,amsfonts,verbatim,graphicx,float}

\begin{document}
\begin{center}
{\bf \Large Universidad Aut\'onoma de San Luis Potos\'i} \\
\end{center}
\vspace{0.5cm}
\begin{figure}[h!]
\center
\includegraphics[width=3cm]{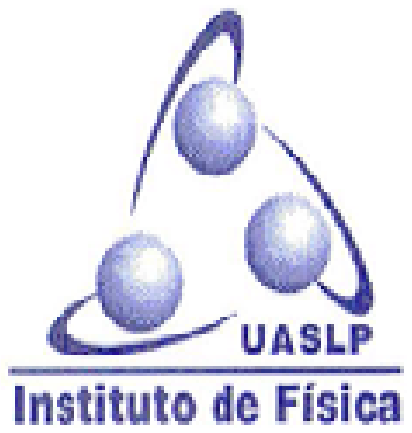}\hspace{1.5cm}{\Large 
Instituto de F\'isica}\hspace{1.5cm}
\includegraphics[width=3cm]{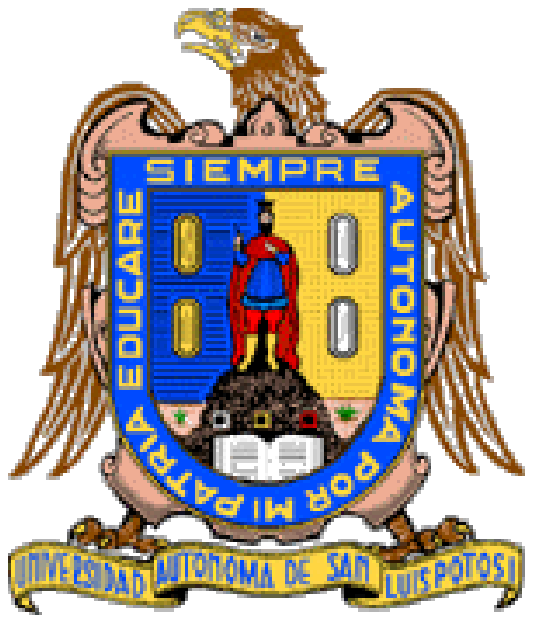}
\end{figure}

\vspace{3.5cm}

\begin{center}
{\bf \Huge Exactly Solvable Potentials and Romanovski Polynomials in 
Quantum Mechanics} \\
\end{center}
\vspace{1.5cm}

\begin{center} 
{\bf \Large  Fis. David Edwin Alvarez Castillo, IF-UASLP}\\
\end{center}
\vspace{0.2cm}
\begin{center}
{\large Supervisors: Dr.\ Mariana Kirchbach (IF-UASLP)}\\
{\large \hspace{2.63cm} Dr.\ Alvaro Per\'ez Raposo (FC-UASLP) }
\end{center}
\vspace{1.8cm}
\begin{center}
March 2007
\end{center}
\newpage

\chapter*{Preface}

This thesis for the University Degree of Master of Science 
is an account of my research at the Institute of Physics
at  the Autonomous University of San Luis Potosi, 
Mexico, during the time period  September 2005---December 2006. It is 
devoted to the exact solutions of the Schr\"odinger equation with the 
hyperbolic Scarf potential
(Scarf II) which finds various applications in physics ranging from
soliton physics to electrodynamics with  non-central
potentials. In the current literature the exact solutions to Scarf II
are written  in terms of Jacobi polynomials of purely imaginary
arguments and parameters that are complex conjugate to each other.
The fact is that the above Jacobi polynomials are proportional to real
orthogonal polynomials  by the purely imaginary phase factor, $(-i)^n$, 
much like the phase relationship 
between the hyperbolic--and the trigonometric functions, i.e. $\sin (ix) =
i\sinh (x)$.
These real polynomials, to be
referred to as the Romanovski polynomials,
have been largely ignored by the standard mathematical textbooks on 
polynomials and mathematical methods of physics texts although 
they are required in exact solutions of several physics problems
ranging from supersymmetric quantum mechanics and quark physics 
to random matrix theory. It is one of the virtues of the present thesis
to draw attention to the Romanovski polynomials and their
importance in the physics with Scarf II.  

I begin with reviewing the five possible real polynomial 
solutions of  the generalized hypergeometric differential equation. 
Three of them are the well known classical orthogonal polynomials
of Hermite, Laguerre and Jacobi, 
but the other two are different with respect to their 
orthogonality properties. The family of polynomials which 
exhibits finite orthogonality (meaning that only
a finite number of them are orthogonal) are the Romanovski polynomials.

Next, I solve the one-dimensional Schr\"odinger equation 
with Scarf II in terms
of the Romanovski polynomials. 
Then I focus on the  problem of an electron within a particular
potential that appears  non-central in the polar angle coordinate
while preserving the rotational invariance with respect to the
azimuthal angle. I report the new observation
that  the (one-dimensional)
Schr\"odinger equation with the hyperbolic Scarf potential defines the polar 
angle part of the respective  wave functions.
The latter define  new non-spherical angular functions.
I furthermore establish a new non-linear relationship between the
Romanovski polynomials to the associated Legendre functions and
employ it to obtain a non-standard orthogonality integral 
between infinite series of polynomials.
This infinite orthogonality does not contradict the finite one because
in the latter case the parameters change with the polynomial 
degree and same does the associated weight function.
This circumstance allows to satisfy the orthogonality condition for 
an infinite number of polynomials.
Finally, I also  solve
the Klein-Gordon equation with scalar and vector potentials 
of equal magnitudes and  given by Scarf II.
I conclude that the Romanovski polynomials are the most adequate
degrees    of freedom in the mathematics of the Scarf II potential.

\newpage
\section*{Acknowledgments}
I would like to thank all the people that surrounded me during the 
time when I was working on this research. Special thanks
to  my supervisor  Mariana Kirchbach who inspired and helped me to 
keep going on with my research with enthusiasm and find the
new results. I am indebted to my second supervisor Alvaro Raposo for his
helpful criticism. 
Thanks also to all my teachers who in one way or another 
taught me knowledge  and discipline. 
To my family and girlfriend, I dedicate this work.

\newpage
\tableofcontents
\chapter{Introduction.}

The exactly solvable Schr\"odinger equations (SE) occupy a pole position
in quantum mechanics in so far as most of them  
directly apply to relevant physical systems. 
Prominent examples are the quantum Kepler-, or, 
Coulomb-potential problem and its
importance in the description of the discrete spectrum of the
hydrogen atom \cite{TC}, the harmonic-oscillator--, the Hulthen--, and
the Morse potentials with their relevance to vibrational spectra
\cite{Hulten}, \cite{Franko}. Another good example is
given by the P\"oschl-Teller potential
\cite{Teller} which appears as an effective mean field in many-body
systems with  $\delta $-interactions \cite{Calogero}. 
Knowing the exact SE solutions is furthermore of interest 
in testing approximative iteration  methods \cite{CHS}. 

There are various methods of finding the exact solutions
of a Schr\"odinger equation for the bound states, 
an issue on which we shall focus in the present work.
\begin{itemize}
\item The traditional method \cite{Manning}  
consists in reducing SE by an
appropriate substitution of the variables to that very form of the
generalized hypergeometric equation \cite{NikUv} whose solutions are
polynomials,  the majority of which, and especially the classical ones,
being well studied.
\item The second method suggests to first unveil the dynamical symmetry of the
potential problem and then employ  the relevant group algebra in
order to construct the solutions as the group representation spaces
\cite{Wybourne}--\cite{JPA_MG_34}.
 \item Finally, there is also the most recent and
powerful method of the super-symmetric quantum
mechanics (SUSYQM)
which considers the special class of Schr\"odinger equations 
(in units of $\hbar =1=2m$) that
allows for a factorization according to \cite{MF}, \cite{Khare}
\begin{eqnarray}
\left( H(z)-e_n\right)\psi_n (z)&=&\left( 
-\frac{d^2}{dz^2} +v(z)- e_n\right)\psi_n(z)=0\, ,\nonumber\\
H(z)&=&A^+ (z)A^-(z) +e_0\, ,\nonumber\\
A^\pm(z) &=& 
\left( \pm \frac{d}{dz}+ U(z)\right)\, .
\end{eqnarray}
Here, $H(z)$ stands for the (one-dimensional) Hamiltonian, 
$U(z)$ is the so called super-potential, and the operators
$A^\pm (z) $ ladder between neighboring solutions.
The super-potential allows to recover the ground state wave function,
$\psi_{\rm gst}(z)$, as
\begin{equation}
\psi_{\rm gst}(z)\sim e^{-\int ^z U(y)dy}\, .
\label{SUSY_1}
\end{equation}
The excited states are then
built up on top of $\psi_{\rm gst}(z)$ through the repeated
action of the $A^+(z)$  operators.
The supersymmetric quantum mechanics manages the family of exactly solvable
potentials presented in Table 1.1. Though this Table contains 10 potentials, 
the variety is in fact not as big. 
The reason is that these potentials split into two classes such that
the potentials belonging to same class can be mapped onto each other 
by means of appropriate point transformations of the coordinates
(so called ``shape invariance'', see Ref.~\cite{RDe,Sukumar} 
and references therein).
The Coulomb, oscillator, and Morse potentials belong to the first class, 
while the second class contains the Rosen-Morse, Eckart, P\"oschl-Teller 
and Scarf potentials.
There are more exactly solvable potentials.
The very popular in nuclear physics Woods-Saxon potential,
\begin{equation}
V_{WS}(r)=-\frac{V_0}{1+e^{\frac{r-R_0}{a}}}
-\frac{ce^{\frac{r-R_0}{a}}}{\left( 1+e^{\frac{r-R_0}{a}}\right)^2}\, ,
\label{WoodsSax}
\end{equation}
where  $V_0$, $R_0$,and $a$ are constant parameters,
has been resolved recently in Refs.~\cite{Berkdemir} 
in terms of the Jacobi polynomials.

\item The above methods refer to potentials 
with unconstrained parameters.
However, there are also potentials
which are exactly solvable only when
the parameters obey certain constraints. Take as example the
sextic anharmonic oscillator,
$ax^6+bx^4+cx^2$ with $a>0$ which allows for exact solvability \cite{Nasser}
only if the parameters are restricted by the condition,
 $3\sqrt{a}-\frac{b^2}{4a} +c=0$. 
In this case, any wave function corresponding to a level of energy $E$
acts as the generating function of a set of 
polynomials in the energy variable
which differ from the solutions of the generalized
hypergeometric equation. For the first time such polynomials have been
reported by Bender and Dunne \cite{Bender} while solving
the one dimensional $(1D)$ Schr\"odinger equation with the potential,
$x^6 -(4s -4J-2)x^2$, for $J$ a positive integer and $s=1/4$, or $s=3/4$.
The two choices of $s$ refer to even-- and odd-parity of the eigen-wave 
functions, respectively. One can find a ``weight function''
with respect to which the Bender--Dunner polynomials 
appear ``orthogonal'' though the ``weight function'' neither needs
to be positive.
\end{itemize}

While the non-relativistic  Schr\"odinger equation is exactly solvable for
the potentials listed in Table 1.1 (and for a few more), 
exact solutions to the relativistic Klein-Gordon-- and Dirac equations 
are quite scarce. Besides the relativistic Coulomb problem,
the Dirac equation seems so far to be  amenable to an exact solution only for
the relativistic oscillator, a result due to Moshinsky and 
Szczepaniak \cite{Moshinsky}.
Klein-Gordon equations with scalar and vector potentials of equal magnitudes
reduce to Schr\"odinger equations and allow for exact solvability
(see Chpt. IV below).
Finally, very recently, low-power potentials of the type,
 $\sim \beta /r^{1+\beta }$
with $-1\not= \beta \not= -2$, have been shown to be
quasi-solvable in the sense that only the 
zero energy solutions are exact \cite{Alhaidari}.
The present study is devoted to the first item above.

\begin{table}
\centering
\begin{tabular}{||r|r|r||}  \hline
Potential & Variable& Wave Function  
\\ \hline \hline
Shifted oscillator:
$\frac{1}{4}\omega^2\left( x-\frac{2b}{\omega }\right)^2 -\frac{\omega}{2}$ &
 $y=\sqrt{\frac{\omega}{2}}\left( x-\frac{2b}{\omega}\right)$&
$\exp \left(-\frac{1}{2}y^2\right)H_n(y)$     \ \\ \hline
3D-oscillator:
$\frac{1}{4}\omega^2 r^2 +\frac{l(l+1)}{r^2}
-\left(l+\frac{3}{2}\right)\omega $ &  $y=\frac{1}{2}\omega r^2$
    & $y^{\frac{1}{2}l(l+1)}\exp \left(-\frac{1}{2}y \right)
L_n^{\left(l+\frac{1}{2}\right) }(y)$   \ \\ \hline
Coulomb: $-\frac{e^2}{r}+\frac{l(l+1)}{r^2}) +\frac{e^4}{4(l+1)^2}$ &  
$y=\frac{re^2}{n+l+1} $   & 
$y^{l+1}\exp \left(-\frac{1}{2}y \right)L_{n-l-1}^{(2l+1)}(y)$ \ \\ \hline
Morse: 
$A^2 +B^2 \exp (-2\alpha x)$ 
&  $y=\frac{2B}{\alpha }\exp (-\alpha x)$ &
 $y^{s-n}\exp \left(-\frac{1}{2}y \right)L_n^{(2(s-n))}(y)$     \ \\ 

$-2B\left(A+\frac{\alpha }{2}\right)
\exp (-\alpha x )$& & \ \\
\hline
Scarf II :  $A^2+(B^2-A^2 -A\alpha ) \mathrm{sech^2}(\alpha x)$
 &  
$y= \mathrm{sinh} (\alpha x)$  &
$i^n(1+y^2)^{-\frac{s}{2}}\exp (-\lambda \tan ^{-1}(y) )$
  \ \\ 
 (P\"oschl-Teller 1) $+
B(2A+\alpha ) \mathrm{sech} (\alpha x) \mathrm{tanh} (\alpha x)$& 
& $\times P_n^{\left(
i\lambda -s -\frac{1}{2},-i\lambda -s -\frac{1}{2}\right) }(iy)$\ \\
\hline
Rosen-Morse II : 
$A^2 +\frac{B^2}{A^2}$ &  $y= \mathrm{tanh}( \alpha x)$    & 
$(1-y)^{\frac{s_1}{2}}
(1+y)^{\frac{s_2}{2}}P_n^{(s_1,s_2)}(y)$
\ \\ 
$ -A(A+\alpha )\,  \mathrm{sech^2}(\alpha x)
 +2B  \mathrm{tanh}(\alpha x)$ & & \ \\
\hline
Eckart:  $A^2 +\frac{B^2}{A^2} -2B  \mathrm{coth}( \alpha x)$
&  $y=  \mathrm{coth}( \alpha x)$    & 
$(y-1)^{\frac{s_3}{2}}(y+1)^{\frac{s_4}{2}}P_n^{(s_3,s_4)}(y)$ 
 \ \\ 
$ +A(A-\alpha )
  \mathrm{csch^2}(\alpha x)$ & & \ \\
\hline
Scarf I:  $-A +(A^2+(A^2+B^2-A\alpha )\sec^2 (\alpha x)$ &  
$y=\sin(\alpha x) $    & 
$ (1-y)^{\frac{s-\lambda}{2}}(1+y)^{\frac{s+\lambda }{2}}$
 \  \\ 
$
+A(2A-\alpha )\tan (\alpha x) \sec (\alpha x) $ & & 
$\times P_n^{\left( 
s-\lambda -\frac{1}{2},-\lambda -s -\frac{1}{2}\right)}(y)$ \ \\
\hline
P\"oschl-Teller 2:  $A^2 +(B^2+A^2+A\alpha ) \,  \mathrm{csch^2} (\alpha x)$ 
&
$y= \mathrm{cosh} (\alpha x)$&
$(y-1)^{\frac{\lambda -s}{2}}(y+1)^{-\frac{\lambda +s}{2}}$ 
\  \\
$ -B(2A+\alpha )\,  \mathrm{coth}( \alpha x)  \mathrm{csch}( \alpha x) $
 & &
$\times P_n^{\left( 
\lambda -s -\frac{1}{2}, -\lambda -s -\frac{1}{2}\right)}(y)$ 
\ \\ \hline
Rosen-Morse I: $A(A-\alpha)\csc^2(\alpha x) $
&
$y=\cot ( \alpha x)$&
$i^n(y^2+1)^{-\frac{s+n}{2}}\exp (a \cot^{-1} (y) )$
\  \\ 
$+2B\cot (\alpha x) -A^2 
+\frac{B^2}{A^2}$ & & $ \times P_n^{\left( -s -n -ia, -s -n +ia\right)}(iy)$
 \ \\
\hline
\end{tabular}
\caption{Exactly solvable shape invariant potentials in one dimension
(according to \cite{Khare,Levai}). 
All potential parameters are supposed to be positive.
Other notations are: $s=A/\alpha$, $\lambda =B/\alpha$, $a=\lambda/(s-n)$,
$s_1=s-n+a$, $s_2=s-n-a$, $s_3=a-n-s$, and $s_4=-(s+n+a)$. 
The orthogonal polynomials of Hermite, Laguerre, and Jacobi have been
denoted in the standard way as $H(x)$, $L_n^{(\alpha )}(x)$, and
$P_n^{(\mu ,\nu )}(x)$, respectively. }
\label{Table}
\end{table}

\section{The goal.}

\begin{quote}

Table 1.1 shows that the hyperbolic-Scarf-- and the trigonometric Rosen-Morse
potentials (termed to as Scarf II, and Rosen-Morse I, respectively)
are solved in terms of Jacobi polynomials of imaginary arguments
and parameters that are complex conjugate to another. 

The goal of this thesis is to solve the Schr\"odinger equation 
with the hyperbolic Scarf potential
\cite{RDe}, \cite{KaSu}--\cite{Bagchi} anew and to  make the case that 
it reduces in a straightforward manner to a particular form of the 
generalized real hypergeometric equation whose solutions 
are given by a \underline{finite set} of real orthogonal polynomials.
In this manner, the finite number of bound states within the
hyperbolic Scarf potential is brought in correspondence
with a finite system of orthogonal polynomials of a new class.
In due course, various new properties of the 
above polynomials are encountered.  
\end{quote}
These polynomials have been discovered in 1884 by 
the English mathematician Sir Edward John Routh
\cite{Routh} and rediscovered 45 years later by the Russian
mathematician Vsevolod Ivanovich Romanovski in 1929 
\cite{Romanovski} within the context of probability distributions. 
 Though they have been studied on few  occasions in the 
current mathematical literature where they are termed to as  
``finite Romanovski'' \cite{Zarzo}--\cite{Neretin},
or, ``Romanovski-Pseudo-Jacobi'' polynomials \cite{Lesky}, they 
have been completely ignored by
the textbooks on mathematical methods in physics,
 and surprisingly enough, by the standard mathematics textbooks 
as well \cite{NikUv}, \cite{MMF}-\cite{ism}.
The notion ``finite'' refers to the observation that for any given set of 
parameters (i.e. in any potential),
only a finite number of polynomials (finite number of bound states)
 appear orthogonal.

The  Romanovski polynomials happen to be equal
(up to a phase factor) to that very Jacobi polynomials with 
imaginary arguments and parameters that are complex 
conjugate to each other \cite{Levai_PL}--\cite{Raposo}, much like the 
$\sinh(z)=i\sin(iz)$ relationship. 
Although one may (\underline{but has not to}) deduce the 
local characteristics of the latter such as 
generating function and recurrence relations from those of the former,
the finite orthogonality theorem is qualitatively new. 
It does not copy none of the properties of the Jacobi polynomials
but requires an independent proof.
For this reason,  in  random matrix theory \cite{Witte} the problem on the 
gap probabilities in the spectrum of the complex circular Jacobi ensemble
is preferably treated in terms of the real Cauchy random ensemble,
a venue that conducts once again to the Romanovski polynomials 
because, as it will be shown below, the weight function of the
Romanovski polynomials for certain values of the parameters
equals the Cauchy distribution.

The thesis contributes two new examples to the circle of 
the typical quantum mechanical problems \cite{Fluegge}.
The techniques used here extend  the 
teachings on the Sturm-Liouville theory of ordinary differential
equations beyond their textbook presentation.  

Before leaving the introduction, a comment is in place on the 
importance of the Scarf II potential.  It
finds various applications in physics ranging from electrodynamics 
 and solid state physics to particle theory.
In solid state physics Scarf II is used in the construction
of more realistic periodic potentials in crystals \cite{Kusnezov} than
those built from the trigonometric Scarf potential (Scarf I)  \cite{Scarf_58}. 
In electrodynamics Scarf II appears in a class of problems with 
non-central potentials \cite{Dutt,Kocak}
(also see section IV below for more details).
In particle physics Scarf II finds application in studies of
the non-perturbative sector of gauge theories by means of 
toy models such as the scalar field theory in 
(1+1) space-time dimensions. Here, one encounters  the so called 
``kink -like'' solutions which are no more but the
static solitons. The spatial derivative of the
kink-like solution is  viewed as the ground state wave function of an 
appropriately constructed Schr\"odinger equation 
which is then employed in the calculation of the quantum corrections 
to first order.
In Ref.~\cite{Hidalgo} it was shown that specifically Scarf II is amenable
to a stable renormalizable scalar field theory.\\

\noindent
The thesis  is organized as follows.

\begin{itemize}
\item
In the next chapter I first
highlight in brief the basics of the generalized hypergeometric
equation, review the classification of its polynomial 
solutions in various schemes,
and construct for completeness of the presentation, the polynomials
of Hermite, Laguerre, Jacobi, Romanovski, and Bessel.
\item
In chapter III I bring  examples for 
potentials  whose exact solutions are given in terms of the
 polynomials presented in the previous chapter.   
\item
The original contribution of thesis is presented in  
chapter IV. There  I 

\begin{itemize}
\item first provide the exact
solution of the $1D$-Schr\"odinger equation with
Scarf II in terms of the finite Romanovski polynomials.

\item Next I  consider the problem of  an electron within a 
particular non-central potential along the line of
Ref.~\cite{Dutt,Kocak}. I made the new observation that
the polar angle part of the electron wave function 
in this problem happens to be defined by
Romanovski polynomials whose 
parameters  depend on the polynomial degree. 
Within this context, the Romanovski polynomials
act as designers of new non--spherical angular functions.
\item I explicitly construct and display graphically
the lowest five non--spherical-functions and compare them with the standard
$Y_l^m (\theta ,\varphi )$ harmonics. 

\item I furthermore establish a new non-linear relationship between the
Romanovski polynomials to the associated Legendre functions and
employ it to obtain a non-standard orthogonality integral 
between infinite series of polynomials.
This infinite orthogonality does not contradict the finite one because
in the latter case the parameters change with the polynomial 
degree and same does the associated weight function.
This circumstance allows to satisfy the orthogonality condition for 
an infinite number of polynomials.

\item
Before closing by a brief summary, I consider a 
Klein-Gordon equation with scalar and vector potentials 
of equal magnitudes which are given by Scarf II.

\end{itemize}

\end{itemize}

\chapter{ Generalized hypergeometric equation. Polynomial
solutions.} 

All classical orthogonal polynomials appear as solutions of the so called
generalized hypergeometric equation (the presentation in this section 
closely follows Ref.~\cite{Koepf}),
\begin{eqnarray}
\sigma(x)y_n^{\prime\prime}(x)+\tau (x)y_n^\prime (x)-\lambda_ny_n(x)&=&0\, ,
\label{hyperg_EQ}\\
\sigma(x)=a x^2+bx+c,\quad \tau(x)=xd +e\,, &&
\lambda_n=n(n-1)a+nd\, .
\end{eqnarray}
There are various methods for finding the polynomial
solution, here denoted by
\begin{eqnarray}
y_n(x)&\equiv &P_n\left(
\begin{array}{ccc}
d&e&\\
a&b&c
\end{array}{\Bigg|}x \right)\, .
\end{eqnarray}
The symbol  $P_n\left(
\begin{array}{ccc}
d&e&\\
a&b&c
\end{array}{\Bigg|}x \right)$
makes the equation parameters explicit and
stands for a polynomial of degree $n$, $\lambda_n$ 
being the eigenvalue parameter, and $ n=0,1,2,...$.

\section{Classification scheme of Koepf-Masjed-Jamei.}

In Ref.~\cite{Koepf} the solutions to Eq.~(\ref{hyperg_EQ}) have been
classified according to the five parameters $a$, $b$, $c$, $d$, and $e$.
Furthermore, a master formula for a generic 
monic polynomial solutions, ${\bar P}_{n}$,
has been derived by Koepf and Masjed-Jamei,
according to them  one finds
\begin{eqnarray}
&&{\bar P}_n\left(
\begin{array}{ccc}
d&e&\\
a&b&c
\end{array}{\Bigg|}x \right)
=\sum_{k=0}^{n}
\left(
\begin{array}{c}
n\\
k
\end{array}
\right) G_k^{(n)}(a,b,c,d,e)x^k\, ,\nonumber\\
&&G_k^{(n)}=
\left( 
\frac{2a}{b+\sqrt{b^2-4ac}}
\right)^{n}\, \, 
_2F_1
\left(
\begin{array}{cc}
(k-n),&\left( 
\frac{2ae -bd}{2a\sqrt{b^2-4ac}}+1-\frac{d}{2a}-n\right) \\
2-\frac{d}{a} -2n&
\end{array}
{\Bigg|}
\frac{
2\sqrt{b^2-4ac}}
{b+\sqrt{b^2-4ac}}
\right).
\nonumber\\
\label{monic_master}
\end{eqnarray}
The $a=0$ case is handled as the $a\to  0$ limit of
Eq.~(\ref{monic_master}) and leads to the appearance of
$\, _2F_0$ in place of $_2F_1$ (see Ref.~\cite{Koepf} for details). 
Though the original derivation of this result
is a bit cumbersome, its verification with the help of 
the symbolic software Maple is straightforward.

\newpage

\section{Classification according to Nikiforov-Uvarov.}
In the method of Nikiforov and Uvarov \cite{NikUv} 
the (polynomial) solutions to the hypergeometric equation 
(\ref{hyperg_EQ}) are classified according to the
so called weight function, $w(x)$, and built up from the Rodrigues formula,
\begin{eqnarray}
\sigma(x)y_n^{\prime\prime}(x)+y_1(x)y_n^\prime (x)-\lambda_ny_n(x)&=&0\, ,
\label{hyperg_NU}\\
 y_1(x)= \frac{N_1}{w(x)}\frac{d}{dx}\lbrack w(x)\sigma (x)\rbrack , &\quad&
\lambda_n=-n(y_1^\prime (x) +\frac{1}{2}(n-1)\sigma ^{\prime\prime} (x))\,,
\end{eqnarray} 
with $N_1$ being the $y_1(x)$ normalization constant.
The weight function is calculated from
integrating the so called Pearson differential equation,
\begin{equation}
\frac{d }{dx}\left( \sigma (x) w(x)\right)=y_1 (x)w(x)\, ,
\label{Pearson}
\end{equation}
and then plugged into the Rodrigues formula to generate the
polynomial solutions as
\begin{equation}
y_n(x)= \frac{N_n}{w(x)}\frac{d^n}{dx^n}\left[ w(x)\sigma^{n} (x)\right] , 
\label{Rds_NikUv}
\end{equation}
with $N_n$ being a normalization constant.
The $y_{n}(x)$'s are normalized polynomials and are orthogonal with 
respect to the weight function $w(x)$ within a given 
interval, $[l_1,l_2]$, provided $\sigma(x)>0$, and 
$w(x)>0$ holds true for all $x\in[l_1,l_2]$, and 
$\sigma(x)w(x)x^{l}|_{x=l_1}=\sigma(x)w(x)x^{l}|_{x=l_2}=0$ 
for any $l$ integer, and with suitable normalization constants,
\begin{eqnarray}\label{integralortonormal} 
\int_{l_1}^{l_2}w (x)y_{n}(x)y_{n'}(x)dx=\delta_{nn'}, &&
\forall  \qquad n ,n'\in\left\lbrace 0,1,2,...\right\rbrace . 
\end{eqnarray}
The Nikforov-Uvarov method combines well with the  
Koepf--Masjed-Jamei scheme. Indeed, the solution to the Pearson equation
can be cast into the form
\begin{equation}
\omega (x)\equiv {\mathcal W}
\left(
\begin{array}{ccc}
d&e&\\
a&b&c
\end{array}
{\Bigg|}x
\right)
=\exp \left(
\int \frac{(d-2a)x+(e-b)}{ax^2+bx+c}dx
\right)\, .
\label{weight_general}
\end{equation}
It shows how one can calculate any weight function 
associated with any parameter set of interest
(the symbol used for the weight function  
 makes again the equation parameters explicit). 
The best strategy in recovering 
the weight function corresponding to a given set of
parameters is to consider its logarithmic derivative,
$\frac{\omega ^\prime (x)}{\omega (x)}$, and then match
the parameters accordingly.
In order to illustrate this procedure, we pick up one of the
examples given in \cite{Koepf} and consider 
the weight function 
\begin{equation}
\omega (x)= (-x^2+3x -2)^{10}=(1-x)^{10}(2-x)^{10}\, .
\label{wx_toy}
\end{equation}
The calculation of the logarithmic derivative gives
\begin{equation}
\frac{\omega ^\prime (x)}{\omega (x)}=
\frac{-20x +30}{-x^2+3x -2}=\frac{(d-2a)x +(e-b)}{ax^2+bx +c}\,,
\label{log_der}
\end{equation}
which matches with the parameters 
$a=-1$, $b=3$, $c=-2$, $d=-22$, and $e=33$.
Therefore, the resulting weight function is
\begin{equation}
\omega(x)=
{\mathcal W}
\left(
\begin{array}{ccc}
-22&33&\\
-1&3&-2
\end{array}
{\Bigg|}x
\right)\,.
\end{equation} 
In the notations of Koepf--Masjed-Jamei 
the Rodrigues formula looks like 
\begin{eqnarray}
 P_n\left(
\begin{array}{ccc}
d&e&\\
a&b&c
\end{array}{\Bigg|}x \right)&=&
\Pi_{k=1}^{k=n}(d +(n+k-2)a){\bar P}_n\left(
\begin{array}{ccc}
d&e&\\
a&b&c
\end{array}{\Bigg|}x \right)=
\frac{1}{
{\mathcal W}
\left(
\begin{array}{ccc}
d&e&\\
a&b&c
\end{array}
{\Bigg|}x
\right)
}\nonumber\\
&\times &
\frac{
d^n
}{dx^n}\left( (ax^2+bx+c)^n
{\mathcal W}
\left(
\begin{array}{ccc}
d&e&\\
a&b&c
\end{array}
{\Bigg|}x
\right)\right)\, .\nonumber\\
\label{Rodrigues}
\end{eqnarray}
The polynomials associated with
the weight function in Eq.~(\ref{wx_toy}) 
can be constructed explicitly by Eq.~(\ref{Rodrigues})
and treated as independent entities and without 
even knowing that they are
no more but Jacobi shifted to  the interval $x\in [1,2]$. 
The great appeal of combining the master formulas
in the respective  Eqs.~(\ref{monic_master}), and (\ref{Rodrigues})
is that they allow for the direct and pragmatic
construction of all the polynomial solutions to
the generalized hypergeometric equation.\\

\noindent
One identifies as special cases 
\begin{itemize}
\item   the
Jacobi polynomials with $a=-1$, $b=0$, $c=1$, 
$d=-\gamma -\delta  -2$, and $e=-\gamma +\delta$,
\item the Laguerre polynomials with
$a=0$, $b=1$, $c=0$, $d=-1$, and $e=\alpha +1$,
\item  the  Hermite polynomials with  $a=b=0$, $c=1$, $d=-2$, 
and $e=0$,
\item the Romanovski polynomials  with
$a=1$, $b=0$, $c=1$, $d=2(1-p)$, and $e=q$ with $p>0$,
\item the Bessel polynomials with $a=1$, $b=0$, $c=0$, $d=\alpha +2$, 
and $e=\beta $.
\end{itemize}

These parametrizations will be referred to as "canonical". 
Any other parametrization can be reduced to one of the above 
sets upon an appropriate shift of the argument.

The  first three  polynomials are the only ones that are 
traditionally  presented in the standard textbooks on mathematical 
methods in physics such like \cite{MMF}--\cite{ism},
while the fourth and fifth seem to have escaped due attention.
Notice, the Legendre--, Gegenbauer--,  and Chebychev polynomials
appear as particular cases of the Jacobi polynomials.
The Bessel polynomials are not orthogonal in the conventional
sense, i.e. over a real interval (see section 3.6 below for details).

Some of the properties of the fourth polynomials have been 
studied in the current mathematical literature such as 
Refs.~\cite{Zarzo},\cite{Mohamed}- \cite{Lesky}.
Their weight function is calculated from Eq.~(\ref{weight_general}) as
\begin{equation}
w^{(p ,q)}(x)=(x^2+1)^{-p}e^{ q\tan^{-1}x}\, .
\label{wafu_Rom}
\end{equation}
This weight function has first been reported by Routh \cite{Routh}, and
independently Romanovski \cite{Romanovski}.
Notice that for a vanishing $q$,  $\omega^{(p,q)}(x)$ 
becomes equal to the student's $t$ distribution \cite{Koepf_t}
and acts as that very probability distribution.
The $\omega ^{(1,0)}(x)$ case stands for the popular
Cauchy (or, Breit-Wigner)  distribution.
Within the light of this discussion, 
the weight function of the Romanovski polynomials has been
interpreted in Ref.~\cite{Koepf_t}
as the most natural extension of the student's 
$t$ distribution.

The polynomials associated with  Eq.~(\ref{wafu_Rom})
are called after Romanovski and will be 
denoted by $R_m^{(p,q)}(x)$. They
have  non-trivial orthogonality properties
over the infinite interval
 $x\in [-\infty, +\infty ]$. 
Indeed, as long as the weight function decreases as 
$x^{-2p}$, hence  integrals of the type
\begin{equation}
\int_{-\infty}^{+\infty}w^{(p,q)}(x)
R_m^{(p,q)}(x) R_{m^\prime}^{(p,q)}(x) dx,
\label{orth_int}
\end{equation}
are convergent only if
\begin{equation}
m+m^\prime< 2p-1\, ,
\label{orth_cond}
\end{equation}
meaning that only a finite number of Romanovski polynomials are orthogonal.
This is the reason for which one speaks of ``finite''Romanovski polynomials.
The orthogonality 
theorem has been proved  in  Refs.~\cite{Mohamed,Raposo}.
The differential equation satisfied by the Romanovski polynomials reads
as
\begin{equation}
(1+x^2)\frac{d ^2 R_n^{(p,q)}(x)}{d^2 x}
+\left( 2(-p+1)x +q\right)\frac{d R_n^{(p,q)}(x)}{dx}
-(n(n-1)+2n(1-p))R_n^{(p,q)}(x)=0\, .
\label{Rom_pol}
\end{equation}
In the next section we shall show that the Schr\"odinger
equation with the hyperbolic Scarf potential reduces precisely to the very 
Eq.~(\ref{Rom_pol}).

\section{Bochner's classification. Explicit polynomial construction. }
The earliest classification of the solutions to the generalized 
hypergeometric equation is due to  Bochner \cite{Bochner}
(see also Refs.~\cite{ism,nicolae,Raposo,Weber07} for more recent works)
and based on the form of $\sigma (x)$ which can be rephrased in
terms of the $\sigma (x)$ roots. The presentation in this
section closely follows Refs.~\cite{nicolae,Raposo}.
Within this scheme one finds the following five polynomial classes :\\

\noindent
1. \emph{ $\sigma(x)$ is a constant:}\\

\begin{figure}
\center
\includegraphics[width=7.5cm]{
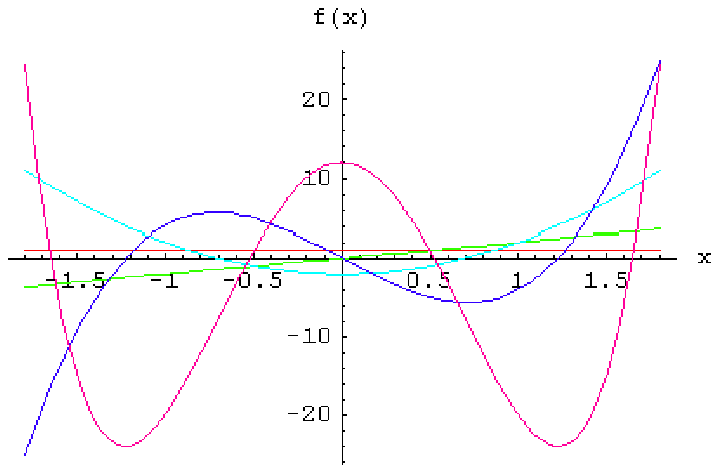}
\caption{Hermite polynomials.}
\label{Hermite}
\end{figure}

\noindent
The canonical form of the generalized hypergeometric equation is
\begin{equation}\label{Hermitegeneral}
H''(x)-2\alpha x H'(x)+\lambda H(x)=0,
\end{equation}
where $\alpha$ is a parameter. The solutions are the 
generalized Hermite polynomials 
$\left\lbrace H^{\alpha }_{n}(x)\right\rbrace$   
($\alpha=1$ characterizes the ordinary  Hermite polynomials), and 
their weight function is
\begin{equation}
\omega^{(\alpha)}(x)=e^{-\alpha x^{2}}.
\end{equation}
The case $\alpha=1$ provides the standard
orthogonality relation,
\begin{equation}\label{ortHermite}
\int_{-\infty}^{\infty}e^{- x^{2}}H_{n}(x)
H_{n'}(x)dx=\delta_{nn'}, \quad
\forall  \qquad n ,n'\in\left\lbrace 0,1,2,...\right\rbrace\, . 
\end{equation}
The lowest Hermite polynomials are then obtained as:
\begin{eqnarray}
&&H_{0}(x)=1, \nonumber\\
&&H_{1}(x)=2x, \nonumber\\
&&H_{2}(x)=-2+4x^{2}, \nonumber\\
&&H_{3}(x)=-12x+8x^{3}, \nonumber\\
&&H_{4}(x)=12-48x^{2}+16x^{3}, 
\end{eqnarray}
and are represented  in Fig. ~\ref{Hermite}.

\vspace{0.5cm}
\noindent
2. \emph{$\sigma(x)$ is of  first degree:}\\

\noindent
The canonical form of the corresponding hypergeometric equation is
obtained as
\begin{equation}\label{lagRaposo}
x L''(x)+y_1(x) L'(x)+\lambda L(x)=0.
\end{equation}
{}For $y_1(x)=-\alpha x+ \beta+ 1$, $\alpha=1$, and
$\beta$ being a real number, 
the latter equation coincides with
the equation for the associated Laguerre polynomials, 
$\left\lbrace L^{(1 ,\beta)}_{n}(x)\right\rbrace$, 
while for $\alpha=1$, and $\beta=0$ it describes the ordinary
Laguerre polynomials. 
The weight  function is given by
\begin{equation}
\omega^{(\alpha,\beta)}(x)=x^{\beta}e^{-\alpha x}.
\end{equation}
{}For $\alpha,\beta>0$ the orthogonality integral  within the interval 
$x\in \lbrack 0,\infty )$ reads
\begin{equation}\label{ortLaguerre}
\int_{0}^{\infty}x^{\beta}e^{-\alpha x}L^{(\alpha,\beta)}_{n}(x)
L^{(\alpha,\beta)}_{n'}(x)dx=\delta_{nn'},
\quad
\forall  \qquad n ,n'\in\left\lbrace 0,1,2,...\right\rbrace . 
\end{equation}
From now onwards we shall suppress the first upper index in 
$L^{(1,\beta)}(x)$ for simplicity. The lowest five  associated 
Laguerre polynomials for, say, $\beta =1$, are now calculated as
\begin{eqnarray}
&&L^{(1)}_{0}(x)=1, \nonumber\\
&& L^{(1)}_{1}(x)=2-x, \nonumber\\
&&L^{(1)}_{2}(x)=\frac{1}{2}(6-6x+x^{2}), \nonumber\\
&&L^{(1)}_{3}(x)=\frac{1}{6}(24-36x+12x^{2}-x^{3}), \nonumber\\
&&L^{(1)}_{4}(x)=\frac{1}{24}(120-240x+120x^{2}-20x^{3}+x^{4}).
\end{eqnarray}

\begin{figure}
\center
\includegraphics[width=7.5cm]{
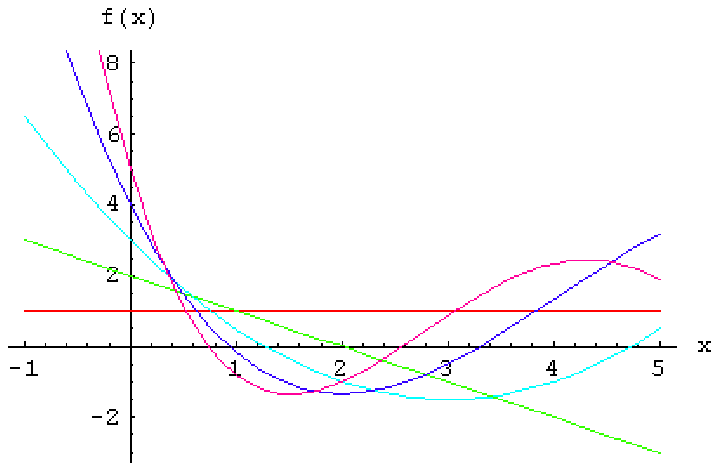}
\caption{Laguerre polynomials for $\beta =1$.}
\label{Laguerre_grph}
\end{figure}
They are shown in Fig. ~\ref{Laguerre_grph}.

\vspace{0.5cm}
\noindent
3. \emph{$\sigma(x)$ is of the second degree, 
with two different real roots:}\\

\noindent
The  canonical form of the corresponding
hypergeometric equation is obtained as
\begin{equation}\label{JacobiDE}
(1-x^{2})P''(x)+y_1(x)P'(x)+\lambda P(x)=0.
\end{equation}
The latter equation coincides with the Jacobi one for
$y_1(x)=\alpha  -\beta -(\alpha+\beta+2)x$ where 
$\alpha , \beta \in \mathbb{R}$ are the polynomial parameters. 
The  Jacobi polynomials, 
$\left\lbrace P^{(\alpha,\beta)}_{n}(x)\right\rbrace$,
are then defined by the Rodrigues formula in terms of
the following weight function: 
\begin{equation}\label{weightFunctionJacobi}
\omega^{(\alpha,\beta)}(x)=(1-x)^{\alpha}(1+x)^{\beta}.
\end{equation}
Furthermore, one has to restrict  the parameters to
$\alpha,\beta > -1$ in order to ensure  orthogonality 
in the interval $x\in [-1,+1]$ according to
\begin{equation}
\int_{-1}^{1}(1-x)^{\alpha}(1+x)^{\beta}P^{(\alpha,\beta)}_{n}(x)
P^{(\alpha,\beta)}_{n'}(x)dx=\delta_{nn'},
\quad 
\forall  \qquad n ,n'\in\left\lbrace 0,1,2,...\right\rbrace. 
\end{equation}
Several particular cases have received their proper names.
These are:  
\begin{itemize}
\item Gegenbauer when $\alpha=\beta$,
\item Chebyshev I and II when $\alpha=\beta= \pm 1/2$, 
\item Legendre when $\alpha=\beta=0$.
\end{itemize}
The lowest five  Jacobi polynomials for the toy values
$\alpha =1$, and $\beta =2$ of the parameters
are explicitly calculated as:
\begin{eqnarray}
&&P^{(1,2)}_{0}(x)=1, \nonumber\\
&&P^{(1,2)}_{1}(x)=\frac{1}{2}(-1+5x), \nonumber\\
&&P^{(1,2)}_{2}(x)=3+9(-1+x)+\frac{21}{4}(-1+x)^{2}, \nonumber\\
&&P^{(1,2)}_{3}(x)=4+21(-1+x)+28(-1+x)^{2}+\frac{21}{2}(-1+x)^{3}, \nonumber\\
&&P^{(1,2)}_{4}(x)=5+40(-1+x)+90(-1+x)^{2}+75(-1+x)^{3}+\frac{165}{8}
(-1+x)^{4}. \nonumber\\
\end{eqnarray}
\begin{figure}
\center
\includegraphics[width=7.5cm]{
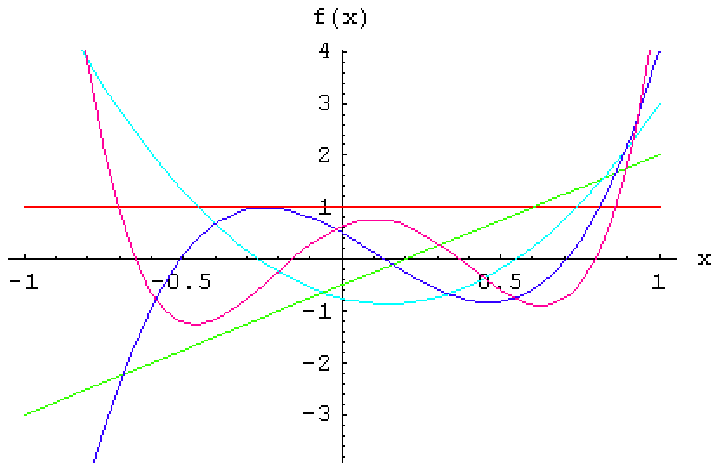}
\caption{Jacobi polynomials for the parameters $\alpha=1$ and $\beta=2$.}
\label{Jacobi_grph}
\end{figure}

\noindent
They are shown in Fig.~\ref{Jacobi_grph}.

\vspace{0.5cm}

\noindent
The lowest Gegenbauer polynomials and for $\alpha =1$ read:
\begin{eqnarray}
&&C^{(1)}_{0}(x)=1, \nonumber\\
&&C^{(1)}_{1}(x)=2x, \nonumber\\
&&C^{(1)}_{2}(x)=-1+4x^{2}, \nonumber\\
&&C^{(1)}_{3}(x)=-4x+8x^{3}, \nonumber\\
&&C^{(1)}_{4}(x)=-1-12x^{2}+16x^{4}.
\end{eqnarray}
\begin{figure}
\center
\includegraphics[width=7.5cm]{
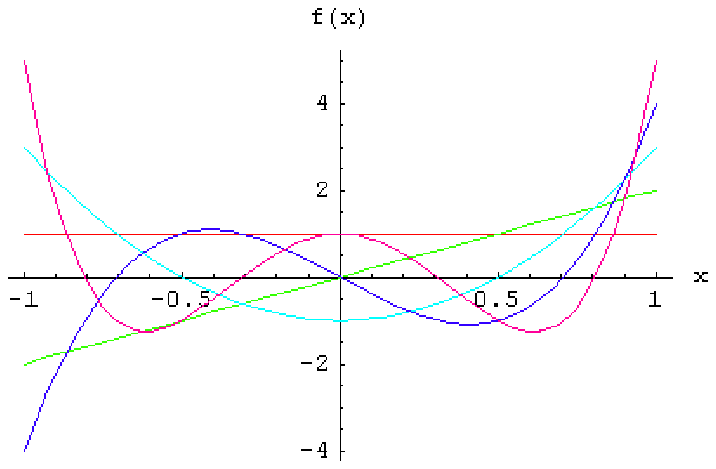}
\caption{Gegenbauer polynomials for $\alpha=1$.}
\label{Gegenbauer}
\end{figure}
They are graphically displayed in  Fig. ~\ref{Gegenbauer}.\\

\noindent
The lowest Chebyshev I polynomials read:
\begin{eqnarray}
&&T_{0}(x)=1, \nonumber\\
&&T_{1}(x)=x, \nonumber\\
&&T_{2}(x)=-1+2x^{2}, \nonumber\\
&&T_{3}(x)=-3x+4x^{3}, \nonumber\\
&&T_{4}(x)=-1-8x^{2}+8x^{4}. 
\end{eqnarray}
\begin{figure}
\center
\includegraphics[width=7.5cm]{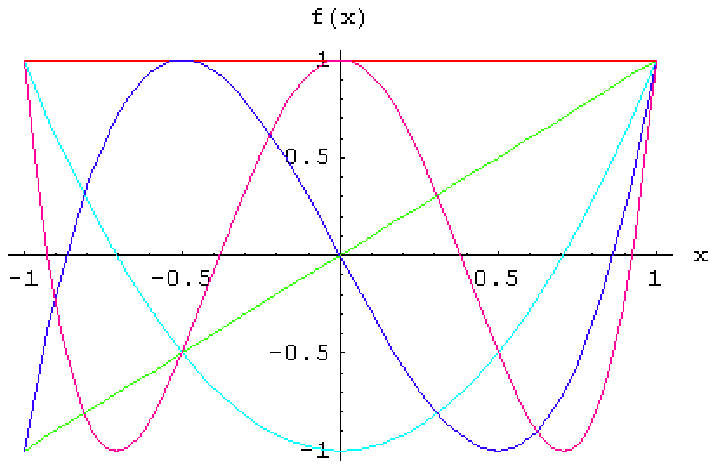}
\caption{Chebyshev I polynomials.}
\label{ChebyshevT}
\end{figure}
They are shown in Fig. ~\ref{ChebyshevT}.\\

\noindent
Next, the  lowest Chebyshev II polynomials are:
\begin{eqnarray}
&&U_{0}(x)=1, \nonumber\\
&&U_{1}(x)=2x, \nonumber\\
&&U_{2}(x)=-1+4x^{2}, \nonumber\\
&&U_{3}(x)=-4x+8x^{3}, \nonumber\\
&&U_{4}(x)=-1-12x^{2}+16x^{4}.
\end{eqnarray}
\begin{figure}
\center
\includegraphics[width=7.5cm]{
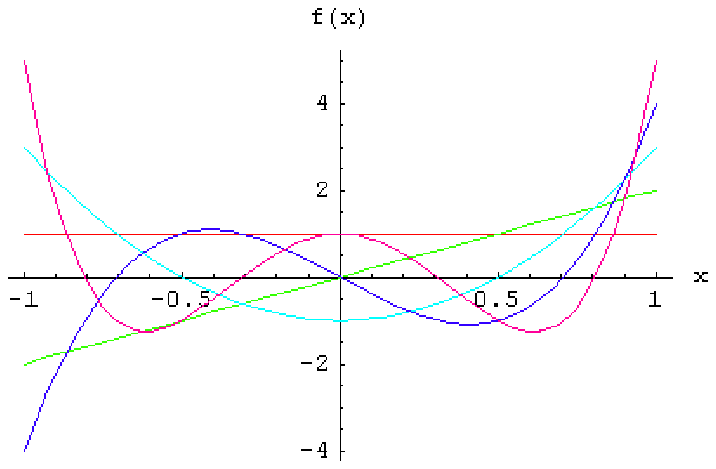}
\caption{Chebyshev II polynomials.}
\label{ChebyshevU}
\end{figure}
They are graphically displayed  in Fig. ~\ref{ChebyshevU}.\\

Finally, one finds the lowest  Legendre polynomials as
\begin{eqnarray}
&&P_{0}(x)=1, \nonumber\\
&&P_{1}(x)=x, \nonumber\\
&&P_{2}(x)=-\frac{1}{2}(1-3x^{2}),\nonumber\\
&&P_{3}(x)=-\frac{1}{2}(3-5x^{3}),\nonumber\\
&&P_{4}(x)=-\frac{1}{8}(-15x+70x^{3}-63x^{5}).
\end{eqnarray}
\begin{figure}
\center
\includegraphics[width=7.5cm]{
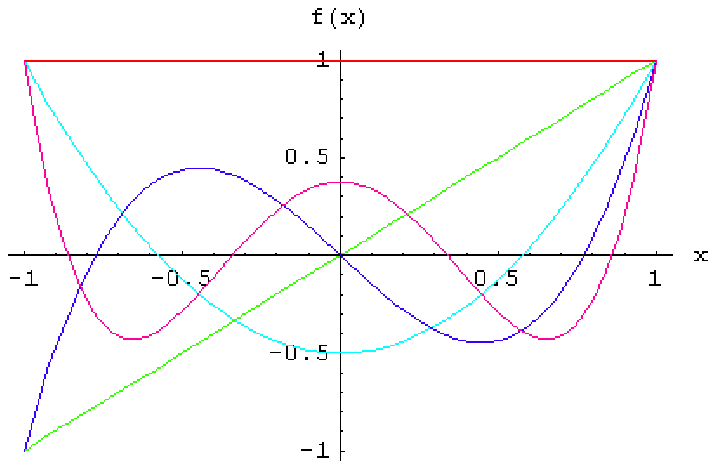}
\caption{Legendre polynomials.}
\label{Legendre}
\end{figure}
They are shown in Fig. ~\ref{Legendre}.\\

\noindent
Related to the Legendre polynomials are the associated Legendre 
functions which are defined as 
\begin{equation}
P^{m}_{n}(x)=(-1)^{m}(1-x^{2})^{\frac{m}{2}}\frac{d^{m}}{dx^{m}}P_{n}(x).
\end{equation}

\vspace{0.5cm}
\noindent
4. \emph{$\sigma(x)$ is of the second degree, 
with one double real root:}\\

\noindent
The canonical form of the corresponding differential
equation is obtained as
\begin{equation}\label{BesselEquation}
x^{2}y''(x)+y_1(x)y'(x)+\lambda y(x)=0.
\end{equation}
For  $y_1(x)=\alpha x+\beta$, with 
$\alpha , \beta \in \mathbb{R}$, the weight function is given by
\begin{equation}\label{WeightFunctionBessel}
\omega^{(\alpha,\beta)}(x)=x^{2-\alpha}e^{-\frac{\beta}{x}}.
\end{equation}
The polynomials $\left\lbrace y^{(\alpha,\beta)}_{n}(x)\right\rbrace$ 
with $\alpha=\beta=2$, are called after  Bessel \cite{Krall}. 
They are not orthogonal in the conventional sense
within a real interval. By considering  $x \in \mathbb{C}$ these polynomials 
have special characteristics.
Only when the integration contour is the unit circle
in the complex plane can one design orthogonality integrals and with 
respect to the
``weight function'' $e^{-\frac{2}{x}}$.
 To see this notice that 
Eq.~(\ref{BesselEquation}) with $\alpha=\beta=2$ can be rewritten to 
give
\begin{equation}
(x^{2}e^{-\frac{2}{x}}y'_{n}(x))'=n(n+1)e^{-\frac{2}{x}}y_{n}(x)\, .
\end{equation}
An integration by parts 
counter-clockwise around the unit circle leads to 
\begin{equation}
n(n+1)\int_{U}y_{m}(x)y_{n}(x)e^{-\frac{2}{x}} dx=\int_{U}(x^{2}
e^{-\frac{2}{x}}y'_{n}(x))'y_{m}(x) dx=-\int_{U}x^{2}e^{-\frac{2}{x}}
y'_{n}(x)y'_{m}(x)dx.
\end{equation}
Upon interchanging  $m$ against  $n$ and a subsequent
subtraction, one arrives at the orthogonality relation
\begin{equation}
\int_{U}y_{m}(x)y_{n}(x)e^{-\frac{2}{x}} dx=0.
\end{equation}
The lowest  Bessel polynomials are:
\begin{eqnarray}
&&y_{0}(x)=1, \nonumber\\
&&y_{1}(x)=1+x, \nonumber\\
&&y_{2}(x)=1+3x+3x^{2},\nonumber\\
&&y_{3}(x)=1+6x+15x^{2}+15x^{3},\nonumber\\
&&y_{4}(x)=1+10x+45x^{2}+105x^{3}+105x^{4}. 
\end{eqnarray}
\begin{figure}
\center
\includegraphics[width=7.5cm]{
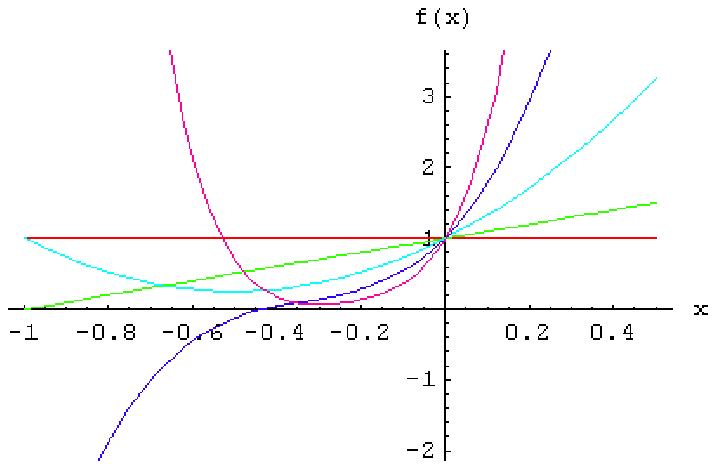}
\caption{Bessel polynomials for $\alpha=\beta =2$.}
\label{Bessel}
\end{figure}
They are displayed in  Fig. ~\ref{Bessel}.\\

\noindent
5. \emph{ $\sigma(x)$ is of the second degree, 
with two complex roots:}\\

\noindent
This is the case of {\bf prime  interest to the thesis\/}.
The generalized hypergeometric equation for this choice
of $\sigma (x)$ takes the form 
\begin{equation}\label{RomanEquation}
(1+x^{2})R''(x)+y_1(x)R'(x)+\lambda R(x)=0\, .
\end{equation}
Taking $y_1(x)$ as $y_1(x)= (2\beta +1) x+\alpha$,  and 
$\alpha , \beta \in \mathbb{R}$, we encounter  a family of 
polynomials which 
we denote by  $\left\lbrace R^{(\beta ,\alpha )}_{n}(x)\right\rbrace$.
These are the Romanovski polynomials. 
The corresponding weight function 
(also mentioned  before in Eq.~(\ref{wafu_Rom})) reads 
\begin{equation}\label{WeightFunctionRomanovsi}
\omega^{(\beta ,\alpha )}(x)=(1+x^{2})^{\beta - \frac{1}{2}}
e^{-\alpha \tan^{-1}(x)}.
\end{equation}
For $\beta =-p+1/2$, and $\alpha =-q$ Eq.~(\ref{WeightFunctionRomanovsi})
coincides with Eq.~(\ref{wafu_Rom}).
The latter polynomials have special orthogonal properties 
and are studied in greater detail in the next chapter.

The lowest  Romanovski polynomials and for the toy values $\beta =-1$,
and $\alpha =1$ are:
\begin{eqnarray}
&&R^{(-1, 1)}_{0}(x)=1, \nonumber\\
&&R^{(-1, 1)}_{1}(x)=-1-9x, \nonumber\\
&&R^{(-1,1)}_{2}(x)=-6+16x+56x^{2},\nonumber\\
&&R^{(-1, 1)}_{3}(x)=16+84x-126x^{2}-210x^{3},\nonumber\\
&&R^{(-1, 1)}_{4}(x)=20-240x-360x^{2}+480x^{3}+360x^{4}. 
\end{eqnarray}
\begin{figure}
\center
\includegraphics[width=7.5cm]{
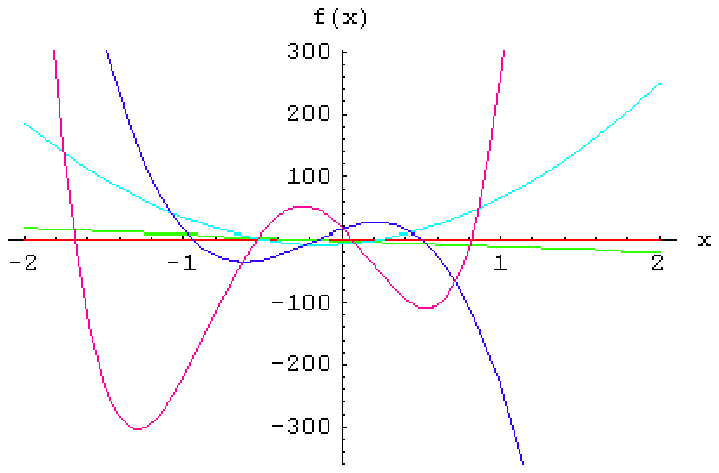}
\caption{Romanovski polynomials for $\alpha =1, \beta =-1$.}
\label{Romanovski}
\end{figure}
Some of them are shown in Fig.~\ref{Romanovski}.
In summary, the polynomial solutions of the
generalized  hypergeometric equation that 
permit for a  Rodrigues representation
fall into five classes.  Three of them 
correspond to the celebrated classical orthogonal polynomials 
of Jacobi, Laguerre, and  Hermite (recall that 
Gegenbauer, Chebyshev I and II, and
 Legendre appeared as special cases of the Jacobi polynomials). 
The principal characteristics of those five classes are collected 
in  Table ~\ref{FirstTable}.

\begin{table}
\centering
\begin{tabular}{||r|r|r|r|r|r||}  \hline
Notion& Symbol & Weight Function: $\omega (x)$ & Density: $\sigma(x)$ & 
Interval &Parameters\ \\ \hline \hline
 Generalized Hermite    &$ H^{(\alpha)}_{n}(x)$ & $e^{-\alpha x^{2}}$   & 
$1$  & $(-\infty,\infty)$ &                \ \\ \hline
 Generalized Laguerre  & $L^{(\alpha,\beta)}_{n}(x)$  &  
$x^{\beta}e^{-\alpha x}$    & $x$     & $[0,\infty)$  & \ \\ \hline
 Jacobi	& $P^{(\alpha,\beta)}_{n}(x)$ &  $(1-x)^{\alpha}(1+x)^{\beta} $   & 
$(1-x^{2}) $&  [-1,1] & $(\alpha,\beta >-1)  $   \ \\ \hline
 Legendre   & $P_{n}(x)$ &  $1$    & $(1-x^{2})$ & [-1,1]  
      &  \ \\ \hline
 Gegenbauer   & $C^{(\lambda)}_{n}(x)$ &  $(1-x^{2})^{\lambda-\frac{1}{2}}$ &
$(1-x^{2})$ & [-1,1] &$(\lambda >-1/2)  $   \ \\ \hline
Chebyshev I  & $T_{n}(x)$ &  $(1-x^{2})^{-\frac{1}{2}}$    & $(1-x^{2})$ & 
[-1,1]        &  \ \\ \hline
Chebyshev II  & $U_{n}(x)$ &  $(1-x^{2})^{\frac{1}{2}}$    & $(1-x^{2})$ & 
[-1,1]        &  \ \\ \hline
Generalized Bessel  & $y^{(\alpha,\beta)}_{n}(x)$ &  
$x^{\alpha}e^{-\frac{\beta}{x}} $    & $x^{2}$ &complex &  \ \\ \hline
Romanovski &$R^{(\beta, \alpha )}_{n}(x)$&$(1+x^{2})^{\beta - \frac{1}{2}}
e^{-\alpha \tan^{-1}(x)}$&$(1+x^{2})$ &$(-\infty,\infty)$&$-\beta >n$ 
\\ \hline
\end{tabular}
\caption{Comparison among the different families of polynomials. }
\label{FirstTable}
\end{table}


\chapter{Orthogonal polynomials in quantum mechanics.}

As already mentioned in the introduction,
the classical orthogonal polynomials shape the  exact
solutions  of a variety of quantum mechanical potentials. 
This is so because the  Schr\"odinger equation   
with one of those potentials can be 
transformed into Eq.~(\ref{hyperg_EQ}) 
upon an appropriate change of the variables.
This section is devoted to the one-dimensional
Schr\"odinger equation (which equally well can be the radial
part of an $S$ wave  $3D$ problem). 
In the following, we shall assign each
polynomial in Table ~\ref{FirstTable} to 
a potential that is exactly solvable in terms of that very polynomial.

\section{The Schr\"odinger equation with a radial potential.}

Two particles  within  a potential that  
depends only  on the relative distance, $\boldsymbol{r}$,
\begin{equation}
\label{potentialSchRadial}
V(\boldsymbol{r}_{1},\boldsymbol{r}_{2})=V(\boldsymbol{r)},
\quad 
\boldsymbol{r}=\vert \boldsymbol{r}_{1} - \boldsymbol{r}_{2}\vert , 
\end{equation}
with $\boldsymbol{r}_{1}$, and $\boldsymbol{r}_{2}$ being
the respective coordinates of first and second  particle, 
are described by a time-independent Schr\"odinger equation 
of  the form of a single particle within the center of mass frame,
\begin{equation}
-\frac{\hbar^{2}}{2\mu}\nabla^{2}\Psi(\boldsymbol{r})+
V(\boldsymbol{r})\Psi(\boldsymbol{r})=E\Psi(\boldsymbol{r}),
\end{equation}
where $\mu$ stands for the reduced mass. 
This equation can be solved by standard techniques like
separation of the variables, in which case
 the solution factorizes into a radial function, $u(r)$, 
and an angular part, $\Theta (\theta)\Phi (\varphi )$,
which for central potentials is given by the spherical harmonics,
$\Theta (\theta)\Phi (\varphi )=Y_{lm}(\theta,\varphi)$, i.e.
\begin{equation}
\label{solProduct}
 \Psi(\boldsymbol{r})=u(r)Y_{lm}(\theta,\varphi).
\end{equation}
In substituting the latter wave function in the three-dimensional
Schr\"odinger equation, amounts to 
the well known one-dimensional radial equation
\begin{equation}
\label{SchrodingerEnu}
-\frac{\hbar^{2}}{2\mu}\frac{1}{r^{2}}\frac{d}{dr}\left( r^{2}\frac{d}{dr}u(r)
\right) +\frac{\hbar^{2}}{2\mu}\frac{l(l+1)}{r^{2}}u(r)+V(r)u(r)=Eu(r).
\end{equation}
In making the coordinate and the wave functions dimensionless upon rescaling, 
\begin{equation}\centering	
z \equiv \frac{r}{d}, \qquad	u(r)\equiv\frac{\psi(z)}{d^{3/2}},
\end{equation}
where $d$ is an appropriate length scale,
Eq.~(\ref {SchrodingerEnu}) takes the form
\begin{equation}
\label{SchrodingerEnPsi}
-\frac{1}{z^{2}}\frac{d}{dz }\left( z^{2}\frac{d}{dz}\psi(z)\right)
+\frac{l(l+1)}{z^{2}}\psi(z)+v(z)\psi(z)=\epsilon\psi(z).
\end{equation}
Here,
\begin{equation}
v(z)\equiv V(r)/(\hbar^{2}/2\mu d^{2}),
\quad 
\epsilon \equiv E/(\hbar^{2}/2\mu d^{2}),
\label{DefEpsilon}
\end{equation}
and the normalization is defined  by the integral 
\begin{equation}
\int_{0}^{\infty} z^{2} \psi(z)\psi^{*}(z)dz=1.
\end{equation}
In terms of the new variable, $R(z)$, introduced via
\begin{equation}
\label{DefPsi-R}
\psi(z) \equiv \frac{R(z)}{z},
\end{equation}
Eq.~(\ref {SchrodingerEnPsi}) becomes
\begin{equation}
\label{SchrodingerEnR}
-\frac{d^{2}}{dz^{2} }R(z)+\left( \frac{l(l+1)}{z^{2}}+v(z)-
\epsilon\right) R(z)=0.
\end{equation}
With that the normalization integral changes to 
\begin{equation}
\int_{0}^{\infty} R(z)R^{*}(z)dz=1.
\end{equation} 
Casting the three-dimensional Schr\"odinger equation
into the form of Eq.~(\ref {SchrodingerEnR}),
brings two important advantages. 
The first one is that all variables are dimensionless,
which allows for an easier mathematical treatment.
The second advantage is that  Eq.~(\ref{SchrodingerEnR})
takes the form of $1D$ 
Schr\"odinger equation with an effective potential 
defined as
\begin{equation}
v_{1d}(z)=v(z)+\frac{l(l+1)}{z^{2}}.
\end{equation}
Finally, not to forget the $R(0)=0$ boundary condition, telling that the wave 
function has to vanish at the origin if it is not to penetrate the 
centrifugal barrier. 
In this fashion,  the $3D$  problem has been replaced by
an equivalent $1D$  problem.

\section{Orthogonal polynomials in exact wave functions.}

The Schr\"odinger wave functions, $\psi (z)$,
in diagonalizing an Hermitian
differential operator, are known to constitute a complete  
orthogonal set.
One can try to reduce the radial
Schr\"odinger equation (written in
 the variable $z$) to an
appropriate polynomial equation (written in the variable $x$) by means of 
the following change of variables: 
\begin{equation}
z=f(x),\quad R \left(f(x)\right)=g(x),\quad
g(x)=\sqrt{ 
\omega^{
(\alpha,\beta)
}(x)  
}\,\,
F_n^{(\alpha ,\beta) }(x)\, 
\frac{1}{
\sqrt{  
\frac{df(x)}{dx}
} }.
\label{substi}
\end{equation}
If this substitution turns out to be successful, the  differential equation 
for $F^{(\alpha ,\beta )}_n(x)$ will be a version of the generalized 
hypergeometric
equation whose solutions have been presented in the
previous chapter.
 When expressed in terms of polynomials,
the orthogonality of the Schr\"odinger wave functions 
translates into orthogonality between the involved polynomials
according to
\begin{eqnarray}
\int _{0}^{+\infty }R_n(z)R_{n^\prime }(z)dz
&=& \int_{l_1}^{l_2}g_n(x)g_{n'}(x)df(x)\nonumber\\
&=&\int_{l_1}^{l_2} \omega ^{(\alpha, \beta ) }(x)F_n^{(\alpha ,\beta )}(x)
F_{n^\prime }^{(\alpha, \beta ) }(x)dx.
\label{ortho_gonality}
\end{eqnarray} 
 
The parameters $\alpha$, and $\beta $ are some functions of the
potential parameters. Occasionally, the price of casting the
Schr\"odinger equation in the form of Eq.~(\ref{hyperg_NU}) 
is that $\alpha $ and $\beta $
acquire an $n$-dependence. In such cases, 
as we shall see below, Eq.~(\ref{substi}) may not hold valid and
the orthogonality of the wave functions may not amount to
the orthogonality integral of the polynomials with free parameters.
In the following we employ the idea of Eq.~(\ref{substi}) 
in order to solve the Schr\"odinger equations 
for a variety of potentials.
We show that the Hermite polynomials shape the solutions 
of the $1D$ oscillator, the  Laguerre polynomials define the
wave functions of a particle within the $3D$-oscillator-- and the
Coulomb wells. The Jacobi polynomials solve exactly the hyperbolic
Rosen-Morse (Rosen-Morse II)  potential, while the 
$\sim e^{\alpha r}$ barrier is
solved in terms of the Bessel polynomials.
The results listed in
Table 1.1 above have been obtained pursuing same path.

\section{Hermite polynomials and the harmonic oscillator.}
The simple harmonic oscillator in quantum mechanics is a very well known 
example elaborated  in all the standard textbooks. 
Its solutions are given in terms of the Hermite 
polynomials. In order to see this we write down the $1D$  
Schr\"odinger equation, which reads:
\begin{equation}
\label{onedimHO}
-\frac{\hbar^{2}}{2m}\frac{d^{2}}{dy^{2}}\phi(y)+V(y)\phi(y)=E\phi(y).
\end{equation}
Here, $m$ is the particle's mass, $V(x)$ stands
for the potential, and $E$ 
is the energy of the system. 
The simple  harmonic oscillation potential is given by 
\begin{equation}
V_{HO}(y)=\frac{1}{2}m\omega^{2}y^{2},
\end{equation}
where $\omega$ is the frequency of the oscillations in classical mechanics. 
Substitution of the latter equation into Eq.~(\ref {onedimHO}) 
gives
\begin{equation}
\frac{d^{2}}{dy^{2}}\phi(y)+\frac{2m}{\hbar^{2}}\left(E-\frac{1}{2}m
\omega^{2}y^{2}\right) \phi(y)=0. 
\end{equation}
The two parameters of the problem ($m,\omega$) provide a unit of 
length for the problem:
\begin{equation}
d\equiv \sqrt{\frac{\hbar}{m \omega}}, \quad z=\frac{y}{d},
\quad \phi (y=zd)=\frac{\psi{(z)}}{\sqrt{d}}\, .
\end{equation}
In performing a substitution of the form in Eq.~(\ref{substi}) 
and given by
\begin{equation}
z=f(x)=x, \quad  \psi(z=x)=g(x), \quad
g(x)=N_{n}
\sqrt{
e^{-x^2}}
H\left(x\right)\,,
\end{equation}
where $N_{n}$ is a normalization constant, 
a new equation is obtained (after substitution and reordering):
\begin{equation}
\frac{d^{2}}{dx^{2}}g (x)+\left( \epsilon - x^{2}\right) g (x)=0,
\end{equation}
where $\epsilon \equiv \frac{2md^{2}}{\hbar^{2}}E$ now is dimensionless 
one finds
\begin{equation}
H''(x)-2xH'(x)+(\epsilon -1) H(x)=0,
\label{hermi}
\end{equation}
which is the Hermite equation
\begin{equation}
H''(x)-2xH'(x)+\lambda H(x)=0,
\end{equation}
with $\lambda \in \mathbb{R}$ a constant to be determined. This equation 
admits a polynomial solution of degree $n$, $H_{n}(x)$, only if 
$\lambda=2n$. Since we are interested in such solutions, we conclude
\begin{equation}
\epsilon_n-1=2n, \qquad n \in\left\lbrace 0,1,2, ...\right\rbrace,
\end{equation}
i.e.
\begin{equation}
E_n=\frac{\hbar^{2}}{md^{2}}(n+\frac{1}{2})=\hbar\omega(n+\frac{1}{2}).
\end{equation}
The orthogonality integral between the wave functions amounts to the
orthogonality integral of the Hermite polynomials
\begin{equation}
\int_{\-\infty}^\infty \psi_n(z)\psi_{n^\prime}(z)dz=
\int_{-\infty}^\infty N_nN_{n^\prime}
e^{-\frac{z^2}{d^2}}H_n\left( \frac{z}{d} \right)
H_{n^\prime}\left( \frac{z}{d} \right) d\left(
\frac{z}{d}\right)=\delta_{nn^\prime}.
\end{equation}
\begin{figure}
\center
\includegraphics[width=7 cm]{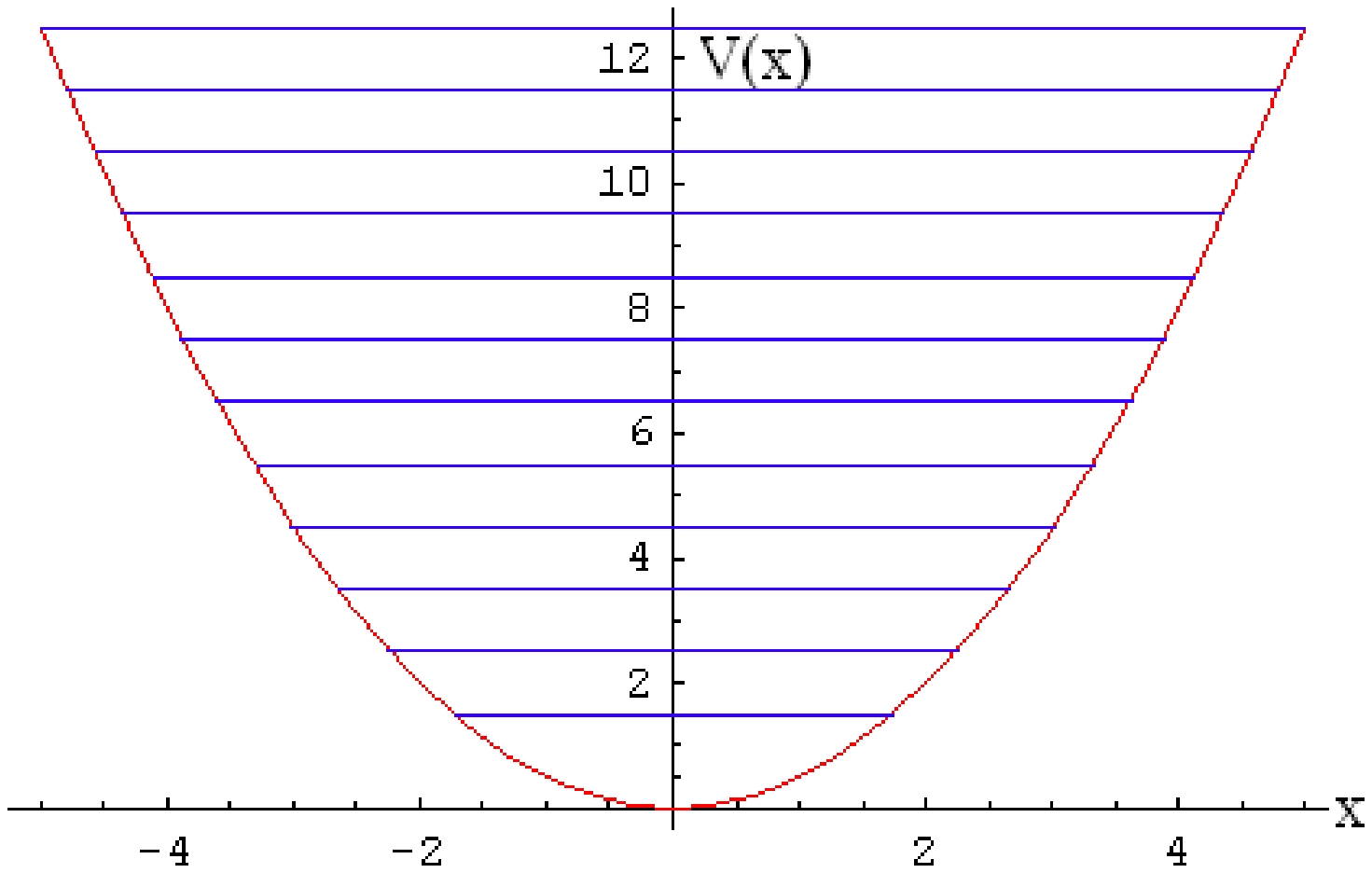}
\caption{The simple $1D$ harmonic oscillator, 
$V(x)=\frac{1}{2} m \omega^{2} x^{2}$. Here 
$m\omega=1$.\label{FigHOscillator1d}}
\end{figure}
The harmonic oscillator potential finds application in the
description of vibrational modes in nuclei,
atoms, molecules, and crystal lattices. 
This potential is shown in Figure~\ref{FigHOscillator1d}.

\section{Laguerre polynomials in  $3D$ oscillator and Coulomb wells.}
\subsection{The $3D$ oscillator.}

\begin{figure}[b]
\center
\includegraphics[width=7 cm]{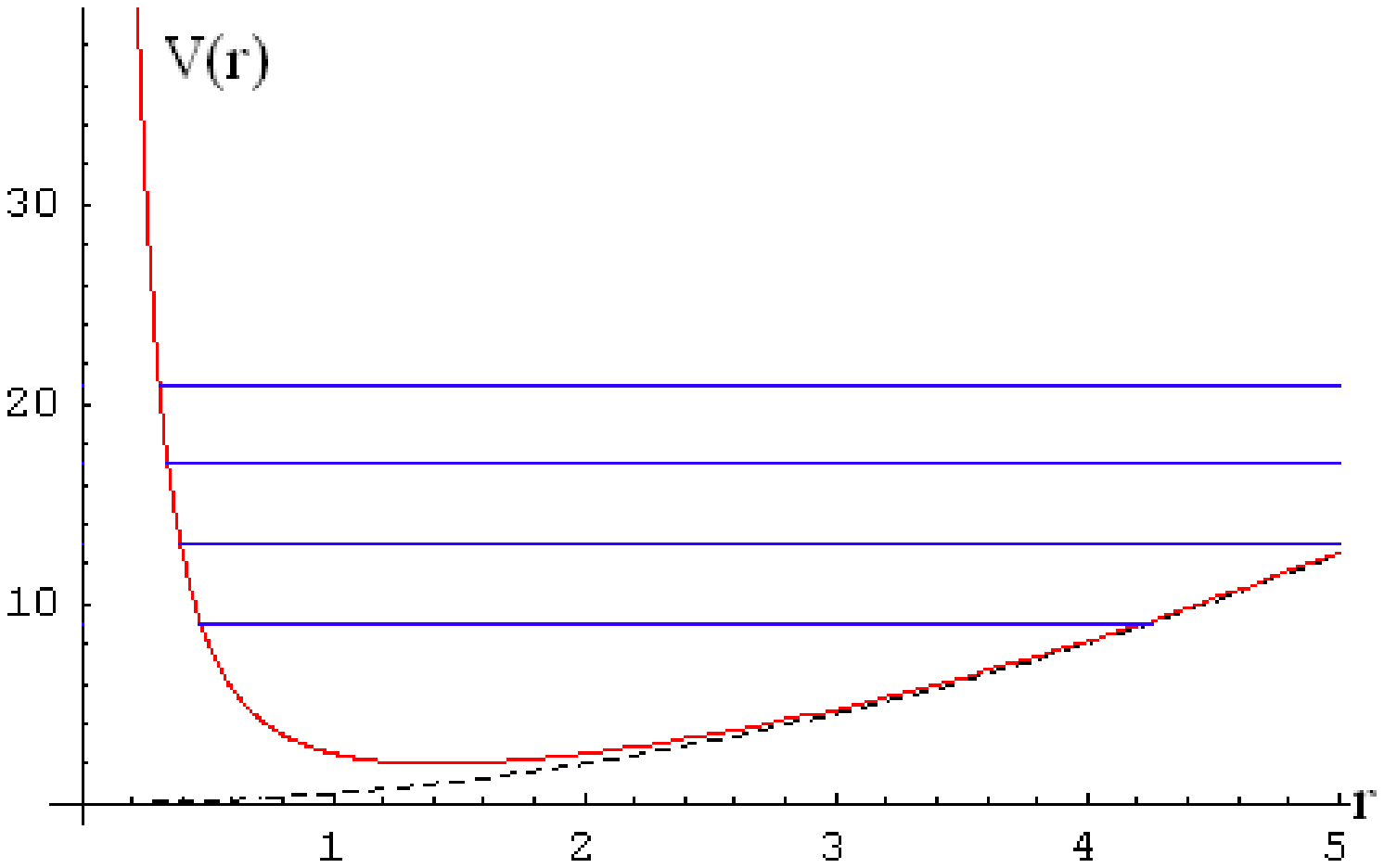}
\caption{Three-dimensional harmonic oscillator potential (dashed line), 
$V(r)=\frac{1}{2} m \omega^{2} r^{2}$. Effective potential (solid line), 
$V_{eff}(r)=\frac{1}{2} m \omega^{2} r^{2}+\frac{l(l+1)}{r^{2}}$ which 
includes the centrifugal barrier. Here $m\omega=1$, and $l=1$.
\label{FigHOscillator3d}}
\end{figure}
Harmonic oscillations in three dimensions lead to a differential 
equation whose solutions are the associated Laguerre polynomials. One 
way to solve this problem is to consider the three dimensional oscillation 
as the result of oscillators in $x$, $y$, and $z$ directions. 
The solutions can then be expressed in terms of
the  product of three Hermite polynomials. 
However, given the rotational symmetry of the problem, the
most natural coordinate choice are the spherical coordinates
in which case the complete  solution factorizes in 
a radial function and spherical harmonics in
accord with Eq.~(\ref{solProduct}). 
The  $3D$ oscillator potential reads
\begin{equation}
V(r)=\frac{1}{2}m\omega^{2}r^{2},
 \end{equation}
and is shown in Fig.~\ref{FigHOscillator3d}.
Changing variables to
$z= r/d $ and substituting into 
Eq.~(\ref {SchrodingerEnPsi})  results in
\begin{equation}
\label {SchrodingerEnPsi2}
\left( -\frac{d^{2}}{dz^{2}}-\frac{2}{z}\frac{d}{dz}+
\frac{l(l+1)}{z^{2}}+z^{2}\right) \psi(z)= \epsilon \psi(z).
\end{equation}
 The latter equation has to have well behaved asymptotic solutions
at the origin and infinity. For $z\longrightarrow \infty$
it reduces to
\begin{equation}
\left( -\frac{d^{2}}{dz^{2}}+z^{2}\right) \psi(z)= \epsilon \psi(z),
\end{equation}
meaning that  $\psi \approx e^{-\frac{z^{2}}{2}}$ would be 
a good approximation. 
Next, we inspect the behavior of Eq.~(\ref{SchrodingerEnPsi2}) 
at the origin,   $z\rightarrow 0$, where it reduces to
\begin{equation}
\label{3dHOztoZero}
\left( -\frac{d^{2}}{dz^{2}}-\frac{2}{z}\frac{d}{dz}+
\frac{l(l+1)}{z^{2}}\right) \psi(z)=0.
\end{equation}
Taking  $\psi(z)=z^{t}$ as a test function,  we find that it would
be a solution for $t=l$, or, $t=-l-1$. 
The former value is the correct one since it is finite at the origin. 
Now we shall find the exact expression for the solution applying the 
standard techniques we used before.\\
The above asymptotic behaviors suggest to try as an ansatz 
\begin{equation}
\label{sust3dHarmonic}
z\psi (z)=R(z)=K_n z^{\beta+\frac{1}{2}}e^{-\frac{z^{2}}{2}}F(z),
\end{equation}
where $K_n$ is a normalization constant. Insertion of 
the latter expression  into Eq.~(\ref{SchrodingerEnPsi2})
leads to
\begin{equation}
(-1+2z^{2}-2\beta)F'(z)-zF''(z)+\left( \frac{1+4l+4l^{2}-4\beta^{2}}{4z}+
z(2+2\beta-\epsilon)\right) F(z)=0,
\end{equation}
where $\beta$ is a free parameter of a choice suitable for  
simplifying the resulting equation.
Next we notice that in terms of the new variable, $x$, introduced as
$z=f(x)=\sqrt{x}$, the ansatz in  Eq.~(\ref{sust3dHarmonic}) 
can be converted to  $R (\sqrt{x})\equiv g(x)=
N_n\,  x^{\frac{\beta}{2}+\frac{1}{4}}e^{-\frac{x}{2}}L(x)$
with $F(z=\sqrt{x})=L(x) $, and $N_n$ a new normalization constant. 
This expression is of the type in Eq.~(\ref{substi})
in so far as
\begin{equation}
g(x)=
N_n \sqrt{x^{\beta }e^{-x}}L(x) 
/\sqrt{\frac{d}{dx}\sqrt{x}},                               
\end{equation}
where one recognizes the square root of the weight function of the associated 
Laguerre polynomials in front of $L(x)$. 
The difference between $N_n$ and $K_n$ accounts for possible
constants emerging from the inverse of the derivative of $f(x)$.
In effect, one arrives at 
\begin{equation}
xL''(x)+(\beta+1-x)L'(x)+\left( \frac{4\beta^{2}-4l(l+1)-1}{4x}+
\frac{\epsilon}{4} -\frac{\beta}{2}-\frac{1}{2}      \right)L(x)=0. 
\label{Lag_1}
\end{equation}
Now we make use of the freedom in $\beta$ to nullify the singularity by setting
$\beta= l+ \frac{1}{2}$ and thus, we are left with the equation for the 
associated Laguerre polynomials, equation (\ref {lagRaposo}). Therefore
\begin{equation}
\frac{\epsilon}{4} -\frac{\beta}{2}-\frac{1}{2}=n,
\quad
4\beta^{2}-4l(l+1)-1=0,
\label{Lag_2}
\end{equation}
allows to identify Eq.~(\ref{Lag_1})  with the
equation for the associated Laguerre polynomials,
\begin{equation}
xL''_{n}(x)+(l+\frac{1}{2}+1-x)L'_{n}(x)+nL_{n}(x)=0,
\label{Lag_3} 
\end{equation}
whose solutions are $L_{n}(x)= L^{\left( l+\frac{1}{2}\right)}_{n}(x)$.
In effect,  the radial part of the Schr\"odinger wave function
for the $3D$ oscillator is given as
\begin{equation} 
g_n(x)=N_n\sqrt{x^{l+\frac{1}{2}}e^{-x}}L_n^{\left(l+\frac{1}{2}\right)}(x)
/\sqrt{\frac{d}{dx} \sqrt{x}}.
\end{equation}
Back to the $z$ variable, the Schr\"odinger wave function is now obtained in
its final form as
\begin{equation}
\psi_n(z)=\frac{R(z)}{z}=
K_n z^{l}e^{-\frac{z^{2}}{2}}L^{(l+\frac{1}{2})}_{n}(z^{2}).
\end{equation}
The orthogonality integral between the wave functions recovers the
orthogonality between  the Laguerre polynomials according to
\begin{equation}
 \int_0^\infty R_n(z) R_{n}(z)dz=
\int_0^\infty K_n K_{n^\prime}
\sqrt{x^{l+\frac{1}{2}}e^{-x}}L_n^{\left(l+\frac{1}{2}\right)}(x)
\sqrt{x^{l+\frac{1}{2}}e^{-x}}L_{n^\prime}^{\left(l+\frac{1}{2}\right)}(x)
dx =\delta_{nn^\prime}\, ,
\label{orth_3d_HO}
\end{equation}
where $dz=d\sqrt{x}$.
The energies are now given by $\epsilon_n=2(2n+l+\frac{3}{2})$.
The reason for which the associated, and not the ordinary
Laguerre polynomials  appeared in the solution of the $3D$ oscillator is
that the angular momentum, $l$, requires a polynomial parameter.

\subsection{The hydrogen atom.}

\begin{figure}[b]
\center
\includegraphics[width=7 cm]{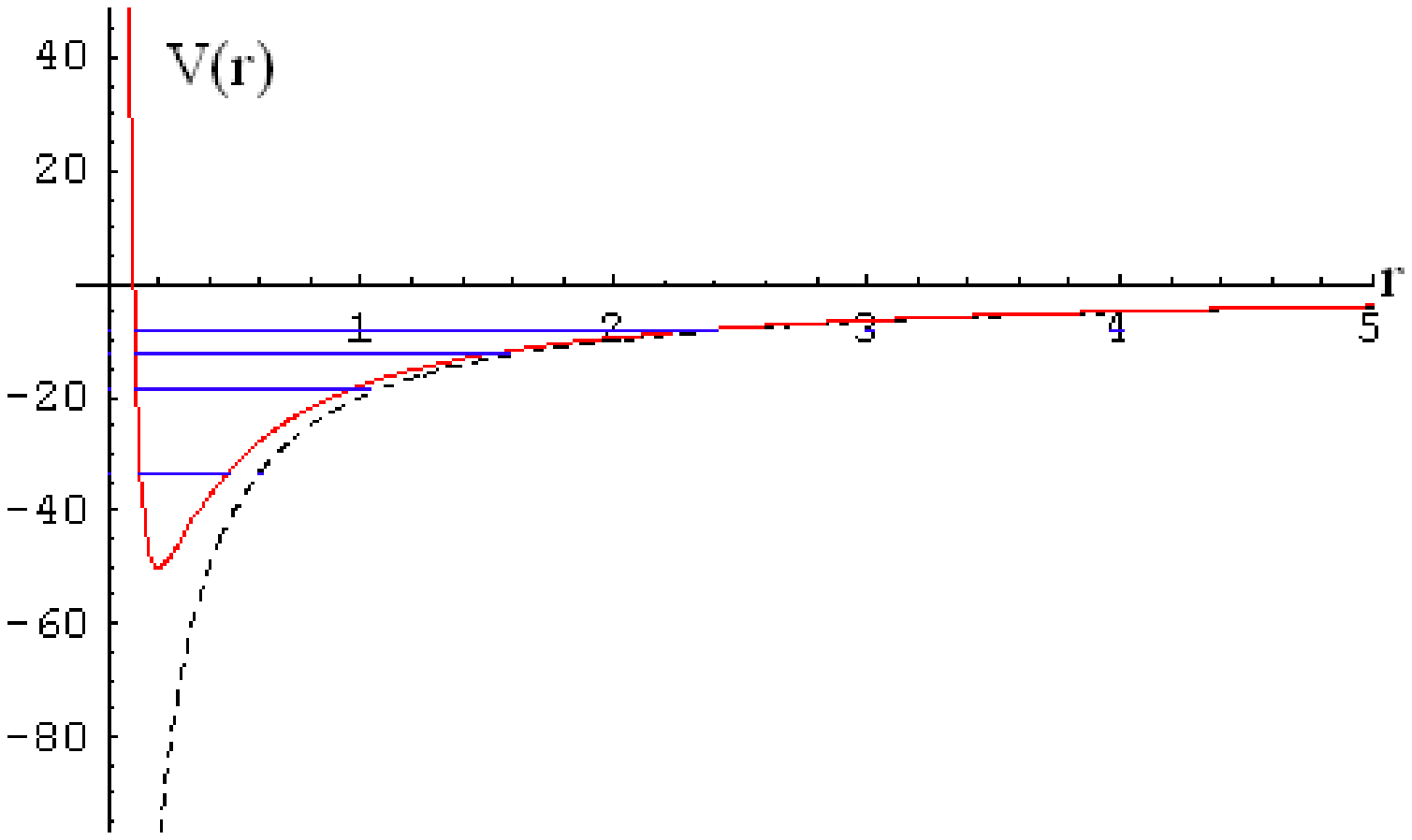}
\caption{Coulomb potential (dashed line), $V(r)=-\frac{Ze^{2}}{r}$. 
The effective potential (solid line), 
$V_{eff}(r)=-\frac{Ze^{2}}{r}+\frac{l(l+1)}{r^{2}}$   
includes the centrifugal barrier. Here $Ze^{2}=10$ and
$l=1$.\label{PotencialCoulomb}}
\end{figure}
In this subsection we  study the hydrogen atom, which is a two-body system
consisting of one proton and one electron. The interaction is governed 
by the Coulomb potential, 
\begin{equation}
\label{CoulombPotential}
V(r)=-\frac{Ze^{2}}{r}, \quad Z=1,
\end{equation}
as displayed in Fig.~\ref{PotencialCoulomb}.
Substituting the latter equation for the potential
in Eq.~(\ref{SchrodingerEnR}) leads to
\begin{equation}
\label{EqCoulomb}
-\frac{d^{2}}{dz^{2} }R(z)+\left( \frac{l(l+1)}{z^{2}}-
\frac{2}{z}-\epsilon\right) R(z)=0,\quad z=rd,
\end{equation}
where the length scale $d$, has been chosen as  the Bohr radius,
$d=h^2/(\mu  e^2)$, and $\epsilon$ stands for $\epsilon=2Ed/e^2$. 
We are going to solve this equation by means of an appropriate variable
substitution. 
In order to find it  we first study the asymptotic behavior
of the solutions at both origin and infinity.
For finite   $z$ the latter equation is equivalently rewritten as
 \begin{equation}
-z \frac{d^{2}}{dz^{2}}R(z)+\left( \frac{l(l+1)}{z}- 2 -
\epsilon z\right) R(z)=0,
\end{equation}
while near the origin, where $R(0)=0$, and $1/z<<1/z^2$ one finds
\begin{equation}
\left( \frac{d^{2}}{dz^{2} }- \frac{l(l+1)}{z^{2}}\right) R(z)\approx 0.
\end{equation}
The latter equation can be solved by  a function of the type,
$R(z)=z^{t}$.
This would restrict $t$ to  $t=l+1$, and $t=-l$, respectively.
The second solution is not acceptable because it is not finite at the origin, 
and we are  left  with the first one.
Next we study the $z\rightarrow\infty$ asymptotic solution.  In
this case Eq.~(\ref{EqCoulomb}) becomes
\begin{equation}
-\frac{d^{2}}{dz^{2}}R(z)=\epsilon R(z).
\end{equation}
 As long as we here are interested in the bound states, 
$ \epsilon < 0$, which amounts to
\begin{equation}
\frac{d^{2}}{dz^{2}}R(z)- |\epsilon| R(z)=0.
\end{equation}
The latter equation is solved by
$R(z)=Ce^{\sqrt{|\epsilon |}z}$ and 
$R(z)=Ce^{-\sqrt{|\epsilon |}z}$. 
As long as we want the wave function to go to  
zero for large $z$, the first solution has to be dismissed.
Next one changes variables in Eq.~(\ref{EqCoulomb}) 
according to 
\begin{equation}
z=f(x)=\frac{x}{\kappa  }, \quad \kappa  =2\sqrt{|\epsilon|}\, ,
\quad R\left( \frac{x}{\kappa }\right)=g(x)\, , 
\label{fx_Hydr}
\end{equation}
where the factor $2$ has been taken  
for convenience. This yields
\begin{equation}
\frac{d^{2}}{dx^{2} }g(x)+\left( - \frac{l(l+1)}{x^{2}}+\frac{1}
{\sqrt{|\epsilon|} x}-\frac{1}{4}\right) g(x)=0.
\label{x_factor}
\end{equation}
Taking into account the correct
asymptotic behavior revealed above,
the solution of the latter equation can be assumed as 
$g(x)=x^{l+1}e^{-\frac{x}{2}}L(x)$.
In result, one  arrives at
\begin{equation}
\label{EqInRs}
xL''(x)+(2l+1+1-x)L'(x)+L(x)\left( \frac{1}
{\sqrt{|\epsilon |}}-l-1\right) =0\,,
\end{equation}
which coincides with the equation (\ref{Lag_3}) for the
associated Laguerre polynomials provided $\alpha =1$, $\beta =2l+1$,
and $\lambda =n$. With this in mind 
it is easy to verify that 
\begin{equation}
\epsilon_{nl}=-\frac{1}{(n+l+1)^{2}}.
\end{equation}
Recalling the definition of 
($\epsilon = 2Ed/e^2$), we obtain
the spectrum as
\begin{equation}
\label{EnergyHydAtomWithn}
E_{nl}=-\frac{e^{4}\mu}{2\hbar^ {2}(n+l+1)^{2}}.
\end{equation}
At that stage one defines $N=n+l+1$ as the principal quantum number, 
and realizes  that while $n$ can be any non-negative integer,   
the angular momentum is restricted to a finite number of values
according to $l=N-1-n$,
\begin{equation}
l=0,1,2,..., N-1.
\end{equation}
In effect, one finds the well known degeneracy patterns in the
spectrum of the hydrogen atom,
 $E_{nl}=-\frac{e^{4}\mu}{2\hbar^{2}N^{2}}$,
meaning that $\kappa $ depends on $n$.
The solution for the radial equation ~(\ref{EqInRs}) is then
concluded as 
\begin{equation}
\frac{g(x)}{x}\equiv 
\psi_{n l}(x)=C_{n l} x^{l}e^{-\frac{x}{2}}L^{(2l+1)}_{N-l-1}(x),
\end{equation}
where use has been made of  Eq.~(\ref{DefPsi-R}).
The latter expression equivalently rewrites to
\begin{equation}
\psi_{n l}(x)=C_{n l} 
\sqrt{x^{2l+1}e^{-x}} L^{(2l+1)}_{n}(x)/\sqrt{x}\, .
\end{equation}
Therefore,  $ \psi_{nl}(x)$, is given in terms of 
the associated Laguerre polynomials,
$\left\lbrace L^{(2l+1)}_{n}(x)\right\rbrace$,
with $C_{n l}$ being the  normalization constant. 
However, the wave function under consideration is {\bf not}
of the form in Eq.~(\ref{substi}) because $\sqrt{x}\not= (x/\kappa)^\prime $. 
The consequence will be
that the orthogonality integral of the wave functions 
\begin{equation}
\int_0^\infty C_{nl}C_{n^\prime l }
\sqrt{(\kappa_n z)^{2l+1}e^{-\kappa_n z}} L^{(2l+1)}_{n}(\kappa_n z )
\sqrt{(\kappa_{n'}z)^{2l +1}e^{-\kappa_{n'}z}} 
L^{(2l +1)}_{n^\prime}(\kappa _{n'}z)
z^2\frac{dz}{z}=\delta_{nn^\prime}\,,
\label{first_sign}
\end{equation}
{\bf does not coincide with the orthogonality integral of the
Laguerre polynomials with free parameters\/}  because 
it contains the additional 
first power of $z$. It is important to be aware of the fact
Eq.~(\ref{first_sign})  describes orthogonality between states bound
within {\bf different} potentials 
corresponding to different factors,
$\frac{Ze^{2}}{2\, \sqrt{|\epsilon_n|}}$ versus 
$\frac{Ze^{2}}
{2\, \sqrt{|\epsilon_{n^\prime}|}}$,  of $1/x$ in Eq.~(\ref{x_factor}).
In the textbook on Mathematical
Methods in Physics by Arfken and Weber (second reference in
~\cite{MMF}) this phenomenon has been also attributed to the 
dependence of $\kappa $ on the degree of the polynomial via the
energy (see Exercise 13.2.11 there). As we shall see below,
such a behavior is much more general and can occur also when
$x$ is neat but the parameters carry
an $n$ dependence.

\section{Jacobi polynomials in  Rosen-Morse II.}

\begin{figure}[b]
\center
\includegraphics[width=7 cm]{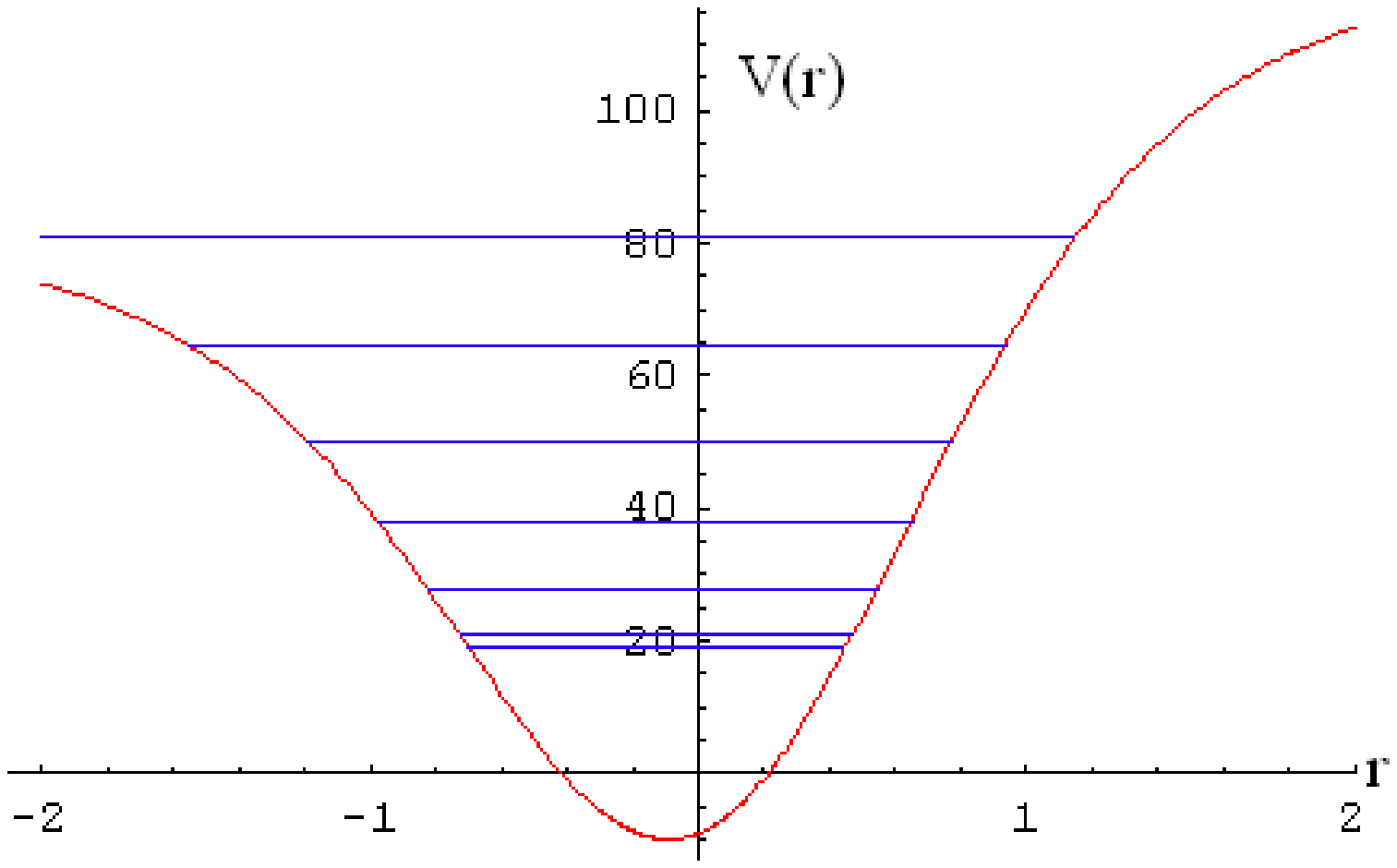}
\caption{The Rosen Morse II (hyperbolic) potential, 
$V(r)=a^{2}+\frac{b^{2}}{a^{2}}-a(a+1) 
\mathrm{sech^{2}}(r)+2b \mathrm{tanh}(r)$. 
Here $a=10$ and $b=10$.\label{Rosen-MorseII}}
\end{figure}
The Jacobi polynomials appear in several physics problems
ranging from classical electrodynamics to quantum mechanics. 
We here focus on the hyperbolic Rosen-Morse potential given by
\begin{equation}
\label{rosen-morseHyper}
v(z)=a^{2}+\frac{b^{2}}{a^{2}}-a(a+1)\,
 \mathrm{sech^{2}}(z)+2b\,\,  \mathrm{tanh}(z),
\end{equation}
and displayed in Fig.~\ref{Rosen-MorseII}.
The corresponding $1D$ Schr\"odinger equation is
\begin{equation}
\frac{d^{2}R(z)}{dz^{2}}+\left( -\frac{b^{2}}{a^{2}}+
a(a+1)(1- \mathrm{tanh^{2}}(z))-2 b\,  \mathrm{tanh}(z)+e \right) R(z)=0,
\end{equation}
with $e=a^{2}-\epsilon$, and  $ \mathrm{sech^{2}}(z)=1- \mathrm{tanh^{2}}(z)$.
Changing variable to $x= \mathrm{tanh}(z)$
meaning $z=f(x)$ with $f(x)=\mathrm{tanh}^{-1}(x)$,results in
\begin{equation}
(1-x^{2})\frac{d^{2}g(x)}{dx^{2}}-2x\frac{dg(x)}{dx}+
\left( -\frac{b^{2}}{a^{2}(1-x^{2})}+a(a+1)-
\frac{2bx+e}{1-x^{2}}\right) g(x)=0,
\end{equation}
with $g(x)$ being defined as $g(x)=R(\mathrm{tanh}^{-1}x)$.
By means of the substitution, 
 $g(x)=(1+x)^{
\frac{\beta}{2}
}(1-x)^{
\frac{\alpha}{2}}
P(x)$, 
the latter equation becomes
\begin{eqnarray}
&&(1-x^{2})\frac{d^{2}P(x)}{dx^{2}}  + \left( \beta - \alpha -
x(2+\alpha+\beta) \right) \frac{dP(x)}{dx}+ \nonumber\\ 
&&\left(a(a+1)-\frac{\alpha(1+\beta)}{2}-\frac{\alpha^{2}}{4}-
\frac{\beta}{2}-\frac{\beta^{2}}{4}+\frac{x(2b-\frac{\alpha^{2}}{2}+
\frac{\beta^{2}}{2})+(\frac{b^{2}}{a^{2}}-\frac{\alpha^{2}}{2}-
\frac{\beta^{2}}{2}-e)}{x^{2}-1}\right) P(x)=0,\nonumber\\
\label{RMI-Schr}
\end{eqnarray}
where $\alpha,\beta$ are free parameters to be used to simplify the 
equation. Specifically, one makes use of the freedom in 
$\alpha,\beta$ to nullify 
the singular term which restricts the parameters to:
\begin{equation}
2b-\frac{\alpha^{2}}{2}+\frac{\beta^{2}}{2}=0,
\end{equation}
\begin{equation}
\frac{b^{2}}{a^{2}}-\frac{\alpha^{2}}{2}-\frac{\beta^{2}}{2}-e=0.
\end{equation}
Next one requires the constant multiplying $P(x)$
to be of the form
\begin{equation}
a(a+1)-\frac{\alpha(1+\beta)}{2}-\frac{\alpha^{2}}{4}-
\frac{\beta}{2}-\frac{\beta^{2}}{4}=\lambda_n=n(1+n+\alpha+\beta).
\end{equation}
The latter equations are resolved b
\begin{equation}
\beta=a-n+\frac{b}{n-a}\equiv \mu_n,
\end{equation}
\begin{equation}
\alpha=a-n-\frac{b}{n-a}\equiv \nu_n,
\end{equation}
and
\begin{equation}
e_n=\frac{b^{2}}{a^{2}}-(a-n)^{2}-\frac{b^{2}}{(a-n)^{2}}.
\end{equation}
With that, equation (\ref{RMI-Schr}) can be identified with the
Jacobi form of the generalized hypergeometric  equation,
\begin{equation}
(1-x^{2})\frac{d^{2}P(x)}{dx^{2}}+\left( \alpha - 
\beta  -x(2+\alpha+\beta) \right) \frac{dP(x)}{dx}+
n(n+\alpha+\beta+1)P(x)=0,
\end{equation}
whose solutions are the Jacobi polynomials 
$\left\lbrace P^{(\alpha,\beta)}_{n}(x)\right\rbrace$.
In effect, the hyperbolic Rosen-Morse potential is solved exactly by
\begin{equation}
g_n(x)=\sqrt{(1-x)^{\mu_n}
(1+x)^{\nu_n}}
P_n^{(\mu_n, \nu_n )}(x),
\end{equation}
and in accord with Table 1.1 (when translated to our notations).
This wave function is {\bf not\/} of the form in Eq.~(\ref{substi}). 
As a consequence,
the orthogonality integral between the wave functions 
{\bf does not recover the orthogonality between the Jacobi polynomials
with free parameters \/}
as visible from
\begin{eqnarray}
\int_{-\infty}^{+\infty}N_nN_{n^\prime}R_n(z)R_{n^\prime}(z)dz
&=&
\int_{-1}^{+1}
\sqrt{(1-x)^{ \mu_n}
(1+x)^{\nu_n}}P_n^
{(\mu_n, \nu_n)}(x)\nonumber\\
&&\sqrt{(1-x)^{\mu_{n'}}(1+x)^{\mu_{n'}}}
P_{n'}^
{ (\mu_{n'} ,\nu_{n'})}(x)
d\, \mathrm{tanh}^{-1} ( x)\, .
\end{eqnarray}
The culprits for this are  the $n$ dependent polynomial parameters. 
This is not to remain the only example for such an anomalous behavior.

\section{The Bessel polynomials in spherical waves phenomena.}

\begin{figure}[b]
\center
\includegraphics[width=7 cm]{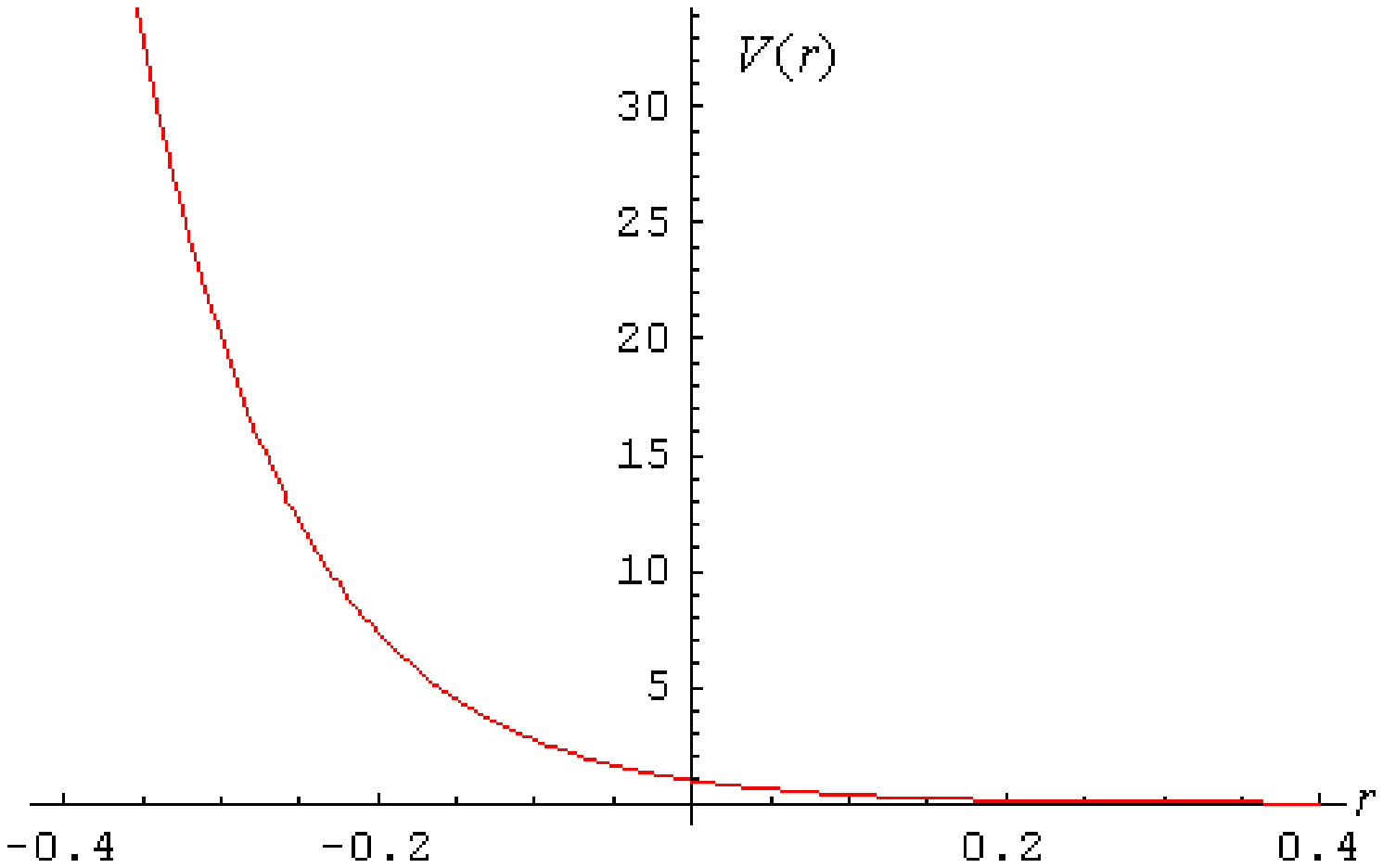}
\caption{A barrier $V(r)=Ae^{ar}$ that leads to a 
Schr\"odinger equation with solutions in terms of Bessel polynomial's
and for the toy values  $a=-10$, and $A=1$.\label{FigBarrier}}
\end{figure}

The Bessel polynomials are special in the sense that their orthogonality is 
achieved by  integration over a contour in the complex plane.
In the context of quantum mechanics an infinite barrier 
(Fig.~\ref{FigBarrier}) of the type $Ae^{az}$ ($A,a$ being constants)
 will lead to Bessel's differential equation. Using this barrier in 
Eq.~(\ref{SchrodingerEnR}) for the case $l=0$ results into
\begin{equation}
-\frac{d^{2}R(z)}{dz^{2}}+\left( \epsilon + Ae^{az}\right) R(z)=0.
\end{equation}
Changing variables to  $x=e^{-az}$ leads to
\begin{equation}
x^{2}\frac{d^{2}g(x)}{dx^{2}}+x\frac{dg(x)}{dx}-
\left( e+\frac{b}{x}\right) g(x)=0,
\end{equation}
\begin{equation}
e=\frac{\epsilon}{a^{2}} \qquad b=\frac{A}{a^{2}}.
\end{equation}
A substitution of the type $g(x)=e^{-\frac{1}{x}}x^{\frac{1}{2}}y(x)$ 
leads to 
\begin{equation}
x^{2}\frac{d^{2}y(x)}{dx^{2}}+2(x+1)\frac{dy(x)}{dx}+
\left( \frac{1-4e}{4}+\frac{4-4b}{4x^{2}}\right) y(x)=0.
\label{step_itrm}
\end{equation}
If the singular term is to vanish, then 
\begin{equation}
4-4b=0,
\end{equation}
\begin{equation}
\frac{1-4e}{4}=-n(n+1).
\end{equation}
From the above conditions we find $b=1$, $e=\frac{1}{4}+n(n+1)$
and Eq.~(\ref{step_itrm}) becomes the  Bessel differential equation:
\begin{equation}
x^{2}\frac{d^{2}y(x)}{dx^{2}}+2(x+1)\frac{dy(x)}{dx}=n(n+1)y(x).
\end{equation}
Here, $y(x)$ stand for the Bessel polynomials.
Take notice that contrary to the previous examples the barrier potential 
is exactly solvable only when the parameters have been fixed by the 
condition $\frac{A}{a^{2}}=1$. This is an example for a potential that is 
exactly solvable when the parameters obey constraints.
The Bessel polynomials have been brought to attention
by  Krall and Frink \cite{Krall} 
and are related to the following wave equation 
(in spherical coordinates)
\begin{eqnarray}
&&\frac{1}{r^{2}}{\Big(} r\frac{\partial^{2}}{\partial r^{2}}
(ru(r, \theta, \varphi, t))+\frac{1}{\sin(\theta)}
\frac{\partial}{\partial\theta}\left( \sin(\theta)
\frac{\partial u(r, \theta, \varphi, t)}{\partial\theta}\right)\nonumber\\
 &+&
\frac{1}{\sin^{2}(\theta)}\frac{\partial^{2}u(r, \theta, \varphi, t )}
{\partial\varphi^{2}}{\Big)} 
=\frac{1}{c^{2}}\frac{\partial^{2}u(r, \theta, \varphi, t)}{\partial t^{2}}.
\end{eqnarray}

If the latter equation  is solved by separation of variables, 
the radial part $f(r)$ is found to satisfy the 
following differential equation:
\begin{equation}
\label{eqBesselWave}
r^{2}\frac{d^{2}f(r)}{dr^{2}}+2r\frac{df(r)}{dr}+k^{2}r^{2}f(r)=n(n+1)f(r).
\end{equation}
Here $kc=\omega$, and $e^{i\omega t}$ is a plane wave.
For  $r=ks$, Eq.~(\ref{eqBesselWave}) becomes
\begin{equation}
\label{besselEqWaveEq}
s^{2}\frac{d^{2}\widetilde{f}(s)}{ds^{2}}+2s
\frac{d\widetilde{f}(s)}{ds}+s^{2}\widetilde{f}(s)=n(n+1)
\widetilde{f}(s), \quad \widetilde{f}(s)=f\left(r=\frac{s}{k}\right). 
\end{equation}
The latter equation can be transformed into Bessel's equation by 
the substitution $f(s)=s^{-\frac{1}{2}}J(s)$, which yields:
\begin{equation}
\label{BesselFunctionsEq}
s^{2}\frac{d^{2}J(s)}{ds^{2}}+s\frac{dJ(s)}{ds}+s^{2}J(s)=(n+1/2)^{2}J(s).
\end{equation}
For integer $n$, the solutions of Eq.~(\ref{BesselFunctionsEq}) 
are the Bessel functions of  half-integral order and are
well known.
However, Eq.~(\ref{besselEqWaveEq}) can also be subjected to the 
transformation $f(s)=w(s)/s$, giving
\begin{equation}
\label{BesselEqReal}
s^{2}\left( \frac{d^{2}w(s)}{ds^{2}}+w(s)\right) =n(n+1)w(s).
\end{equation}
Upon introducing the variable, $z=ikr=is$,  imaginary and 
real parts of the solution represent 
traveling waves. Because of that we admit in 
Eq.~(\ref{BesselEqReal})  $z=is=ikr $,
and $w(s)=e^{-z}y(z)=e^{-is}y(is)$, and obtain
\begin{equation}
\label{EqBesselCompleja}
z^{2}\left( \frac{d^{2}y(z)}{dz^{2}}-2\frac{dy(z)}{dz}\right) =n(n+1)y(z).
\end{equation}
For integer values of $n$, Eq.~(\ref{EqBesselCompleja}) has 
solutions which are polynomials in $1/z$. Therefore the final 
substitution should be, $x=1/z=1/ikr$, which allows to obtain
the differential equation for the Bessel polynomials 
$\left\lbrace y_{n}(x)\right\rbrace$,
\begin{equation}
x\frac{d^{2}y_n(x)}{dx^{2}}+(2x+2)\frac{dy_n(x)}{x}=n(n+1)y_n(x).
\end{equation}
 The full solution of the spherical  wave equation is then
given by
\begin{equation}
\label{solutionWaveEq}
u(r, \theta, \varphi, t)=r^{-1}P^{m}_{n}(cos(\theta)) 
\sin(m\varphi-\alpha) e^{i(\omega t -kr)} y_{n}(1/ikr),
\end{equation}
where $y_{n}(x)=y_{n}(1/ikr)$ is a Bessel polynomial, and $kc=\omega$. 
The real and imaginary parts of~(\ref{solutionWaveEq}) describe waves 
traveling in the radial direction with velocity $c$. In conclusion, 
spherical waves are equivalently
described either in terms of a class of  
polynomials orthogonal over the unit circle,
or in the standard way in terms of Bessel functions.

\section{Romanovski polynomials in  Rosen-Morse I.}
As already mentioned in the introduction, the Table 1.1 contains
one more potential whose exact solutions require the Romanovski polynomials
and  this is the trigonometric Rosen-Morse potential (Rosen-Morse I). 
This case has been considered in great detail in Refs.~\cite{CK},\cite{Cliff}
and will not be repeated here. Instead, I would prefer to briefly
review  the main properties of its solutions
within the context of its relevance in quark physics (the presentation on this
section closely follows Ref.~\cite{Raposo}).
This aspect appears especially  important to me because
the Romanovski polynomials were found for the first time in the solutions of
that very potential while searching to construct a quark model that
matches reality on  nucleon excitations as part of
the research project ``Dynamics of baryon resonances''
run by our group.
The subject of the present thesis continues the study of the Romanovski
polynomials started in Refs.~\cite{CK,Cliff} and
extends knowledge on  their properties by various new observations.

The great appeal of Rosen-Morse I is that
\begin{quote}
{\it besides the Coulomb potential, the  trigonometric Rosen-Morse potential 
is to the best of our knowledge the only exactly solvable
 potential that relates to a fundamental massless gauge theory.\/}
\end{quote} 
\begin{quote}
{\tt Recall that the Coulomb potential is no more but the 
image in coordinate space of the propagator, $-1/\mathbf{q}^2$,
 of the photon,
the U(1) gauge boson of elec\-tro\-dynamics,
as it appears in elastic scat\-ter\-ing of charged par\-tic\-les.
In a similar way, as it will be argued below,
Rosen-Morse I can be viewed as the image in co\-or\-di\-na\-te space of 
the propagator of the gluons,
the  $SU(3)_c$ gauge fields of  the fun\-da\-men\-tal field theory of 
strong int\-er\-act\-ions, the
\underline{Q}uantum \underline{C}hromo\underline{D}ynamics (QCD),
as it appears in elastic scattering of quarks.
Within this context, the Romanovski polynomials acquire the special
status of major ingredients of the wave functions of bound quarks.\/}
\end{quote}

\noindent
The quarks, as is well known, are the constituents of strongly
interacting particles, baryons, and mesons.
In what follows, the presentation  will be focused on the
baryons constituted by the so called light flavors,
$u$, $d$, and $s$, as are the nucleon 
$(N)$, the $\Delta$,  the $\Lambda$, and their resonant 
excitations \cite{PART}.
The nucleon is understood as a particle which can exist in two
different states distinguished by their
electric charges, the proton, $p (uud)$, and the neutron, $n(udd)$,
and is said to be a charge-doublet.
The $\Delta $ stands for a particle that can exist in four
different states distinguished by their  charges which are
$\Delta^{++}(uuu)$, $\Delta^+(uud)$, $\Delta^0(udd)$, and $\Delta^-(ddd)$ 
and is termed to as a charge-quadruplet.
The $\Lambda (uds)$ particle is neutral
and  a charge singlet. There are also charge triplets like 
$\Sigma $ represented by $\Sigma^+(uus)$,$\Sigma^0(uds)$, and $\Sigma^-(dds)$, 
and one more charge-doublet given by the 
$\lbrace \Xi^{-}(dss),\Xi^0(uss) \rbrace$ family.
Compared to the $N$-, $\Delta $-, and $\Lambda$-
charge-multiplets,
the $\Sigma$ and $\Xi$ are less known and
will be left out of consideration in the following.  

\begin{figure}[b]
\center
\includegraphics[width=8.9 cm]{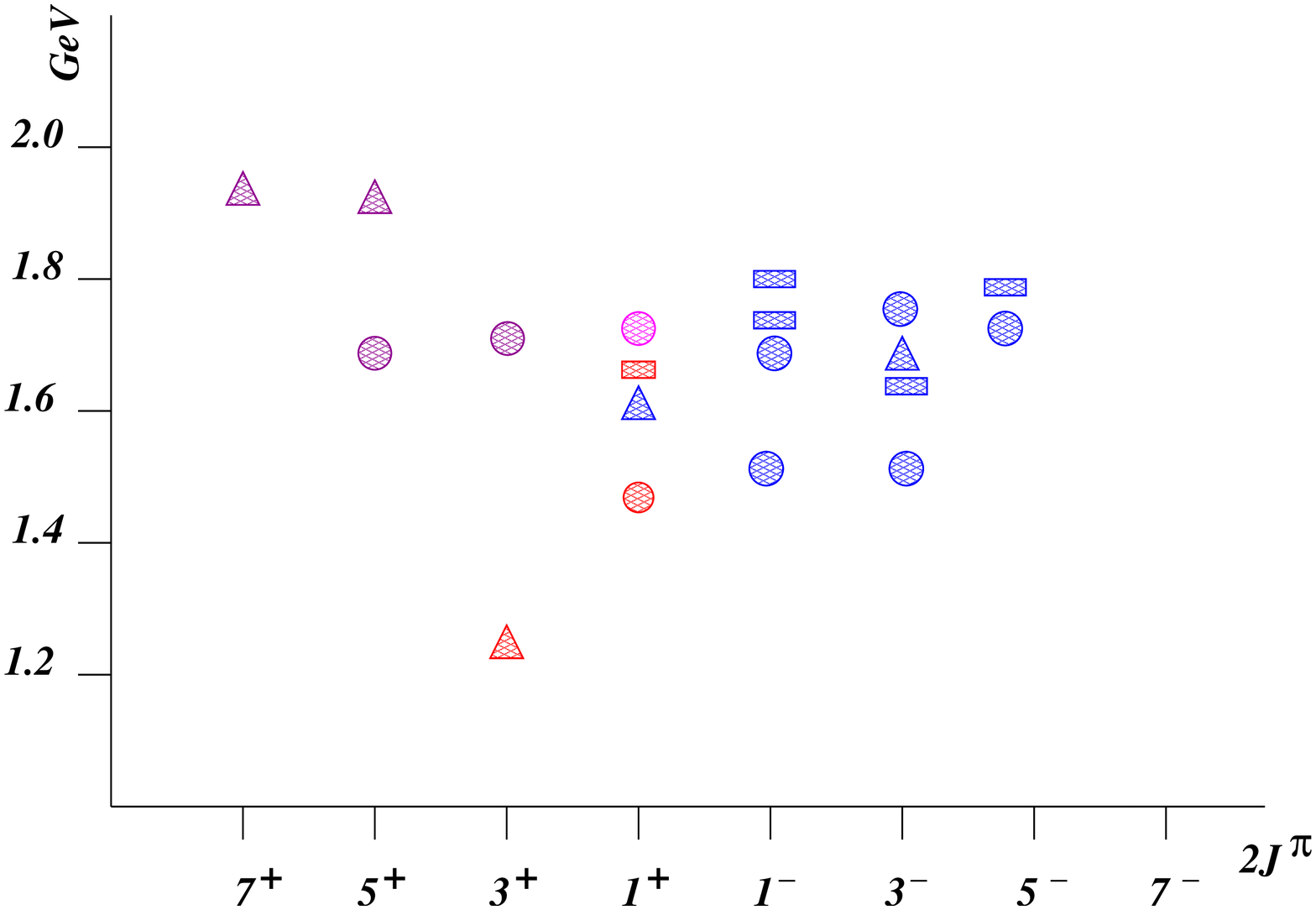}
\caption{Baryon resonances in the traditional quark model.
Circles, bricks, and triangles stand for
nucleon, $\Lambda $, and $\Delta $ states, respectively.
Different colors mark different 
$SU(6)_{SF}\times O(3)_L$ multiplets.
Notice the strong multiplet intertwining and the large
mass separation inside the multiplets.
(Courtesy M.\ Kirchbach)
\label{su6}}
\end{figure}

The problem which one is facing with the baryon resonances 
is the lack of an adequate systematics and the resulting deficits as 
the prediction of a large number of unobserved states. 
In the standard quark models, 
different charge-multiplets  are supposed to join to bigger
families, the so called $SU(6)_{SF}\times O(3)_L$ super-multiplets
as displayed in  Fig.~\ref{su6}.  
This figure reveals a strong overlap between
the different super-multiplets and an apparent  lack of degeneracy
between the states belonging to same multiplet thus questioning
the adequacy of the underlying classification scheme.
In Refs.~\cite{MK-97-2001} the $SU(6)_{SF}\times O(3)_L$
baryon classification scheme has been given up,
the super-multiplets have been decomposed into   
$N$--, $\Delta$--, and $\Lambda$--spectra and have then 
have been studied separately.
 
\begin{figure}[b]
\center
\includegraphics[width=8.9 cm]{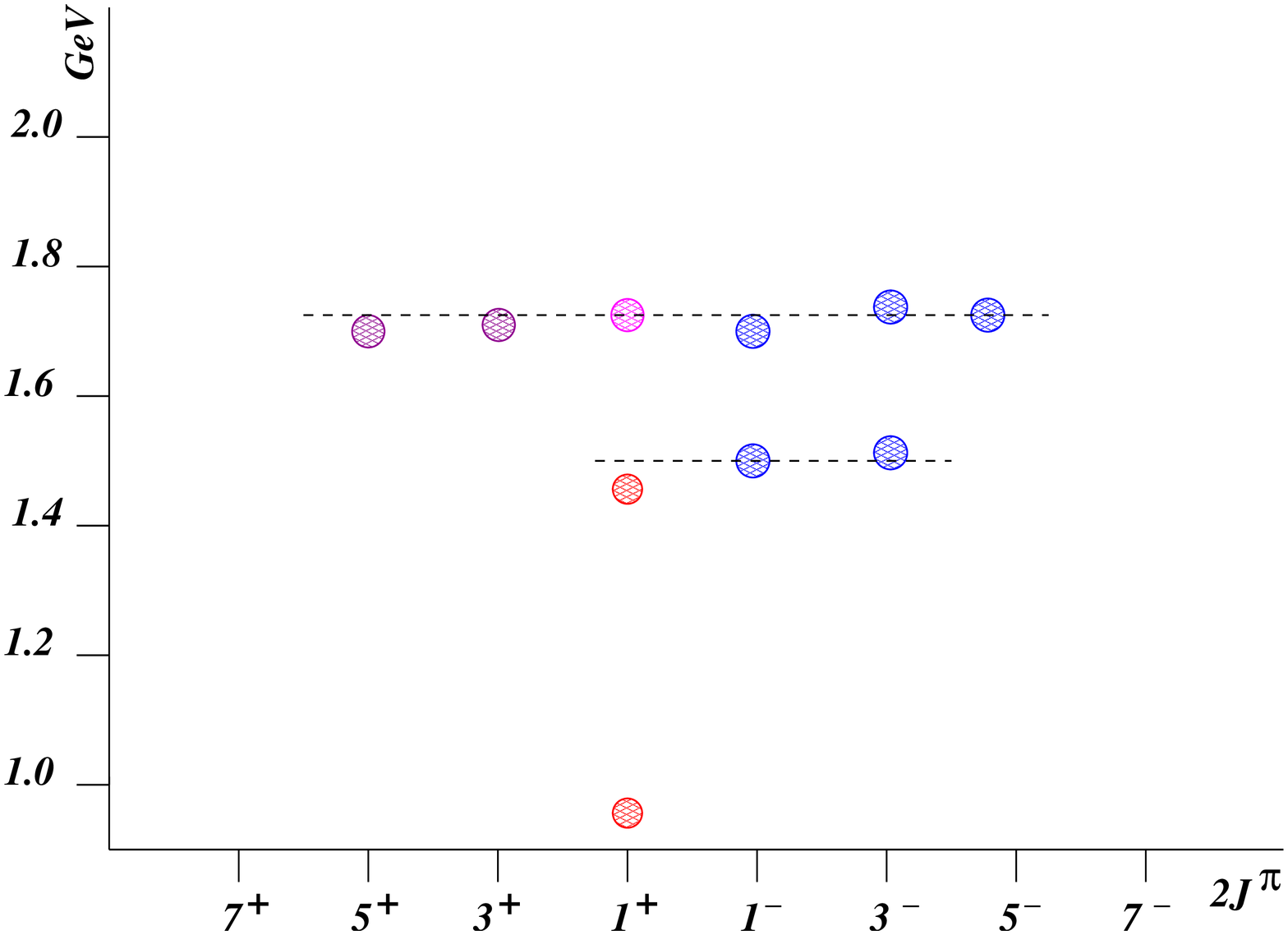}
\caption{The nucleon excitation spectrum below 2 GeV.
(Courtesy M.\ Kirchbach). 
\label{spectra_1}}
\end{figure}

\begin{figure}[b]
\center
\includegraphics[width=8.9 cm]{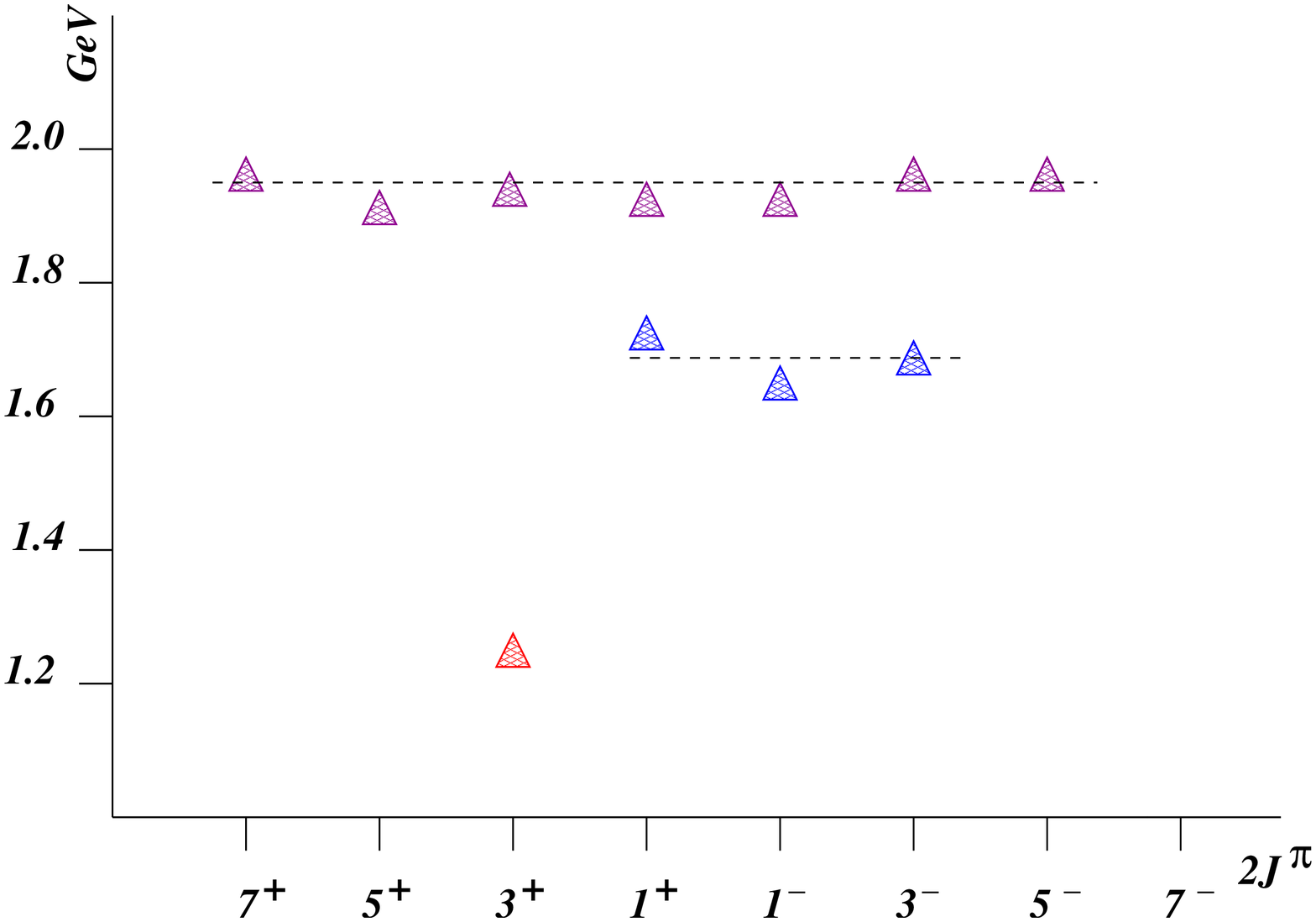}
\caption{The  $\Delta $ excitation spectrum below 2 GeV. 
(Courtesy M.\ Kirchbach)
\label{spectra_2}}
\end{figure}

The result was that to a surprisingly
good accuracy, the nucleon excitation levels carry the same degeneracies 
as the  levels of the electron with spin in the hydrogen atom,
though the splittings of the former are quite different from those
of the latter. Namely, compared to the hydrogen atom, 
the baryon level splittings contain in addition to the Balmer term 
also its  inverse but of opposite sign. Same patterns
are repeated by the 
excitation spectrum of the $\Delta (1232)$ particle, the most 
important baryon after the nucleon 
(see Figs.~\ref{spectra_1},\ref{spectra_2}). 
In this way baryons have been classified according
to $SU(2)_I\times O(4)$ with $I$ standing for isospin
(a number that encodes the dimensionality, $D$, of  the
charge multiplets as $D=2I+1$).
The appeal of the new classification scheme  
lies in the fact that no states drop out of the systematics, on the one side,
and that the number of unobserved (``missing'' )
states predicted by it is significantly
smaller than those of all preceding schemes.

\noindent
The observed degeneracies in the spectra of the light quark
baryons have been attributed in Ref.~\cite{MK-97-2001} to the dominance of
a quark--antiquark configuration in baryon structure. Within the light of 
these findings, the form of the potential in configuration space acquires 
crucial importance. 
\begin{quote}
{\it In Refs.~\cite{CK},\cite{CK-1} a case was made that
precisely the trigonometric Rosen-Morse potential provides the degeneracies
and level splittings that are required by the light quark baryon spectra.\/}
\end{quote}
The success of the trigonometric Rosen-Morse potential in quark
physics is not accidental. It is due to the property of the 
latter to interpolate between the Coulomb- and the infinite
well potentials. In order to understand this virtue one has
to recall the QCD  basics.

\begin{figure}
\begin{center}
\includegraphics[width=70mm,height=70mm]{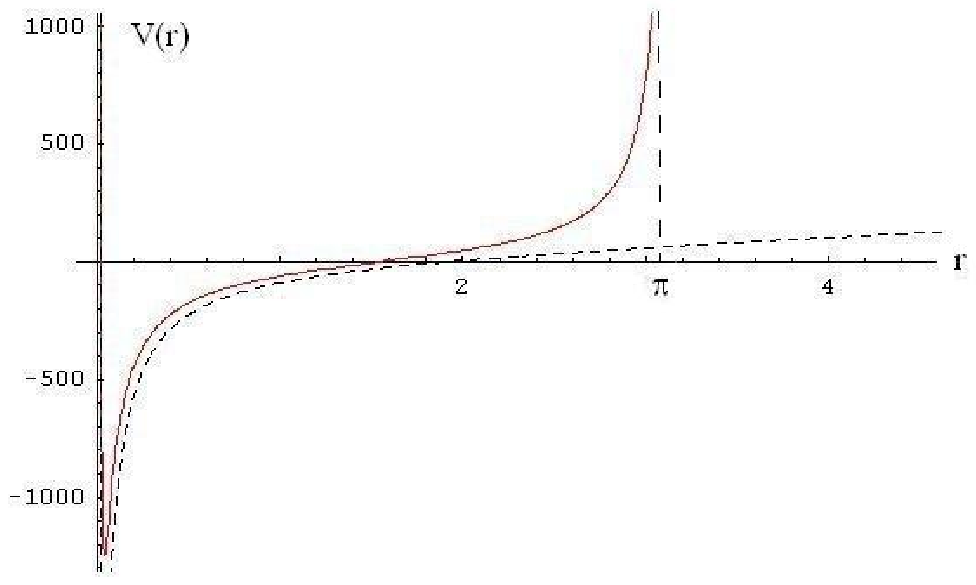}
\caption{ The trigonometric Rosen-Morse potential (solid line) and its
proximity to the Coulomb-- plus lineal potential 
as predicted by lattice QCD (thin dashed line)  for the
toy values $l=1, b=50$ of the parameters.}
\label{fig-RMt}
\end{center}
\end{figure}

Indeed, the strong interaction between quarks within QCD
is governed by interchangings of
massless gauge bosons between quarks of three different ``colors'' , 
(``red'', ``blue'',  ``green''). The above 
gauge theory  predicts that the quark interactions proceed over  
one- or many gluon exchanges including gluon self-interactions
(so called ``non-Abelian'' gauge theory). 
The latter are believed to be 
responsible for the so-called quark confinement, where highly energetic quarks 
remain trapped but behave as (asymptotically) free at small distances. 
The QCD equations are nonlinear and complicated due to the gluonic
 self-interaction processes. Their solution requires employment of highly 
sophisticated techniques, such as discretization of space time, so-called 
lattice QCD. Lattice QCD calculations of the properties of hadrons
associates the one-gluon exchange with a Coulomb like, $\sim 1/r$,
potential, and predicts  a linear confinement 
potential with increasing energy brought about by
gluon-self interactions.
The trigonometric Rosen-Morse potential in the parametrization of
Ref.~\cite{CK-1} where the $a$ parameter has been identified with
$l$, the relative quark--di-quark angular momentum
(in units of $\hbar^2=1=2\mu $, and $d$
a suited length scale making the $z$ variable dimensionless),
\begin{equation}
 v_{tRM}(z)=-2 b \cot (z) +l(l+1)\frac{1}{\sin^2 (z)}\, ,\quad z=\frac{r}{d},
\label{v-RMt}
\end{equation}
has precisely the properties required by lattice QCD.
It captures the essential traits of the QCD quark-gluon dynamics
in interpolating between the Coulomb potential
(associated with the one-gluon exchange) and the infinite well potential
(associated with the trapped but asymptotically free quarks) while
passing through a linear confinement region (as predicted by lattice QCD) 
(see Fig.~\ref{fig-RMt}).
In order to see this one has only  to perform the 
Taylor expansion of the potential of interest,
\begin{equation}
v(z)_{tRM}\approx -\frac{2b}{z} +
\frac{2b}{3}\, z + \frac{l(l+1)}{z^2}\,+ \frac{l(l+1)}{15}z^2\, +...
\label{Taylor_lin_HO}
\end{equation}
This expansion clearly reveals the proximity of the $\cot(z)$ term to the
Coulomb-- plus linear confinement potential, and the proximity
of the $\csc^2(z)$ term to the standard centrifugal barrier.
\begin{quote}
{\it In this sense, Rosen-Morse I can be viewed as the image of 
space-like gluon propagation in coordinate space.\/}
\end{quote}
The great advantage of the trigonometric Rosen-Morse potential
over the linear-- plus Coulomb potentials is that while the latter is
neither especially symmetric, nor exactly soluble, the former is both,
it has the dynamical $O(4)$ symmetry (as the hydrogen atom) and is
exactly soluble.
The exact solutions of the, now three dimensional,
Schr\"odinger equation with $v_{tRM}(z)$ from Eq.~(\ref{v-RMt})
have been constructed in \cite{CK-1} on the basis of the one-dimensional
solutions found in \cite{CK} and read:
\begin{equation}
\psi_n(\cot^{-1} x)=
(1+x^2)^{-\frac{n+l+1}{2}}e^{-\frac{b}{n+l+1}\cot^{-1}( x)}
C_n^{\left( -(n+l) ,\frac{2b}{n+l+1} \right)}(x)\, ,
\label{R_z}
\end{equation}
with $x=\cot (z)$. The $C$ polynomials from \cite{CK} are Romanovski
polynomials but with running parameters attached to the degree of 
the polynomial.
The notations of Ref.~\cite{CK} translate to the present ones as:
\begin{equation}
C_n^{(-(n+l),\frac{2b}{n+l+1})}(x)\equiv R_n^{(p_n, q_n)}(x),
\quad q_n=-\frac{2b}{n+l+1}, \quad p_n=n+l+1, \quad n=0, 1,2,...
\label{identif}
\end{equation}
The wave function in Eq.~(\ref{R_z}) is {\bf not\/} of the type in
Eq.~(\ref{substi}), a reason for which the orthogonality
integral between the wave functions will {\bf not recover the
orthogonality of the Romanovski polynomials with free parameters\/} 
as visible from
\begin{equation}
\int_0^\pi  \psi_{n}(z) \psi _{n'}(z){\rm d}z\ =\int_{-\infty}^\infty
\sqrt{w^{(p_n,q_n)}(x)}
N_nR_n^{(p_n,q_n)}(x)
\sqrt{w^{(p_{n'}, q_{n'})}(x)}
N_{n^\prime}R_{n'}^{(p_{n'},q_{n'})}(x)  \frac{{\rm d}x}{1+x^{2}}=
\delta_{n\ n'},
\label{orto-1}
\end{equation}
where the factor $\frac{1}{1+x^2}$ in the integrand comes from
$\frac{{\rm d} \cot^{-1}( x) }{{\rm d}x}=-1/(1+x^2)$.
In this way, the Romanovski polynomials that enter the solutions
of the Schr\"odinger equation with the trigonometric Rosen-Morse potential
seem to disobey the finite orthogonality prescription. A discussion of
this behavior will be given below.
Finally, the associated energy spectrum is found as
\begin{equation}
\epsilon_n=(n+l+1)^2 -\frac{b^2}{(n+l+1)^2 }\, .
\label{O4_RMI}
\end{equation}
Therefore, the Romanovski polynomials have been shown in Refs.~\cite{CK,CK-1}
to be important ingredients of the wave functions of quarks designed 
in accord with QCD quark-gluon dynamics.

\chapter{Romanovski polynomials in Scarf II.}

This chapter is devoted to the presentation of the original results
obtained in the thesis.
In the first section 4.1 I present the hyperbolic Scarf potential.
The corresponding Schr\"odinger equation is solved 
in terms of the Romanovski polynomials and presented  in subsection 4.1.1.
The finite orthogonality of the Romanovski polynomials is discussed
in section 4.2.
Section 4.3 is devoted to the problem of an electron within a non-central
potential, which is solved in spherical coordinates by separation of 
the variables. There, in sub-section 4.3.2 it is shown that the 
Romanovski polynomials solve exactly the polar angle equation 
and define new non-spherical angular functions.
In subsection 4.3.3 a non-linear relationship between Romanovski
polynomials and associated Legendre functions is established.
Finally, in section 4.4 I solve the Klein-Gordon equation with 
scalar and vector  potentials of same magnitudes and given by Scarf II.

\section{The hyperbolic Scarf potential.}

\begin{figure}[b]
\center
\includegraphics[width=7 cm]{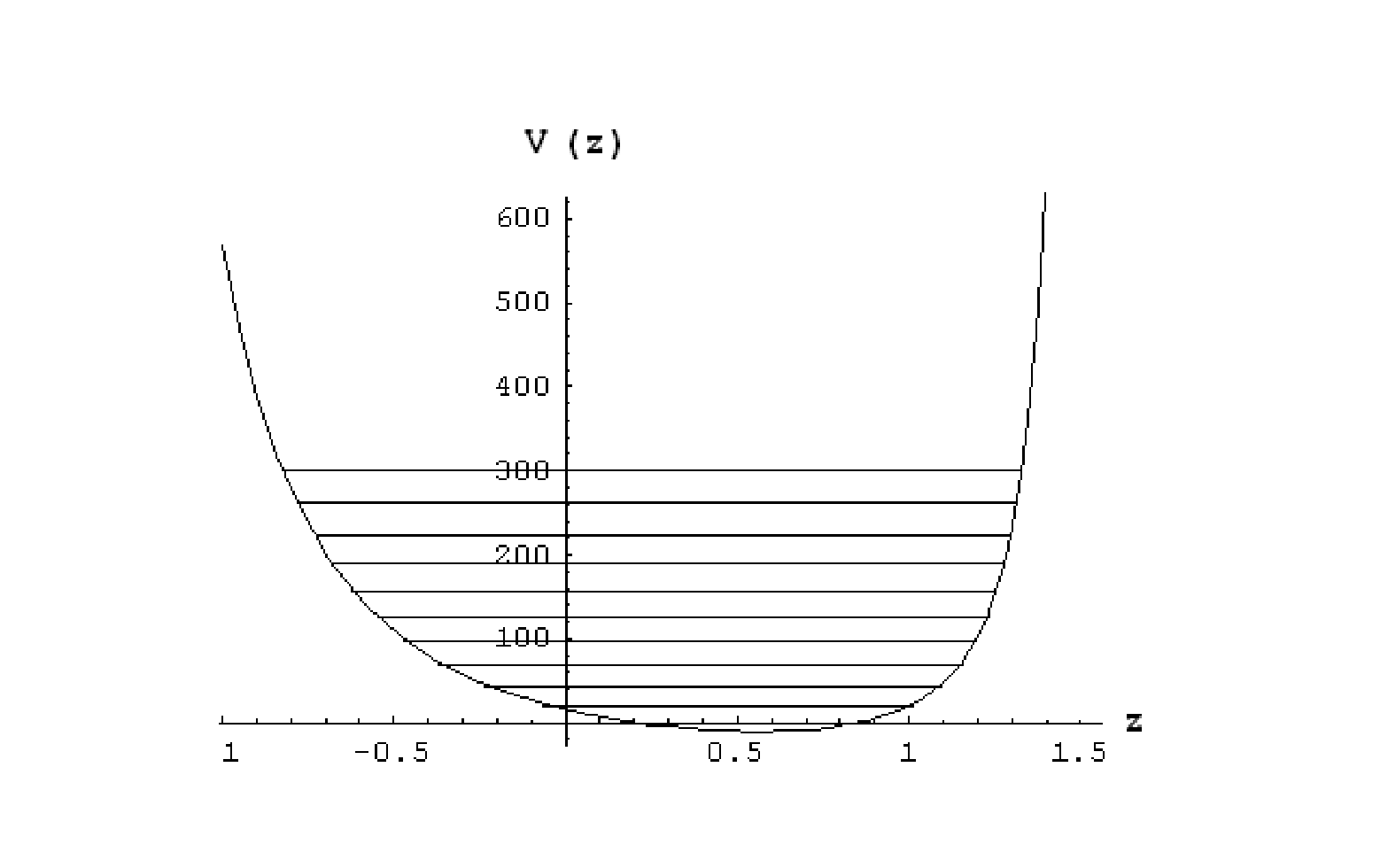}
\caption{The trigonometric Scarf potential (Scarf I) for the toy values
of the parameters, $a=10$, $b=5$, and $\alpha =1$.
The horizontal lines represent the discrete levels.
\label{ScarfiT}}
\end{figure}

\begin{figure}[h!]
\center
\includegraphics[width=10 cm]{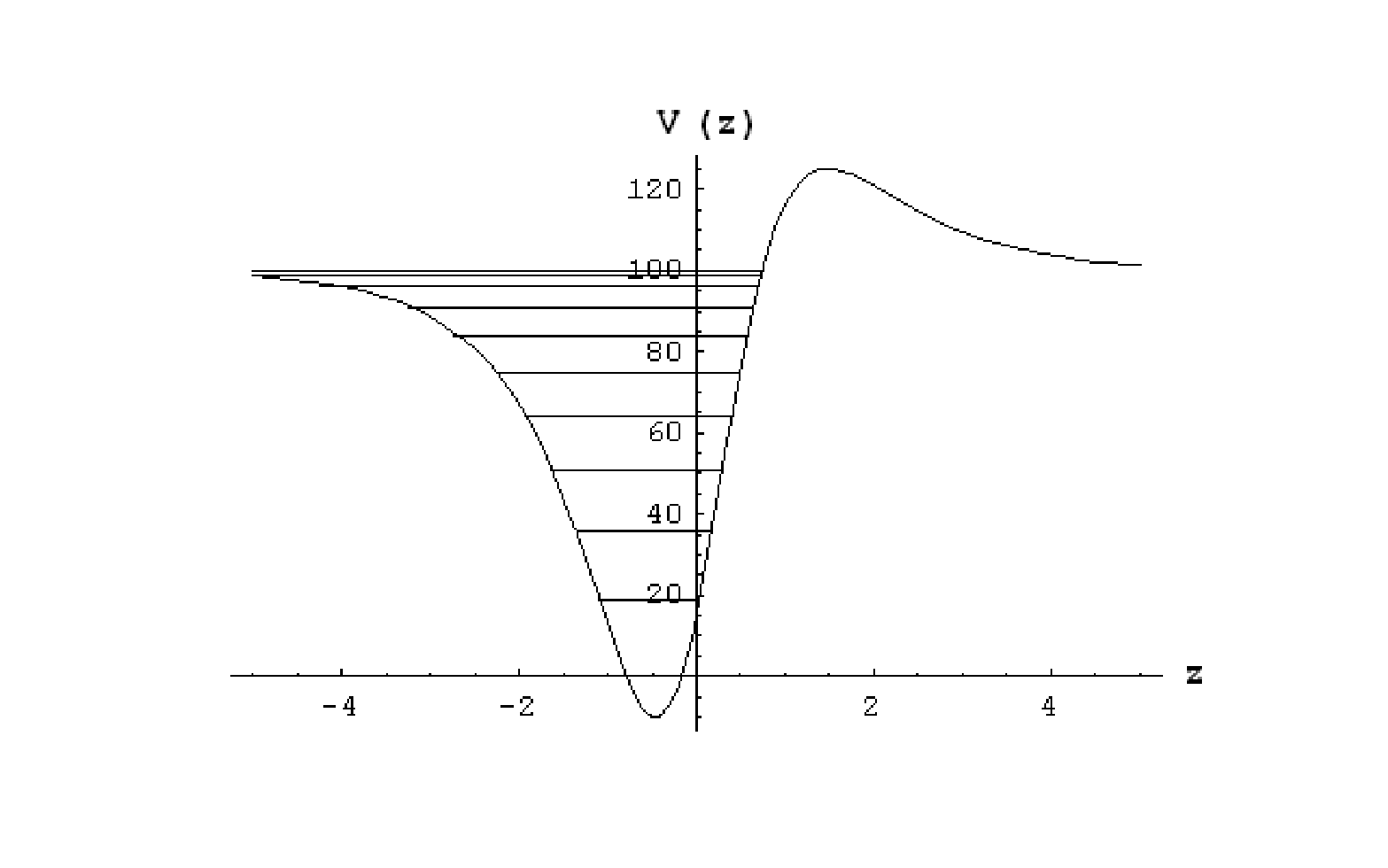}
\caption{The hyperbolic Scarf potential 
$V_{h}(z)=a^2 +(b^2-a^2-a\alpha)\mathrm{sech^2}(\alpha z)
+b(2a+\alpha)\mathrm{sech}(\alpha z)\mathrm{tanh}(\alpha z)$ with 
$a=10$, $b=5$ and 
$\alpha=1$. Energy levels, $e_n=a^2-(a-n\alpha)^2$, are included. 
\label{scarfHypPotential}}
\end{figure}
The hyperbolic Scarf potential can be viewed as the extension of the
$ \mathrm{sech^2}(r)$ potential, or, better, of the original 
P\"oschl-Teller potential
\cite{Teller}. Take notice that in the SUSYQM nomenclature 
(presented in Table 1.1) 
the names of P\"oschl and  Teller are rather associated with the 
extended $ \mathrm{csch^2}(r)$ (here marked as P\"oschl-Teller 2).
It seems that for the first time Scarf II has been constructed in 
Ref.~\cite{Dabrowska}. It also has been encountered independently 
within the framework of the supersymmetric quantum mechanics
~\cite{Khare,Levai,Bagchi} while  exploring the  superpotential 
\begin{equation}
U_m(x)=a\,\,   \mathrm{tanh}(\alpha x) +b\,\,  \mathrm{sech}(\alpha x).
\label{Scarfh_superpot}
\end{equation}
On the other side, it can equally well be
approached from the perspective of the trigonometric Scarf
potential, here denoted by $V_{t}(z)$ and given by
\begin{equation}
V_{t}(z)=-a^2 +(a^2 +b^2-a\alpha )\sec^2 (\alpha z) -
b(2a+\alpha )\tan(\alpha z)\sec(\alpha z).
\label{S_Trigo}
\end{equation}
The exact solution of the Schr\"odinger equation with the
trigonometric Scarf potential (displayed in Fig.~\ref{ScarfiT})   
can be obtained along the line of the concepts
of the previous chapter. It is well known and given in terms of the  Jacobi
polynomials, $P_n^{(\beta ,\delta) }(z)$,  as \cite{Khare},
\begin{equation}
\psi_n(z)=(1-\sin(\alpha z))^{\frac{(a-b)}{2\alpha }}
(1+\sin(\alpha  z))^{\frac{(a+b)}{2\alpha }}
P_n^{\left( (\frac{b}{\alpha}-\frac{a}{\alpha} -\frac{1}{2}),
( -\frac{b}{\alpha}-\frac{a}{\alpha}-
\frac{1}{2})\right)}(z)\, .
\label{Scarf_tr_sol}
\end{equation}

The corresponding energy spectrum is obtained as
\begin{equation}
e_n=(a+\alpha n)^2-a^2\, .
\label{Scarf_tr_en}
\end{equation}
The trigonometric Scarf potential 
can be transformed into its hyperbolic
partner, the so called {\it hyperbolic\/} Scarf potential, here
denoted by $V_{h}(z)$ and given by
\begin{equation}
V_h(z)=a^2 +(b^2-a^2-a\alpha)\mathrm{sech}^2(\alpha z)
+b(2a+\alpha)\mathrm{sech}(\alpha z)\mathrm{tanh} (\alpha z).
\end{equation}
Figure~\ref{scarfHypPotential} visualizes the hyperbolic Scarf 
potential and its discrete spectrum.

The $V_h(z)$ potential has been obtained from $V_t(z)$
in performing the following substitutions in Eq.~(\ref{S_Trigo}) :
\begin{eqnarray}
a\longrightarrow ia\, ,&\quad& \alpha\longrightarrow -i\alpha\, ,\nonumber\\
\frac{a}{\alpha}\longrightarrow -\frac{a}{\alpha},
&\quad& b\longrightarrow b\, .
\label{cmplxf}
\end{eqnarray}

Upon the above substitutions the energy changes to
\begin{equation}
e_n=a^2-(a-n\alpha)^2 .
\label{enrg_hip}
\end{equation}
In the following we shall show that $ n=0,1,2,...< a $ meaning
that the  number of bound states is {\bf finite\/}.
Yet, the most profound changes are suffered by the wave functions.
Substitution of Eqs.~(\ref{cmplxf}) into Eq.~(\ref{Scarf_tr_sol})
results in
\begin{eqnarray}
\psi_n (-i\sinh (z) )=(1+i\sinh (z) )^{-\frac{a}{2}}(
1-i\sinh (z))^{-\frac{a}{2}}&&\left( 
\frac{1-i\sinh(z)}{1+i\sinh(z)}\right)^{-\frac{b}{2}}\nonumber\\
c_nP_n^{(ib+a-\frac{1}{2}), (-ib+ a-\frac{1}{2})}(-i\sinh (z))\,,
\label{cmplx_Jac}
\end{eqnarray}
where  $c_n$ is some state dependent complex phase,
and where we took $ \alpha =1$  for simplicity.
The latter expression can be cast into the form
frequently  mentioned in the literature 
\cite{Khare},\cite{Levai}, \cite{nicolae},\cite{Hidalgo},
\begin{eqnarray}
\psi_n(-ix)&=&(1+x^2)^{-\frac{\rho }{2}}e^{-\sigma  
\tan^{-1} (x)}c_n P_n^{(ib+a-\frac{1}{2}), (-ib+ a-\frac{1}{2})}(-ix)\,,
\nonumber\\
 x&=&\sinh (z)\, ,\quad \rho  =a, \quad \sigma =b\, .
\label{cmplx_Jac_x}
\end{eqnarray}
The latter equation gives the impression that the exact solutions of the 
hyperbolic Scarf potential rely upon Jacobi polynomials with complex
indices and arguments.

\begin{quote}
In this thesis the case is made that this needs not be so and that
the above wave functions can be expressed in terms of
the real Romanovski polynomials. 
\end{quote}

\subsection{The polynomial equation.}
 
The Schr\"odinger equation for the potential of interest when rewritten in
a new  variable, $x$, introduced via an appropriate point canonical 
transformation \cite{Wipf}, \cite{De}, taken by us as
$z=f(x)=\sinh^{-1} x$, is obtained as:
\begin{equation}
(1+x^2)\frac{d^2g(x)}{dx^2}+x\frac{dg(x)}{dx}
+\left( \frac{-b^2 +a(a+1)}{1+x^2} 
-\frac{b(2a+1)}{1+x^2}x
+\epsilon_n  \right)g (x)=0,
\label{Schr_bzero}
\end{equation}
with $g(x)=\psi ( \mathrm{sinh^{-1}} (x) )$.
Inspired by Eq.~(\ref{cmplx_Jac_x})
we now  test the following substitution in Eq.~(\ref{Schr_bzero})
\begin{eqnarray}
g(x)&=&(1+x^2)^{\frac{\beta }{2}}e^{-\frac{\alpha}{2}\tan^{-1}(x)}
D^{(\beta ,\alpha )}(x)\, , \quad x= \mathrm{sinh}(z)\begin{scriptsize}
\end{scriptsize},
\label{azero_1}
\end{eqnarray}
In effect, Eq.~(\ref{azero_1}) reduces to the following equation
for $D^{(\beta,\alpha )}(x)$,
\begin{eqnarray}
(1&+&x^2)
\frac{
d^2D^{( \beta,\alpha  )}(x)}{dx^2}+((2\beta +1)x-\alpha )
\frac{dD^{(\beta ,\alpha  )}(x)}{dx}\nonumber\\
&+&\left(\beta^2 +\epsilon_n +\frac{(a+a^2+\beta -
\beta^2-b^2 +\frac{\alpha^2}{4})+x(-b -2ab 
+\frac{\alpha}{2}-\alpha\beta )}{1+x^2} 
\right)D^{(\beta ,\alpha  )}(x)=0\,.
\nonumber\\
\label{azero_2}
\end{eqnarray}
Making use of the freedom in $\alpha $ and $\beta $, the coefficient 
in front of $1/(1+x^2)$ may nullify,
\begin{eqnarray}
a+a^2-b^2+\frac{\alpha^2}{4} +\beta -\beta^2&=&0\, ,
\label{1s_cond}\\
-b-2ab+\frac{\alpha}{2}-\alpha\beta &=&0\, .
\label{2nd_cond}
\end{eqnarray}
Then, Eq.~(\ref{azero_2}) reduces to  the
Romanovski equation (\ref{Rom_pol}) .
Finally, the identification of the constants 
in Eqs.~(\ref{azero_2}) and (\ref{Rom_pol})
leads to a condition that defines the energy spectrum of the
hyperbolic Scarf potential as
\begin{equation}
\beta^2+\epsilon_n=-n(2\beta +n)\,.
\label{azero_3}
\end{equation}
Resolving the three equations (\ref{1s_cond}), (\ref{2nd_cond}), and 
(\ref{azero_3}) for 
$\alpha$, $\beta$ and $\epsilon_n$ results in
\begin{eqnarray}
\beta =-a\, , &\quad& \alpha =2b\, ,\nonumber\\
\epsilon_n &=&-(a-n)^2\, .
\label{azero_4}
\end{eqnarray}
This  expression for the energy
coincides with Eq.~(\ref{enrg_hip}), as it should be.
In this way it is proved that  the $D^{(\beta ,\alpha ) }(x)$  functions
that enter the solution of the Schr\"odinger equation 
are equal to the Romanovski polynomials.
Therefore, the $D$ functions are polynomials. 
As a result, the wave functions in $x$ space  take the form
\begin{equation}
g_n(x)= (1+x^2)^{-\frac{a}{2}}e^{-b \tan^{-1} (x)} D_n^{(-a,2b)}(x), 
\quad dx=\sqrt{1+x^2}dz\, .
\label{wfu_x}
\end{equation}
The weight function from which the $D$ polynomials are obtained via the
Rodrigues formula is
\begin{eqnarray}
w^{(a+\frac{1}{2} ,-2b )}(x)&=&(1+x^2)^{-a-\frac{1}{2}} 
e^{-2b \tan^{-1}x}.
\label{D_wafu}
\end{eqnarray}
The wave function is now equivalently rewritten to
\begin{eqnarray}
g_n(x)&=&\sqrt{(1+x^2)^{-a+\frac{1}{2}}e^{-2b\mathrm{tan}^{-1} (x)}}
D_n^{(-a,2b )}(x)\frac{1}{
\sqrt{\frac{d\mathrm{sinh}^{-1}(x)}{dx}
}
},
\label{g_n_ScarfII}
\end{eqnarray}
and {\bf is of the type in Eq.~(\ref{substi})\/}.
As a consequence, {\bf
the orthogonality integral between the wave functions will
recover the orthogonality between the polynomials\/} as shown in 
Eq.~(\ref{orth_Schr_wafu})
below.
In order to relate the $\alpha$ and $\beta $ 
parameters to those of the Romanovski polynomials
one can compare the coefficients in front of the first derivatives
in the respective Eqs.(\ref{azero_2}), and (\ref{Rom_pol}),
\begin{equation}
2(-p+1)x+q=(2\beta +1)x-\alpha\, ,
\label{prm_const}
\end{equation}
giving 
\begin{equation}
\beta=-a=-p+\frac{1}{2}, \quad -\alpha=q=-2b.
\label{p_q_par}
\end{equation}
In this way, the  polynomials
that enter the solution of the Schr\"odinger equation will be
\begin{equation}
D_n^{(\beta =-a, \alpha  =2b)}(x)
\equiv R_n^{\left(p=a+\frac{1}{2}, q=-2b\right)}(x).
\label{bong}
\end{equation}
They are obtained by means of the Rodrigues formula from the
weight function $w^{(a+\frac{1}{2}, -2b )}(x)$ as
\begin{eqnarray}
R_n^{(a+\frac{1}{2}, -2b)}(x)&=&
\frac{1}{w^{(a+\frac{1}{2}, -2b)}(x)}\frac{d^n}{dx^n}
(1+x^2)^n {w^{(a+\frac{1}{2}, -2b)}(x)}\,.
\label{Rod_Rom}
\end{eqnarray}
In this fashion, the hyperbolic Scarf potential has been solved in terms of the
real Romanovski polynomials.

The orthogonality integral of the Schr\"odinger wave functions gives rise
to the following orthogonality integral of the Romanovski polynomials,
\begin{equation}
\int_{-\infty}^{+\infty} g_n(x)g_{n^\prime}( x)dx=
\int_{-\infty}^{+\infty}
(1+x^2)^{-a+\frac{1}{2}}e^{-2b\tan^{-1}(x)}R_n^{(a+\frac{1}{2},-2b)}(x)
R_{n^\prime}^{(a+\frac{1}{2},-2b)}(x)dx\, ,
\label{orth_Schr_wafu}
\end{equation}
which coincides in form with the integral in Eq.~(\ref{orth_int}) and
is convergent for $n<a$.

In order to relate the result obtained by us to the  current literature,
it is quite instructive  to compare Eq.~(\ref{Rom_pol})
to the Jacobi equation,
\begin{equation}
(1-x^2)\frac{d^2P_n^{(\gamma,\delta)}(x)}{dx^2}
+(\gamma -\delta  -(\gamma +\delta +2)x)\frac{dP_n^{(\gamma,\delta)}(x)}{dx}
-n(n+\gamma +\delta +1)P_n^{(\gamma, \delta )}(x)=0\, .
\label{Jacobi}
\end{equation}
Upon complexification of the argument, $x\to ix$,
the latter equation transforms into
\begin{equation}
(1+x^2)\frac{d^2P_n^{(\gamma,\delta)}(ix)}{dx^2}
+i(\gamma -\delta  -i(\gamma +\delta +2)x)\frac{dP_n^{(\gamma,\delta)}(ix)}{dx}
+n(n+\gamma +\delta +1)P_n^{(\gamma, \delta )}(ix)=0\, .
\label{Jacobi_cplx}
\end{equation}
{}From a formal point of view, Eq.~(\ref{Jacobi_cplx}) can be made
to coincide with  Eq.~(\ref{Rom_pol}) for the following
parameters:
\begin{equation}
\gamma=-p -\frac{iq}{2}\, ,\quad
\delta =\gamma^\ast .
\label{cplx_csts}
\end{equation}
As long as identical equations have solutions that differ by at most a
phase factor, the Romanovski polynomials are related to the complex 
Jacobi polynomials via
\begin{equation}
R^{(p,q)}_n(x)=
i^nP_n^{\left( -p -i\frac{q}{2}, -p+i\frac{q}{2}\right)}(ix)\, .
\label{Rom_Jac}
\end{equation}
In this sense one relates in the literature the Jacobi polynomials
of complex arguments and indices to the solutions of the 
hyperbolic Scarf potential.
However,this relation is in our opinion misleading because the 
finite orthogonality
makes the real orthogonal
Romanovski  polynomials 
$\left\lbrace R_n^{(p,q)}(x)\right\rbrace$ to 
 a specie that is
fundamentally different from the complex $\left\lbrace P_n^{
\left( -p -\frac{i q}{2},
-p+\frac{iq}{2}\right)}(ix)\right\rbrace$. 
The orthogonality properties of the complex Jacobi polynomials 
depend on the interplay between the integration contour and the 
parameter values and need special care \cite{Jacobi-c}.
Equation (\ref{Rom_Jac}) in combination with 
Eqs.~(\ref{orth_int}), and (\ref{orth_cond}) 
in fact states that the contour over
which Jacobi polynomials of the type $P_n^{(\eta ,\eta ^* )}(ix)$ are
orthogonal is the real axis and not, as one naively would have expected,
the finite interval $[-i,i]$.

\section{Polynomial construction and finite orthogonality.}
The construction of the $R_n^{\left( a+\frac{1}{2},-2b \right)}(x)$
polynomials needed in the exact solutions of Scarf II is now
straightforward and based upon the Rodrigues representation
in Eq.~(\ref{Rodrigues}) where we plug in the weight function from
 Eq.~(\ref{D_wafu}). In carrying out the differentiations we find the
lowest four (unnormalized) polynomials as
\begin{eqnarray}
R^{\left(a+\frac{1}{2}, -2b\right)}_{0}&=&1\, ,
\label{d0}\\
R^{\left(a+\frac{1}{2}, -2b\right)}_{1}(x)&=&-2b +(1-2a)x\, ,
\label{d1}\\
R^{\left(a+\frac{1}{2}, -2b\right)}_{2}(x)&=&3-2a+4b^2-8b(1-a)x +
(6-10a+4a^2)x^2\, ,
\label{d2}\\
R^{(a+\frac{1}{2}, -2b)}_{3}(x)&=&
-266 +12ab -8b^3 +
\lbrack -3(-15 +16a-4a^2) +
12(3-2a)b^2\rbrack x \nonumber\\
&+&(-72b +84ab -24a^2b)x^2
+2(-2+a)(-15 +16a -4a^2)x^3\, 
\,.
\label{d3}
\end{eqnarray}

\begin{figure}[h!]
\center
\includegraphics[width=10 cm]{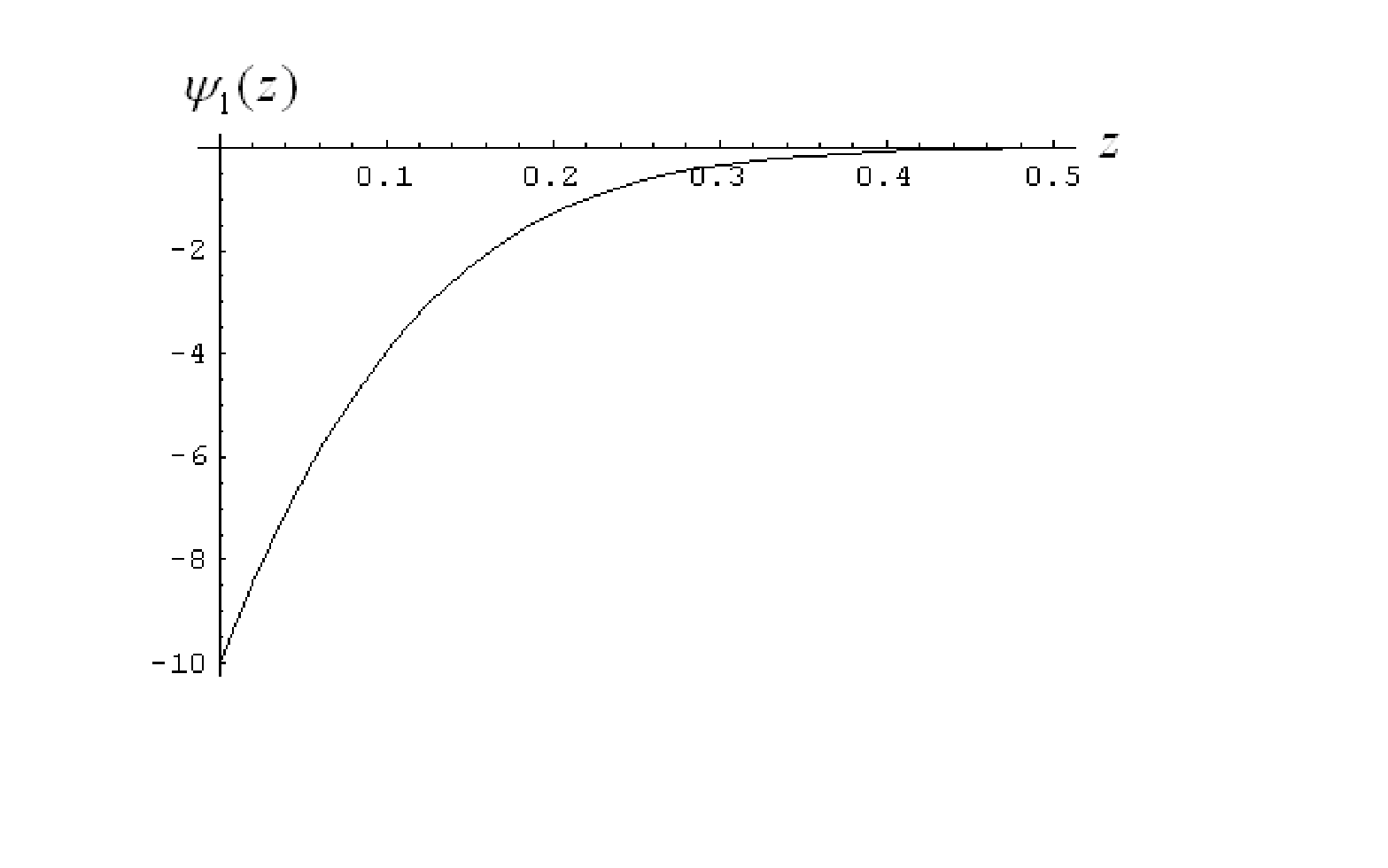}
\caption{The wave function $ \psi_1(z)$. 
\label{level1}}
\end{figure}

\begin{figure}[h!]
\center
\includegraphics[width=10 cm]{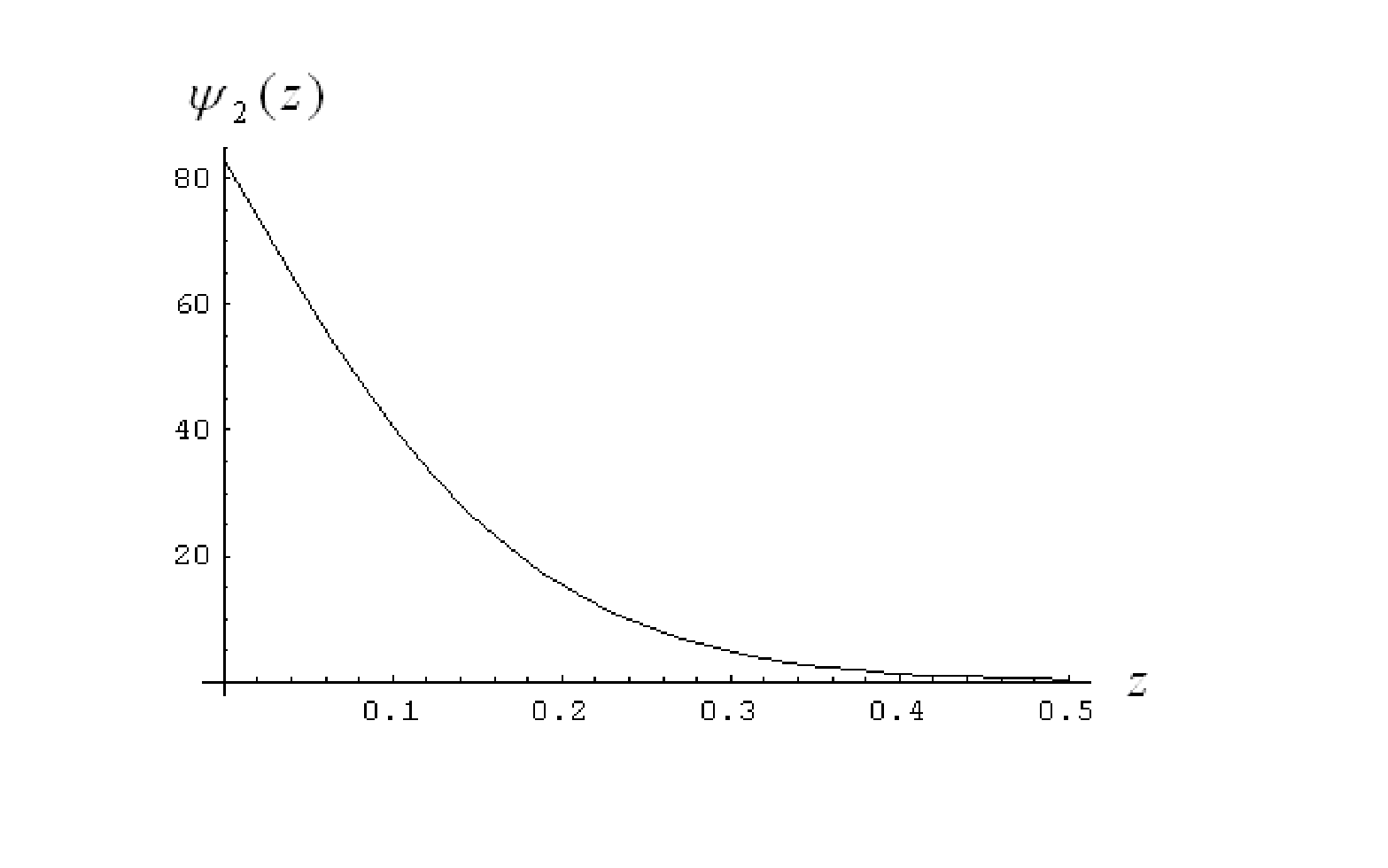}
\caption{The wave function $ \psi_2(z)$. 
\label{level2}}
\end{figure}

\begin{figure}[h!]
\center
\includegraphics[width=10 cm]{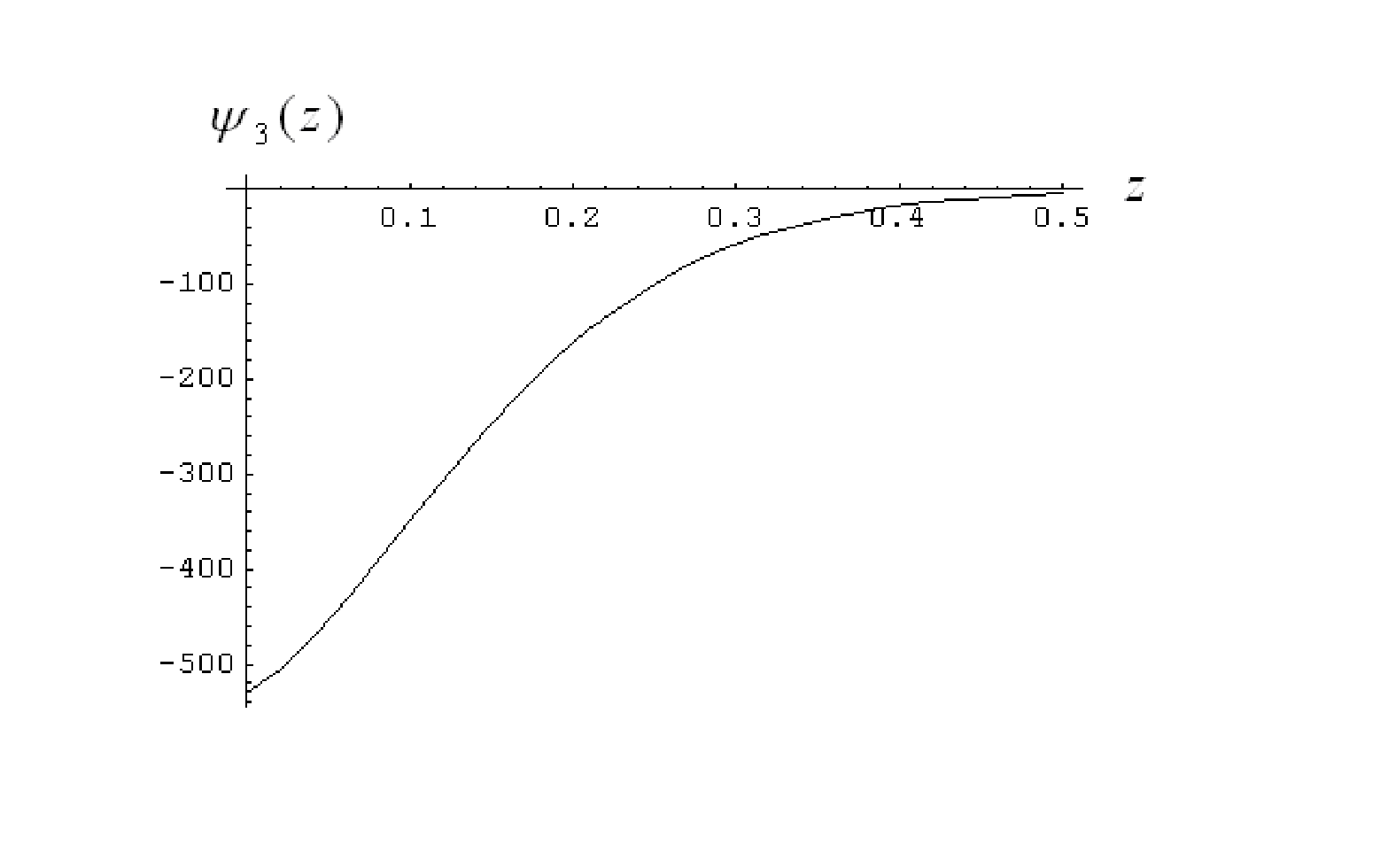}
\caption{The wave function $ \psi_3(z)$. 
\label{level3}}
\end{figure}

The finite orthogonality of the above polynomials has been proved
among others in Refs.~\cite{Mohamed,Raposo}. We here rather shall 
illustrate this property in terms of the polynomial normalization 
constants which becomes especially transparent
in the interesting limiting case of the $ \mathrm{sech^2}(z)$ potential
(it appears  in the non-relativistic reduction of the sine-Gordon
equation (c.f. \cite{Rosu} )) where one
easily finds that the normalization constant, $N^{(a+\frac{1}{2}, 0)}_n$, 
is given by 
\begin{eqnarray}
\left( N^{\left(a+\frac{1}{2}, 0\right)}_1\right)^2&=& \frac{
(2a-1)^2\sqrt{\pi}\Gamma (a-1)}
{2\Gamma(a+\frac{1}{2})}\, , \quad a>1 \, ,\nonumber\\
\left( N^{\left(a+\frac{1}{2}, 0\right)}_2\right)^2&=& 
\frac{2\sqrt{\pi}(a-1)\Gamma (a-2 )}
{\Gamma(a-\frac{1}{2})
} (3-2a)^2,\quad a>2\, ,\nonumber\\
\left( N^{\left(a+\frac{1}{2}, 0\right)}_3\right)^2&=&
\frac{3\sqrt{\pi}(a-2)\Gamma(a-3) }{
\Gamma(a-\frac{1}{2})
}(4a^2-16a +15)^2 ,\quad a>3\,\,\,  {\mbox{etc}}.
\label{norm_4}
\end{eqnarray}
Using symbolic softwares such as Maple and Mathematica is quite useful
in verifying the results reported here.
The latter expressions show  that for positive integer values of the $a$
parameter, $a =n$, only the first $(n-1)$ Romanovski polynomials are 
orthogonal,
as it should be in accord with Eq.~(\ref{orth_cond}), and the comment
after  (\ref{enrg_hip}).
The general expressions for the normalization constants of any Romanovski
polynomial are defined by integrals of the type
$\int_{-\infty}^{+\infty}(1+x^2)^{n-p}e^{q\tan^{-1}(x)}dx$
and are analytic for $(n-p)$ integer or half-integer.


\section{Romanovski polynomials and non--spherical angular functions.}

\begin{figure}[b]
\center
\includegraphics[width=7 cm]{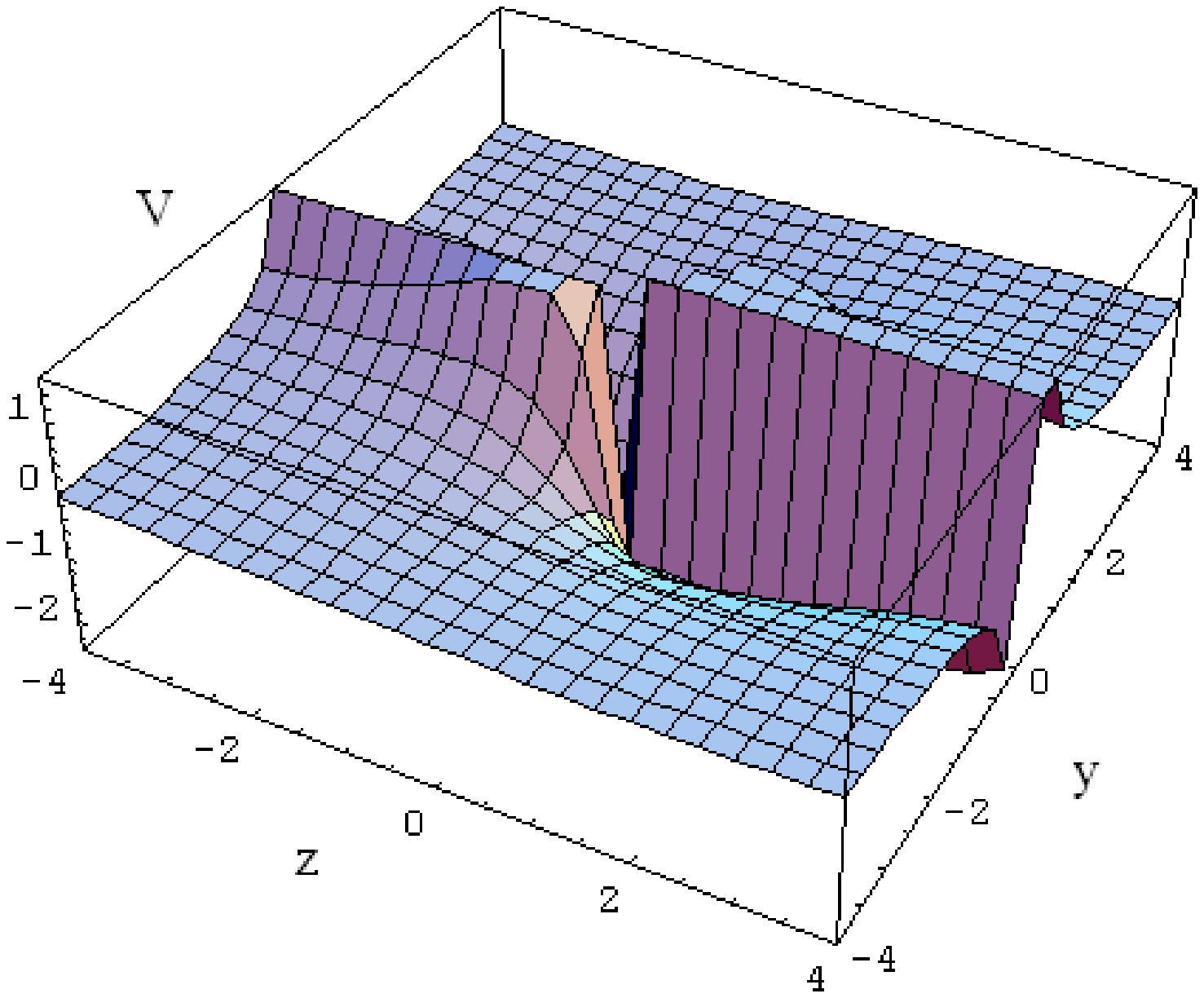}
\caption{
The non-central potential $V(r,\theta)$, for $c=-5$,
here displayed in its intersection with the $x=0$ plane,
i.e. for  $r=\sqrt{y^2 +z^2} $, and 
$\theta =\tan^{-1}\frac{y}{z }$. 
The polar angle part of its exact solutions
is expressed in terms of the Romanovski polynomials.
\label{nctrl}
}\end{figure}

The Romanovski differential equation appears in the problem of a particle
within a non-central scalar potential, a result that can be
concluded  form Ref.~\cite{Dutt}.
In denoting such a potential  by $V(r,\theta )$ , one can make for it the
specific choice  of
\begin{eqnarray}
V(r,\theta )= V_1(r) + \frac{V_2(\theta )}{r^2}\,,
&\quad&  V_2(\theta )= -c\cot (\theta)\, .
\label{non_ctrl_pt}
\end{eqnarray}
An interesting problem is the electrostatic non-central
potential in which case  $V_1(r)$ is the Coulomb potential.
The corresponding Schr\"odinger equation, 

\begin{equation}\label{SEE-SCC}
\left[-\frac{\hbar^{2}}{2\mu}\left[\frac{1}{r^{2}}
\frac{\partial}{\partial r}r^{2}\frac{\partial }{\partial r}+
\frac{1}{r^{2}\sin(\theta)}\frac{\partial}{\partial \theta}\sin(\theta) 
\frac{\partial }{\partial \theta}+\frac{1}{r^{2}\sin^{2}(\theta)}
\frac{\partial^{2}
}{\partial\varphi^{2}}\right]+V(r,\theta )\right]\Psi(r,\theta ,\varphi )
=E\Psi (r,\theta ,\varphi ),
\end{equation}
is solved in the standard way by separating variables,

\begin{equation}\label{SSVV}
\Psi(r,\theta,\varphi)= {\mathcal R}(r) \Theta (\theta)e^{i m \varphi}\, .
\end{equation}

The radial and angular differential equations for ${\mathcal R}(r)$ and
$\Theta ( \theta )$ are then found as

\begin{equation}\label{ecuacionRadial}
\frac{{\rm d}^{2}{\mathcal R}(r)}{{\rm d}r^{2}}+\frac{2}{r}
\frac{{\rm d}{\mathcal R}(r)}{{\rm d}r}+\left[\frac{2\mu}{\hbar^{2}}
(V_1(r) + E) -\frac{l(l+1)}{ r^{2}}\right]{\mathcal R}(r)=0,
\end{equation}
and
\begin{equation}\label{ecuacionAngular}
\frac{{\rm d}^{2}\Theta ( \theta )}{{\rm d}\theta ^{2}}+\cot(\theta)
\frac{{\rm d}\Theta  ( \theta )}{{\rm d}\theta}+\left[l(l+1)-\frac{2\mu
V_2(\theta)}{\hbar^{2}}-\frac{m^{2}}{\sin^2 (\theta) }\right]
\Theta  ( \theta )=0\, ,
\end{equation}
with $l(l+1)$ being the separation constant.

\subsection{Radial equation.}

The radial equation ~(\ref{ecuacionRadial}), in the $\rho $ variable
introduced as 
$r\equiv \sqrt{\frac{\hbar^{2}}{8\mu\mid E
\mid}\rho }$ and for E$<$0 (bound states) reads
\begin{equation}\label{radial1}
\frac{2}{\rho}\frac{d\mathcal R (\rho )}{d\rho}
+\frac{d^{2}\mathcal R (\rho )}{d\rho^{2}}-\frac{l(l+1)}{\rho^{2}}
\mathcal R (\rho )+
(\frac{k}{\rho}-\frac{1}{4})\mathcal R (\rho )=0,
\end{equation}
with
\begin{equation}\label{k}
k\equiv\frac{Z e^{2}}{\hbar}\sqrt{\frac{\mu}{2\mid E \mid}}.
\end{equation}
This differential equation is identical to the radial equation
of  the Schr\"odinger equation for the hydrogen atom. 
For this reason, the substitution
$\mathcal R(\rho)=e^{-\frac{\rho}{2}}\rho^{l}G(\rho)$ seems
convenient. It leads to
\begin{equation}\label{AssLaguerre}
\rho\frac{d^{2}G (\rho )}{d\rho^{2}}+
[(2l+2)-\rho]\frac{dG (\rho )}{d\rho}+(k-l-1)G (\rho )=0,
\end{equation}
which coincides with the associated
Laguerre differential equation, 
\begin{equation}\label{Laguerre}
x\frac{d^{2}L (x)}{dx^{2}}+(\beta+1-x)\frac{dL(x)}{dx}
+\lambda L(x)=0,
\end{equation}
upon setting $\beta=2l+1$. For $\lambda$, the following condition
holds valid, 
\begin{equation}
\lambda_{n_r}=-n_r\left(N_1\frac{d L_1(x)}{dx}+\frac{1}{2}
(n_r-1)\frac{d^2\sigma (x)}{dx^2}\right)\, ,
\label{lambda}
\end{equation}
which gives $\lambda_{n_r}=k-l-1=n_r$. This implies  $k=n_r+l+1$, 
and with $k$ from Eq.~(\ref{k}), the energy is obtained as
 \begin{equation}\label{EnergySTheta}
E_{n_rl}=-\frac{Z^{2}e^{4}\mu}{2\hbar^{2}}\frac{1}{(n_r+l+1)^{2}}.
\end{equation}

\subsection{Angular equation.}
Before proceeding further we wish to notice that 
for $V_2(\theta )=0,$ and upon changing variables from
$\theta $ to $\cos(\theta),$ Eq.~(\ref{ecuacionAngular}) transforms into
the associated Legendre equation and correspondingly

\begin{displaymath}
\Theta (\theta) \xrightarrow{V_2(\theta ) \to 0} P_{l}^{m}(\cos(\theta)) ,
\label{ass_Leg}
\end{displaymath}
an observation that will become important below.
Following Ref.~\cite{Dutt} one begins by substituting the polar 
angle variable
by  a new variable, $z$, introduced via $\theta \equiv f(z)$.
This leads to the new equation
\begin{equation}\label{angular1}
\left[ \frac{{\rm d}^{2}}{{\rm d}z^{2}}+\left[
-\frac{f''(z)}{f'(z)}+
f'(z)\cot (f(z))\right]
\frac{{\rm d} }{{\rm d}z }
+\left[-\frac{2\mu}{\hbar^{2}}V_2(f(z))+l(l+1)-
\frac{m^{2}}{\sin^2 (f (z))}\right]f'^{2}(z)\right] \psi (z)\ =0,
\end{equation}
with $f'(z)\equiv \frac{{\rm d}f (z)}{{\rm d}z}$, and $\psi (z)$ defined as
 $\psi (z)\equiv \Theta (f(z))$.
Next one can require that $f^\prime (z)$ approaches zero at $z=0$
like $\sin(z)$, meaning,
 $\lim_{z\to 0}f^\prime (z)/\sin (z)\to 1$, and define $f(z)$ via
\begin{equation}\label{angular2}
\frac{f''(z)}{f'(z)}=f'(z)\cot (f(z))\, .
\end{equation}
In considering the left hand side of the latter equation as the
logarithmic derivative of $f^\prime(z)$, and the right hand side as
the derivative of $\ln \sin (f(z))$, 
Eq.~(\ref{angular2}) equivalently rewrites to
\begin{equation}
{\Big(}\ln f^\prime (z){\Big)}^\prime={\Big(}  \ln(\sin( f(z))){\Big)}^\prime,
\end{equation}
leading to
\begin{equation}
\frac{df(z)}{dz}=\sin(f(z)).
\end{equation}
{}From this equation one concludes 
\begin{equation}
dz=\frac{df(z)}{\sin f(z)}.
\end{equation}
The latter expression can be integrated by means of
the so called $\lambda $ function and is given by \cite{Dwight},
\begin{equation}
z=\ln \vert \csc(f(z))-\cot(f(z))\vert 
=\lambda \left( f(z)-\frac{\pi}{2}\right).
\end{equation}
Finally, upon exponentiation one arrives at
\begin{eqnarray}
e^{z}&=&
|\csc(f(z))-\cot(f(z)) | \nonumber\\
&=& \frac{
1-\cos (f(z))
}{
\sin (f(z))
}\, .
\label{guau}
\end{eqnarray}
In using $\sin (f(z))=2\sin \left( \frac{f(z)}{2}\right)
\cos\left(  \frac{f(z)}{2}\right) $
and $1-\cos (f(z)) =2\sin^2\left( \frac{f(z)}{2}\right)$, 
Eq.~(\ref{guau}) simplifies to
\begin{equation}
e^{z}=\tan\left(\frac{f(z)}{2}\right)
\end{equation}
With that the function $f(z)$ is obtained as
\begin{equation}
\theta \equiv f(z)=2\tan^{-1} (e^{z}).
\end{equation}
From this equation one obtains $e^z$, and $e^{-z}$ as
\begin{equation}
e^z=\tan \left( \frac{\theta}{2}\right)\,,
\quad
e^{-z}=\cot \left( \frac{\theta}{2}\right)\,,
\end{equation}
and  finds $ \mathrm{cosh}(z)$ as
\begin{eqnarray}
 \mathrm{cosh}(z) &=&\frac{e^z+e^{-z}}{2}\nonumber\\ 
 &=&\frac{\tan \left(\frac{f(z)}{2}\right) +\cot 
\left(\frac{f(z)}{2}\right)}{2}\nonumber\\
&=& \frac{1}{\sin (f(z))}.
\end{eqnarray}
Getting back to $f(z)=\theta $ results in
\begin{equation}
 \mathrm{cosh}(z) =\frac{1}{\sin (\theta )}\, ,
\quad  \mathrm{sinh}(z)=-\sqrt{ \mathrm{cosh ^2}(z)-1}=-\cot (\theta ) . 
\label{mapping}
\end{equation}
Correspondingly, 
\begin{equation}
\cos (\theta )=- \mathrm{tanh}(z).
\end{equation}
Now, the derivative of $f(z)$ is calculated as
$f'(z)=\sin (f (z))= \mathrm{sech}(z) $.
Upon substituting the last relations  into Eqs.~(\ref{non_ctrl_pt}),
and (\ref{angular1}), one arrives at
\begin{equation}\label{angular3}
\frac{{\rm d}^2 \psi (z)}{{\rm d}z^2 }+
\left[
l(l+1)\frac{1}{ \mathrm{cosh^2}(z)} - \frac{2\mu}{\hbar^2}
c\,\,  \mathrm{tanh (z)}\frac{1}{ \mathrm{cosh}(z)}  
- m^{2}\right] \psi (z)=0\,.
\end{equation}
Defining $\tilde{c} \equiv \frac{2\mu}{\hbar^{2}}c$  
the latter equation takes the form:
\begin{equation}
\label{EqScarfInTheta}
\frac{d^{2}\psi(z)}{dz^{2}}+\left[ l(l+1)
 \mathrm{sech^{2}}(z)-\tilde{c}\,\,  \mathrm{sech}(z) 
\mathrm{tanh}(z)-m^{2}\right] \psi(z)=0.
\end{equation}
This equation has same form as the radial Schr\"odinger equation
with Scarf II, i.e.   
\begin{equation}
\frac{d^{2}\psi(z)}{dz^{2}}+\left[ 
(-b^{2}+a(a+1)) \mathrm{sech^{2}}(z)-b(2a+1) \mathrm{sech}(z) 
\mathrm{tanh}(z)+\epsilon\right] \psi(z)=0\,,
\label{Scarf_hip}
\end{equation}
provided,
\begin{itemize}
\item $l(l+1)$ plays the role of  $-(b^2 -a(a+1))$,
\item  $m^2$ plays the role of  $ -\epsilon$,
\item $\tilde{c}$ plays  the role of $-b(2a+1)$.
\end{itemize}
Recall that the solution to eq.~(\ref{Scarf_hip})
was obtained in subsection 4.1.1 of the present chapter as, 
\begin{eqnarray}
\psi_n( z )&=&N_n (1+ \mathrm{sinh^2}(z))^{-\frac{a}{2} }
e^{-b \tan^{-1} (\mathrm{sinh} (z)) }
R^{( a+\frac{1}{2} , -2b )}_n(\mathrm{sinh} (z) )\, ,  \nonumber\\
 \epsilon_n =-(a-n)^2\, ,
&\quad& -\infty <x<+\infty\,,
\end{eqnarray}
with $N_n$ being a normalization constant. Back to the $\theta $ variable and 
in making use of the equality $\sinh (z) =-\cot ( \theta ) $, we find
\begin{equation}
\Theta(\theta) =\psi_n(\mathrm{sinh}^{-1}(-\cot (\theta) )  )=N_n 
(1+\cot^2 (\theta) )^{-\frac{a}{2} } e^{-\frac{b }{2} \tan^{-1} 
(- \cot (\theta) ) }
R^{( a+\frac{1}{2} , -2b  )}_n(-\cot (\theta)  )\, ,
\label{The_the}
\end{equation} 
showing that the angular part of the exact solution to the
non-central potential under consideration is defined 
by the Romanovski polynomials. 
The two parameters of the Romanovski polynomials have to
be determined from the system of three equations,
\begin{equation}
-b^{2}+a(a+1)=l(l+1),
\end{equation}
\begin{equation}
-b(2a+1)=\tilde{c},
\end{equation}
\begin{equation}
\epsilon_n=-(a-n)^2=-m^{2},\quad a>n, \quad m>0,
\end{equation}
meaning that the $l$, $m$, and $\widetilde{c}$ constants
can not be independent. 
There exist various choices for $a$ and $b$.
If defined on the basis of the first two equations, one encounters
\begin{eqnarray}
\left( a+\frac{1}{2}\right)^2&=&
\frac{1}{2}\left(
\left(l+\frac{1}{2}\right)^2
+\sqrt{\left(l+\frac{1}{2}\right)^4 +\widetilde{c}^2}\, 
\right),\nonumber\\
b^2&=&
\frac{1}{2}\left(
-\left(l+\frac{1}{2}\right)^2
+\sqrt{\left(l+\frac{1}{2}\right)^4 +\widetilde{c}^2}\,
\right)\, .
\label{set_1}
\end{eqnarray}
Substitution of $a$ into the third equation imposes a constraint
on $l$ as a function of $m$, $\widetilde{c}$, and $n$.
A second choice for $a$ and $b$ is obtained by 
expressing $a$ from the  third equation in terms of $m$, and $n$
as $a=m+n$ and substituting in the second equation
to obtain $b$ as  
\begin{eqnarray}
b&=&-\frac{\widetilde{c}}{2(m+n)+1}.
\label{set_2_b}
\end{eqnarray}
Then the first equation imposes the following restriction on $l$
\begin{eqnarray}
X\stackrel{\mathrm{def}}{:=}(b^2-a(a+1)),  &\quad&
l= -\frac{1}{4} +\sqrt{\frac{1}{4}+X}\, .
\label{parmts}
\end{eqnarray}
This $l$ value which is not necessarily integer, 
can be plugged into Eq.~(\ref{EnergySTheta}) leading to
a (discrete) spectrum that no longer bears any resemblance to
the $O(4)$ degeneracy.
This is the path pursued  by Ref.~\cite{Dutt}.
I here instead take a third chance and express 
$a$, $b$, and $\widetilde{c}$ as functions of
$l$ alone according to
\begin{eqnarray}
a=b= l(l+1), &\quad& n=a-m=l(l+1)-m,\quad
\widetilde{c}=-b(2a+1).
\label{parmts_2}
\end{eqnarray}
This choice allows to consider integer $l$ values.
In making use of  Eqs.~(\ref{p_q_par}),(\ref{bong}),
the polar angle part of the wave function in this case
becomes 
\begin{equation}
\psi_{n=l(l+1)-m}(\mathrm{sinh}^{-1}( -\cot(\theta)))=
(1+\cot(\theta)^{2})^{-\frac{l(l+1)}{2}}e^{-l(l+1)
\tan^{-1} (-\cot(\theta))}
R^{\left( l(l+1)+\frac{1}{2},-2l(l+1)\right)}_{l(l+1)-m}(-\cot(\theta)).
\end{equation}
The complete angular wave function now can be labeled by $l$ and $m$
(as a tribute to the spherical harmonics) and is given by 
\begin{multline}
Z^{m}_{l}(\theta,\varphi)=\psi_{n=l(l+1)-m}(
\mathrm{sinh}^{-1}(-\cot(\theta)))e^{im\varphi}= \\
(1+\cot^{2}(\theta))^{-\frac{l(l+1)}{2}}e^{-l(l+1)\tan^{-1}(-\cot(\theta))}
R^{\left( l(l+1)+\frac{1}{2},-2l(l+1)\right)}_{l(l+1)-m}(-\cot(\theta))
e^{i m\varphi}\, . 
\end{multline}
It reduces to the spherical harmonics $Y^{m}_{l}(\theta,\varphi)$ 
for $a=b=0$. In this way, the Romanovski polynomials shape the
angular part of the wave function in the problem under consideration.
In the following, we shall refer to $Z_l^m(\theta ,\varphi )$ as
``non-spherical angular functions".  
In the appendix I present a side by side comparison of 
 $\mid Y^{m}_{l}(\theta,\varphi)\mid$ and  
$\mid Z^{m}_{l}(\theta,\varphi)\mid$.
A comment is in order on  $|Z_l^m(\theta ,\varphi )|$. 
In that regard, it is important to become aware of the fact 
already mentioned above that
the Scarf II potential possesses $su(1,1)$ as a potential algebra,
a result reported by Refs.~\cite{JPA_MG_34,Suk_SU1_1} among others.
There, it was pointed out that the respective Hamiltonian, $H$, equals 
$H=-C-\frac{1}{4}$, with $C$ being the $su(1,1)$ Casimir operator,
whose eigenvalues in our convention are $j(j-1)$ with $j>0$
versus $j(j+1)$ and $j<0$ in the convention of \cite{JPA_MG_34,Suk_SU1_1}.
As a consequence, the Scarf II solutions can be viewed as 
representation spaces of irreducible 
$SU(1,1)$ representations. Specifically, in the case under consideration
the represenation is dicrete, unitary and of infinite dimensionality.
It is the one denoted by
$\lbrace D^+_j\, ^ {(m^\prime)}(\theta ,\varphi )\rbrace $,
with $m^\prime =j,j+1,j+2,....$.  
The $SU(1,1)$ labels  $m^\prime $, and $j$ are mapped onto ours via
\begin{equation}
m^\prime =a+\frac{1}{2}=l(l+1)+\frac{1}{2},
\quad j=m^\prime -n\,, \quad m^\prime =j, j+1, j+2, ....
\end{equation}
meaning that both $j$ and $m^\prime $ are a half-integer.
In terms of these labels the energy rewrites as 
$\epsilon_n=-(j-\frac{1}{2})^2$. The condition $a>n$ translates now as
$j>\frac{1}{2}$. In result,  
$\Theta (\theta )$ becomes
\begin{equation}
\Theta (\theta )= \psi_{n=m^\prime -j}\left(\sinh^{-1}(-\cot \theta)\right)=
\sqrt{(1+\cot^2\theta )^{-m^\prime +\frac{1}{2} }
e^{-2b\tan^{-1} (-\cot \theta )}}
R_{m^\prime -j}^{( m^\prime ,-2b)}(-\cot\theta ) \, .
\label{su_1_1}
\end{equation}
Here we kept the parameter  $b$ general because its value does not affect
the $SU(1,1)$ symmetry.
Within this context, the  $|\psi_ {m^\prime -j}\left(\sinh^{-1}(-\cot\theta 
\right)|$'s can be viewed as absolute values of  
$\lbrace D^+_j\, ^ {(m^\prime)}(\theta ,\varphi )\rbrace $ 
eigenvector components \cite{Wybourne} and
realized in terms of the Romanovski polynomials.
The $|Z_l^m(\theta ,\varphi )|$ functions  are then images in 
polar coordinate space of 
$\lbrace D^+_{j=m+\frac{1}{2}}\, ^{\left( m^\prime =l(l+1)+\frac{1}{2}\right)}
(\theta , \varphi)\rbrace $ eigenvector components.
The representations are infinite because for a fixed $j$ value,  
$m^\prime$  is bound from below to
$m^\prime _{\mathrm{min}}=j$, but it is not bound from above.
{}For example,   $|Z_1^1(\theta ,\varphi )|$ refers to $
D^+_{j=\frac{3}{2}}\, ^{\left( m^\prime=\frac{5}{2}\right)} 
(\theta ,\varphi )$, $|Z_2^1(\theta ,\varphi )|$ refers to
$
D^+_{j=\frac{3}{2}}\, ^{\left( m^\prime=\frac{13}{2}\right)} 
(\theta ,\varphi )$ etc.

\subsection{Romanovski polynomials and associated Legendre functions.}
Next, it is quite instructive to consider the case of a vanishing
$V_2(\theta )$, i.e. $c=0$, and compare Eq.~(\ref{angular3}) to
Eq.~(\ref{Scarf_hip}) for $b=0$.
In this case, and in accordance with Eq.~(\ref{p_q_par}) 
\begin{eqnarray}
l=a=p-\frac{1}{2}\, , &\quad & m^2=(l-n)^2\, ,
\quad q=-2b=0,
\end{eqnarray}
which allows one to relate $n$ to $l$ and $m$ as $m=l-n$.
As long as the two equations are equivalent, their solutions differ
at most by a constant factor. This allows to establish a relationship
between the associated Legendre functions and the Scarf II wave functions.
In taking into account Eqs.~(\ref{azero_1}),and (\ref{wfu_x}) together with
Eqs.~(\ref{mapping}), one finds
 $\cot(\theta) = - \mathrm{sinh}(z) $ which  produces the
following new relationship between the associated Legendre
functions and the Romanovski polynomials 
\begin{equation}
P_l^{m}( \cos (\theta )  )\sim 
(1+\cot^2 ( \theta ) )^{-\frac{l}{2}}
R_{l-m}^{(l+\frac{1}{2}, 0 )}(-\cot ( \theta ) )\, , \quad l-m=n=0,1,2,...l.
\label{Ass_Leg_Rom}
\end{equation}
In substituting the latter expression  into the orthogonality integral between
the associated Legendre functions,
\begin{equation}
\int_{-1}^{1} P_l^m(\cos (\theta ) )
P_{l^\prime}^m(\cos (\theta ) ){\rm d}\cos (\theta )\,
=0\, , \quad l\not=l^\prime ,
\label{orth_assc_Leg}
\end{equation}
results in the following integral 
\begin{equation}
\int_{-1}^{1} (1+\cot^2 (\theta ) )^{-\frac{l+l^\prime }{2}} 
R_{l-m}^{\left(l+\frac{1}{2}, 0\right)}(-\cot (\theta ) )
R_{l^\prime -m}^{\left(l^\prime +\frac{1}{2}, 0 \right)}
(-\cot (\theta ) )
{\rm d}\cos  (\theta )\,
= 0\, , \quad l\not= l^\prime .
\label{orth_as_Leg}
\end{equation}
When rewritten to conventional notations, the latter expression becomes
\begin{eqnarray}
\int _{-\infty }^{+\infty }
\sqrt{
w^{
\left(l+ \frac{1}{2}, 0 \right) }
(x)} R^
{(l+\frac{1}{2}, 0)}_{n=l-m}(x) \sqrt{
w^{\left( l^\prime + \frac{1}{2}, 0 \right) }
(x)} R^{(l^\prime +\frac{1}{2}, 0 )}_{n^\prime =l^\prime -m }(x ) 
\frac{{\rm d}x}{1+x^2}
&=& 0\, , \quad l\not= l^\prime \,,\nonumber\\
x=\sinh (z )\, , \quad l-n=l^\prime -n^\prime &=&m\ge 0\, .
\label{Leg_Rom_orth}
\end{eqnarray}
This integral  describes  orthogonality between an {\it infinite\/} 
set of Romanovski
polynomials with  {\it different polynomial parameters\/} (they would
define wave functions of states bound in {\it different potentials\/}).
This new orthogonality relationship
does not contradict the finite orthogonality in Eq.~(\ref{orth_cond}) which
is valid for states belonging to {\it same potential\/} 
({\it equal polynomial parameters\/}).
Rather, for different potentials, Eq.~(\ref{orth_cond})
can be fulfilled for an infinite number of states. 
To see this let us consider, for simplicity , $n=n^\prime =l-m $, i.e.
$l=l^\prime$.  Given $p=l+\frac{1}{2}$,
the condition in Eq.~(\ref{orth_cond}) defines normalizability and
takes the form
\begin{eqnarray}
2(l-m)<2(l+\frac{1}{2})-1=2l\, ,
\end{eqnarray}
which is automatically fulfilled for any $m>0$. The presence of the
additional factor of $(1+x^2)^{-1}$ guarantees convergence also
for $m=0$.
Equation (\ref{Leg_Rom_orth})
reveals that for parameters attached to the degree of the polynomial, 
an infinite number of  Romanovski polynomials can appear orthogonal,
although not precisely with respect
to the weight function that defines  their Rodrigues
representation.  
The study  presented here is  kindred to Ref.~\cite{CK}
and Eq.~(\ref{orto-1}) from above. Also
there, the exact solutions of the Schr\"odinger equation with the 
trigonometric Rosen-Morse  potential 
(employed as quark-di-quark interaction) have been expressed
in terms of Romanovski polynomials (not identified as such at that time)
and also with parameters that depended on the degree of the polynomial.
Also in this case, the $n$-dependence of the parameters, and the
corresponding varying  weight function allowed to fulfill
Eq.~(\ref{orth_cond})  for  infinitely many polynomials.

\section{Hyperbolic Scarf potential in the Klein-Gordon equation.}
The final example to be considered is the case of the relativistic
Klein-Gordon equation, 
\begin{equation}
(P^2-\mu^2)\Psi  =0\, , \quad P_\nu=i\partial_{\nu},
\label{Kl_Go}
\end{equation}
where $\mu $ is the mass.
One can introduce two different potentials in this equation.
The first is a vector potential, $A_\nu$,
introduced via minimal coupling as $P_\nu -gA_\nu $, with $g$ being a constant,
and the second is a scalar potential, $S$, introduced via
$m\longrightarrow \mu +S$. The vector potential in the so called Coulomb gauge
satisfies $\vec{\nabla }\cdot \vec {A}=0$ but on many occasions one simplifies
the problem in choosing $\vec{A}=0$ in which case only the time like component
of the vector potential, $gA_0$, denoted by
$V$ in the following, enters the equation.
In effect, the Klein-Gordon equation with vector and scalar potential
(in units of $c=\hbar=1$) takes the form
\begin{equation}\label{1}
[(i\frac{\partial }{\partial t}-V(\boldsymbol{r}))^{2}+
\nabla^{2}- (S(\boldsymbol{r})+\mu)^{2}]\psi(\boldsymbol{r})=0.
\end{equation}
The latter equation simplifies significantly when $S$ and $V$ are equal.
It has been shown in Refs.~\cite{Alhaidari_KG,Wen-Chao} that in this case 
the solution of Eq.~(\ref{1}) can be found
from those of an associated  Schr\"odinger equation. 
Indeed,  for time-independent potentials,  
the total wave function can be written as
$\Psi(\boldsymbol{r}, t)=e^{-iEt}\psi(\boldsymbol{r})$, 
with $E$ being the relativistic energy. This
substitution results in:
\begin{equation}\label{2}
[\nabla^{2}+(V(\boldsymbol{r})-E)^{2}-(S(\boldsymbol{r})+\mu)^{2}]
\psi(\boldsymbol{r})=0.
\end{equation}
From now onward I shall focus  on the special case of equal
scalar and vector potentials, i.e. 
$V(\boldsymbol{r})=S(\boldsymbol{r})$. In the following we change 
variable to $V(\boldsymbol{r})\rightarrow \frac{v(\boldsymbol{r})}{2}$ .
In this case, the Klein-Gordon equation then rewrites to
\begin{equation}\label{3}
[\nabla^{2}+\left(E-\frac{v(\boldsymbol{r})}{2}\right)^{2}-
\left(\frac{v(\boldsymbol{r})}{2}+\mu\right)^{2}]\psi(\boldsymbol{r})=0.
\end{equation}
In the following  $v(\boldsymbol{r})$ is taken as  
the central hyperbolic Scarf potential. 
Separating variables in polar coordinates,
$\psi(\boldsymbol{r})=R(r)H(\theta )K(\varphi )$, leads to
\begin{equation}\label{4}
K''(\varphi)+m^{2}K(\varphi)=0,
\end{equation}
\begin{equation}\label{5}
H''(\theta)+\cot(\theta)H'(\theta)-[m^{2}\csc^{2}(\theta)-s(s+1)]H(\theta)=0,
\end{equation}
and
\begin{equation}\label{6}
(r^{2}R'(r))'-[s(s+1)+(E+\mu)r^{2}v(r)-(E^{2}-\mu^{2})r^{2}]R(r)=0,
\end{equation}
where $s$ and $m$ are the separation constants. 
The angular equation for the azimuthal coordinate, $\varphi$, has
solutions satisfying periodical conditions: 
$K(\varphi)=\frac{1}{\sqrt{2\pi}}e^{im\varphi}$, $m=0,\pm1, \pm2,...$
The equation for the polar angle, $\theta$, can be transformed by means of
$x=\cos(\theta)$ and becomes
\begin{equation}\label{7}
(1-x^{2})\frac{d^{2}f(x)}{dx^{2}}-2x\frac{df(x)}{dx}+[-
\frac{m^{2}}{1-x^{2}}+s(s+1)]f(x)=0.
\end{equation}
It can be identified with the associated Legendre differential 
equation whose solutions are
$f(\cos( \theta ) )\equiv P^{m}_{s}(\cos(\theta))$.
The radial equation can be solved for $s=0$.
{}For $R(r)=D(r)/r $ it takes the form
\begin{eqnarray}\label{8}
\frac{d^{2}D(r)}{dr^{2}}-
{\Big(}(E +\mu)A^{2}&+&
(E +\mu)(B^{2}-A^{2}-A) \mathrm{sech^{2}}(r)\nonumber\\
&+&(E +\mu)B(2A+1)
 \mathrm{sech}(r) \mathrm{tanh}(r)-(E +\mu)(E -\mu){\Big)}D(r)=0.
\nonumber\\
\end{eqnarray}
Upon naming  
\begin{itemize}
\item  $(E +\mu)(-A^{2}+E-\mu)$ as $\epsilon$, 
\item  $(E +\mu)A^{2}$ as $a^{2}$,
\item  $(E +\mu)(B^{2}-A^{2}-A)$ as $b^{2}-a^{2}-a$,
\item  $(E +\mu)B(2A+1)$ as $b(2a+1)$ , 
\end{itemize}
and in changing variables to  
$x= \mathrm{sinh}(r)$, $D(r)\longrightarrow f(x)$,
amounts  to
\begin{equation}\label{9}
(1+x^{2})\frac{d^{2}f(x)}{dx^{2}}+x\frac{df(x)}{dx}+
\left(\frac{-b^{2}+a(a+1)}{1+x^{2}}+\frac{-b(2a+1)}{1+x^{2}}x+\epsilon
\right)f(x)=0.
\end{equation}
The latter equation is equal to the polynomial form of
the $1d$-Schr\"odinger equation for the 
Scarf II in Eq.~(\ref{Schr_bzero}) whose solutions have been 
explicitly constructed in Eq.~(\ref{g_n_ScarfII}) above.
Matching parameters leads to

\begin{eqnarray}
\alpha=2b\, , &\quad &
\beta=-a\, , \qquad  \epsilon_n=-(a-n)^{2}.
\end{eqnarray}

The energies are then found as

\begin{eqnarray}
E^{1}_{n}&=&
\frac{A^{2}+2An-2A^{2}\mu+\sqrt{A^{4}+4A^{3}n-4n^{2}+
4A(A+2n)\mu+4\mu^{2}}}{2(1+A^{2})}\, ,
\nonumber\\
E^{2}_{n}&=&
\frac{A^{2}+2An-2A^{2}\mu-\sqrt{A^{4}+
4A^{3}n-4n^{2}+4A(A+2n)\mu+4\mu^{2}}}{2(1+A^{2})}\,  .
\end{eqnarray}
The two values for the energies correspond to  particles, 
and antiparticles,  as  expected from the relativistic
Klein-Gordon equation.

\chapter{Conclusions and outlooks.}

In this thesis I presented  

\begin{itemize}
\item the classification of the (orthogonal) polynomial
solutions to the generalized hypergeometric equation in the 
respective schemes of
Koepf--Masjed-Jamei \cite{Koepf}, Nikiforov-Uvarov \cite{NikUv}, 
and Bochner \cite{Bochner},

\item the explicit construction of the Hermite,
Laguerre, Jacobi, Bessel and  Romanovski polynomials,

\item the solutions of the Schr\"odinger equations with
the respective one- and three dimensional oscillator, the
Coulomb- and the hyperbolic Rosen-Morse potentials, and the
$\sim e^{\alpha x}$ barrier in terms of one of
the above polynomials.

\end{itemize}
As new results I report
\begin{itemize}

\item the exact wave functions of the  bound states within
the hyperbolic Scarf potential in terms of the Romanovski 
polynomials,

\item the  finite orthogonality of the Romanovski polynomials
in terms of their normalization constants,
a property that allowed to map the finite number of bound states 
within Scarf II onto a finite set of polynomials,
 
\item the Romanovski polynomials as main designers  of
 non--spherical angular functions of a new type,
which we identified with components of the eigenvectors  
of the  infinite discrete unitary SU(1,1) representation,
$\lbrace D^+_{j=m+\frac{1}{2}}\, ^{\left( m^\prime =l(l+1)+\frac{1}{2}\right)}
(\theta ,\varphi)\rbrace $,

\item a non-linear relationship between Romanovski polynomials
with parameters attached to the degree of the polynomial
and the associated Legendre functions which lead to a new
orthogonality integral for an infinite series of such
Romanovski polynomials,

\item the solution of the Klein-Gordon equation with
equal scalar and vector potentials, taken
as the hyperbolic Scarf potential.
\end{itemize}
I conclude that the Romanovski polynomials
represent the most adequate
degrees of freedom in the mathematics of the hyperbolic 
Scarf potential.\\

\noindent
Further conclusions are: 

\begin{itemize}

\item The orthogonality integral of
Schr\"odinger wave function of the form given in 
Eq.~(\ref{substi}) and free parameters always recovers the orthogonality
of the involved polynomials. 

\item Equation (\ref{substi})  did not hold valid on several occasions 
in which  the parameters happened to  depend on the degree of the polynomials,
in which case the orthogonality of the wave functions failed in
recovering the orthogonality of the polynomials with the free parameters. 

\item As a rule, polynomials with parameters running with the degree of
the polynomial, appear orthogonal with respect to
a weight function which is altered in comparison with the one that
enters the Rodrigues formula. 

\item The case of the hydrogen atom was special in so far as there
the anomalous orthogonality integral between Laguerre polynomials
was observed  for polynomial parameters which  
{\bf did not depend on the degree of the polynomial \/}, however
the variable $x$ did.
\end{itemize}

\noindent
In future research one can
\begin{itemize}
\item
exploit the relationship between the Romanovski polynomials 
and the associated  Legendre functions to
 write down various new recurrence relations for the former,
\item employ  their weight function as an extension of the
student's $t$ distribution and test it in statistical problems
of estimating  standard deviations from data,
\item study symmetry relationships between the non-spherical angular
functions which lead to  equality between such functions of
different parameters  as visible by inspection from the Appendix.
\end{itemize}

\noindent
The Romanovski polynomials are interesting mathematical entities in their own
and future  research is expected to shed more light on their properties and 
physics applications.

\chapter{Appendix.}
In this Appendix I present a graphical side by side comparison of the
absolute values of the
spherical harmonics $\mid Y^{m}_{l}(\theta,\varphi)\mid$ and 
the  $\mid Z^{m}_{l}(\theta,\varphi)\mid$ functions. 

\begin{figure}[!h]
\center
\includegraphics[width=7cm]{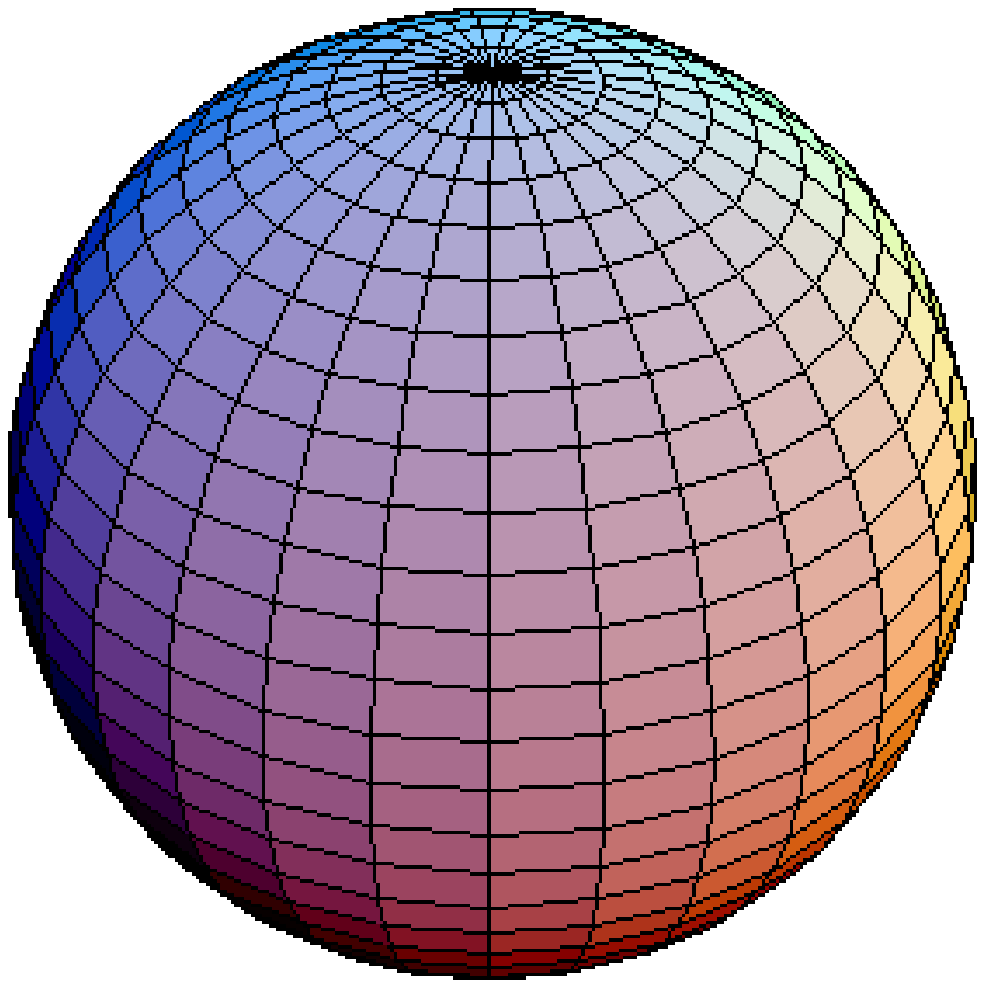}
\includegraphics[width=7cm]{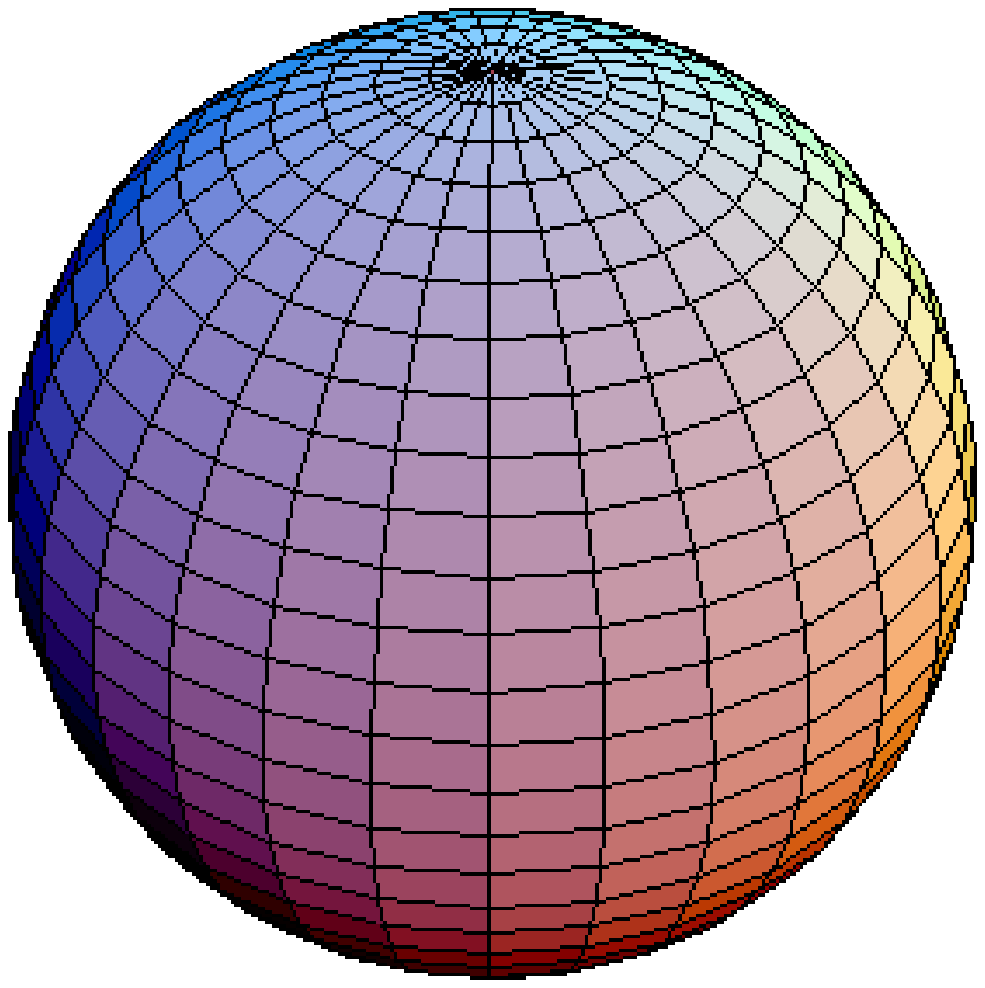}
\caption{$\mid Y^{0}_{0}(\theta,\varphi)\mid$ vs $\mid Z^{0}_{0}(\theta,\varphi)\mid$}
\end{figure}
\begin{figure}[!h]
\center
\includegraphics[width=7cm]{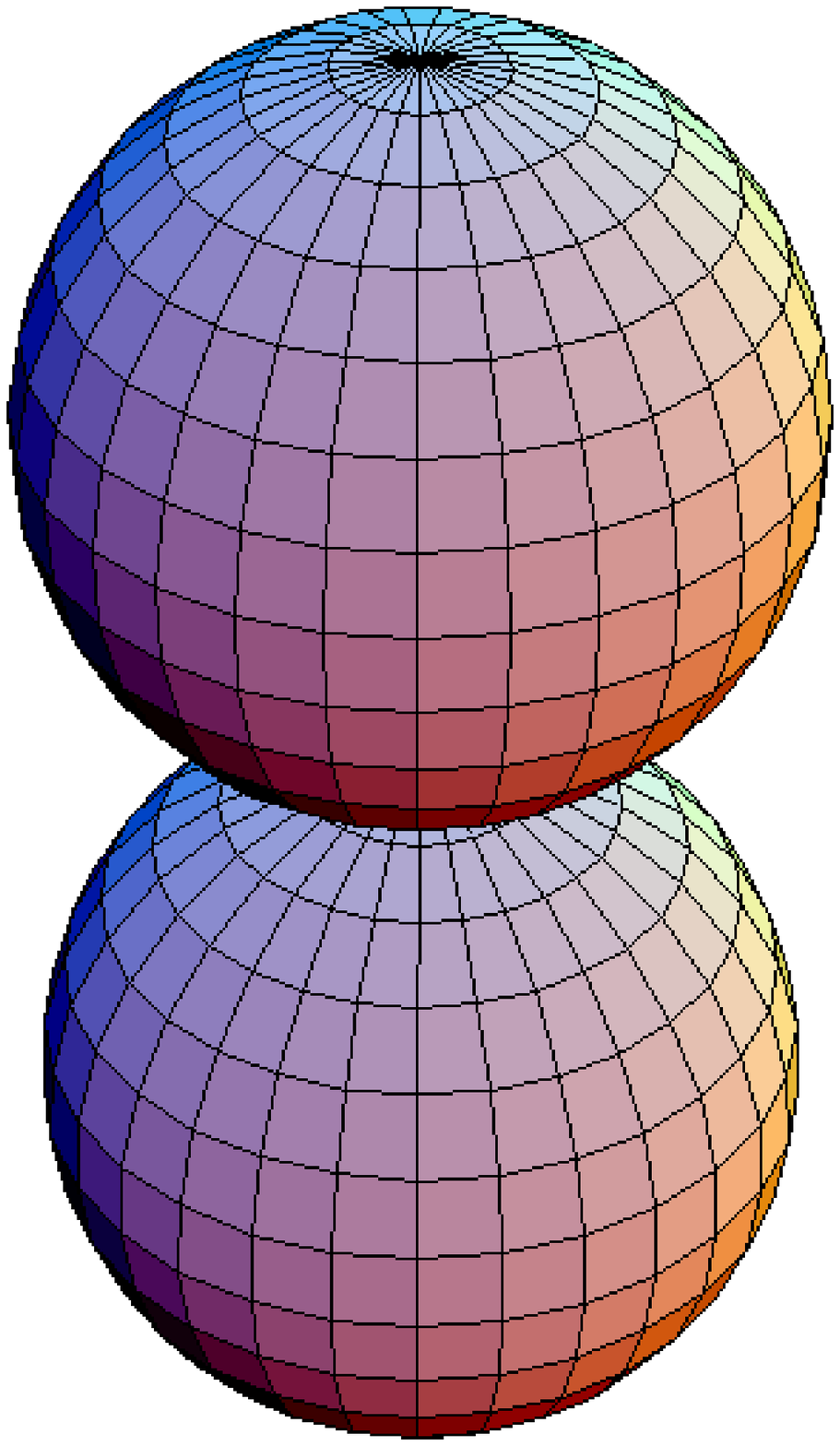}
\includegraphics[width=7cm]{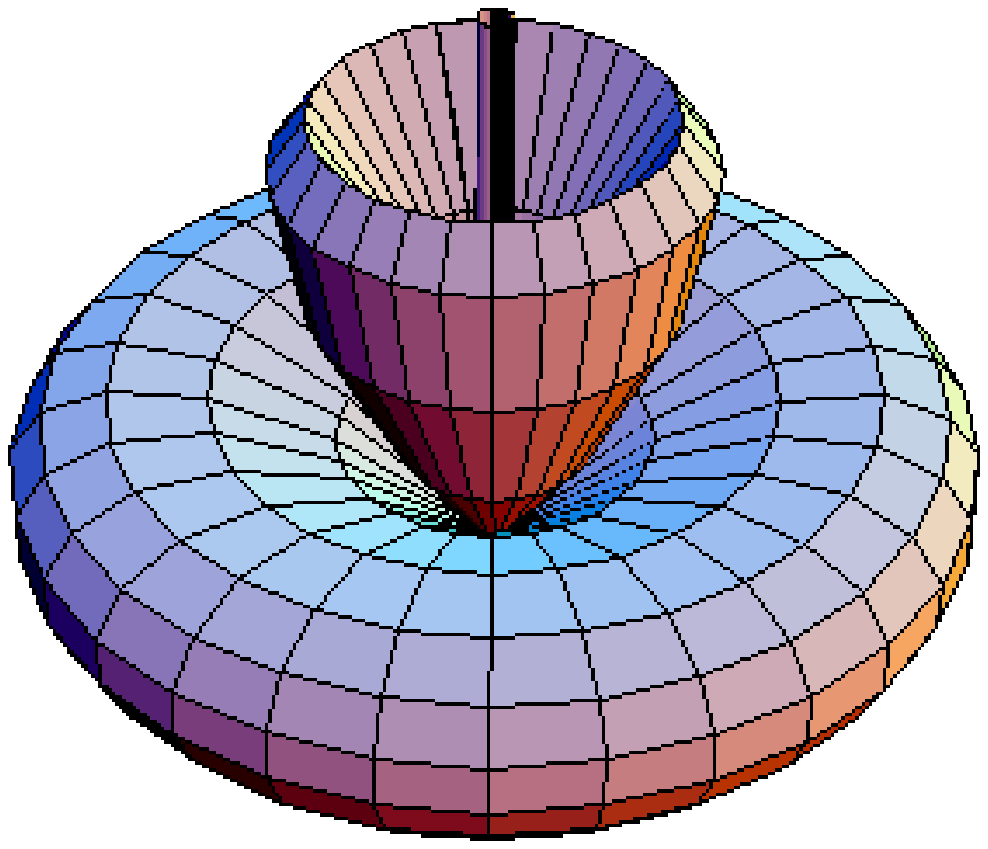}
\caption{$\mid Y^{0}_{1}(\theta,\varphi)\mid$ vs $\mid Z^{0}_{1}(\theta,\varphi)\mid$}
\end{figure}
\begin{figure}
\center
\includegraphics[width=7cm]{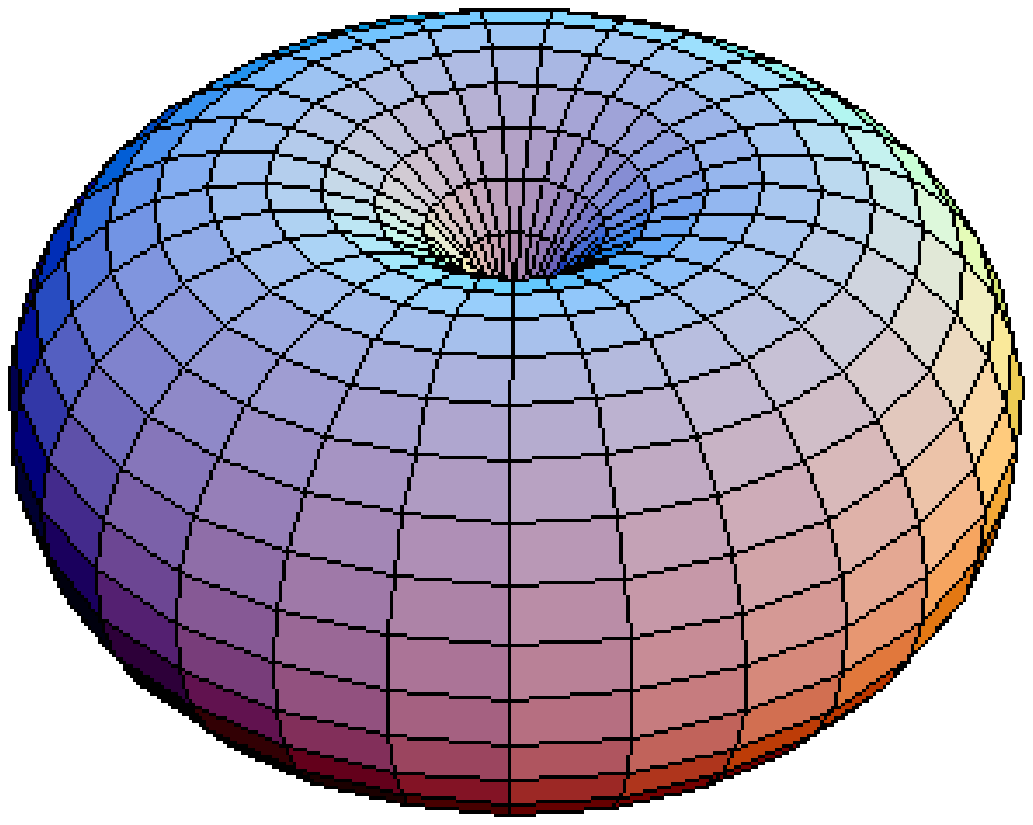}
\includegraphics[width=7cm]{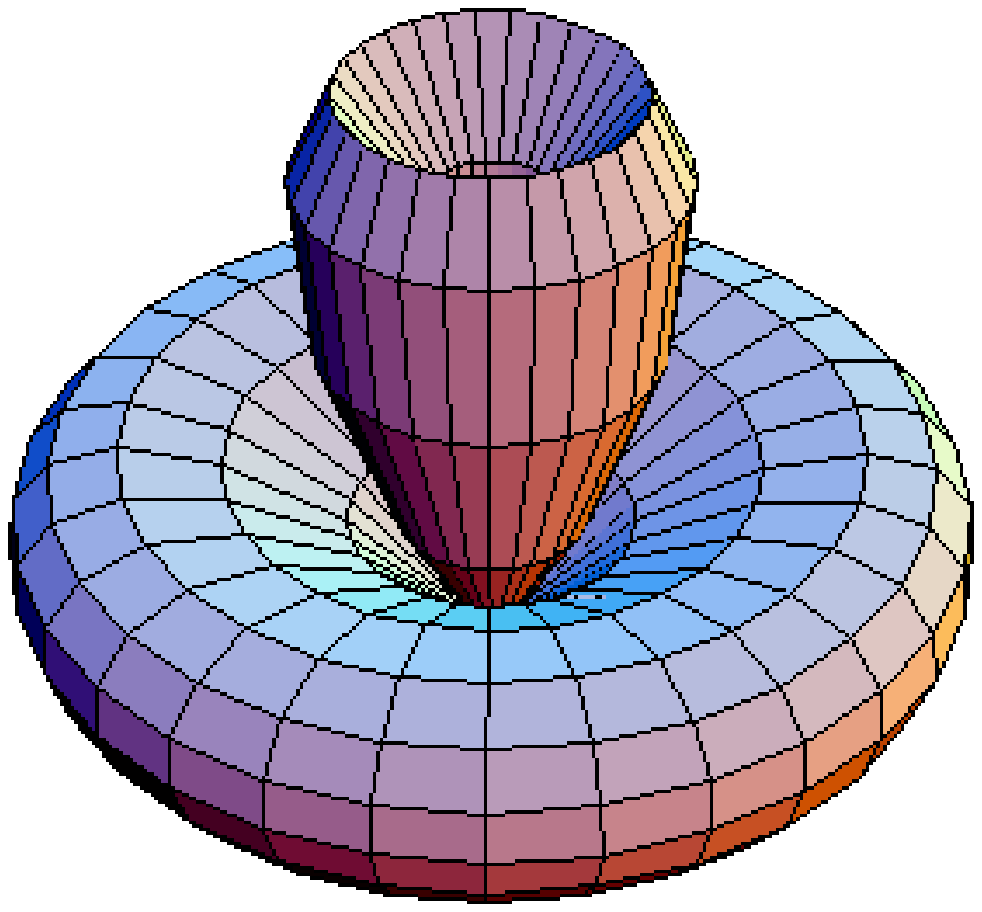}
\caption{$\mid Y^{\pm1}_{1}(\theta,\varphi)\mid$ vs $\mid Z^{\pm1}_{1}(\theta,\varphi)\mid$}
\end{figure}
\begin{figure}
\center
\includegraphics[width=7cm]{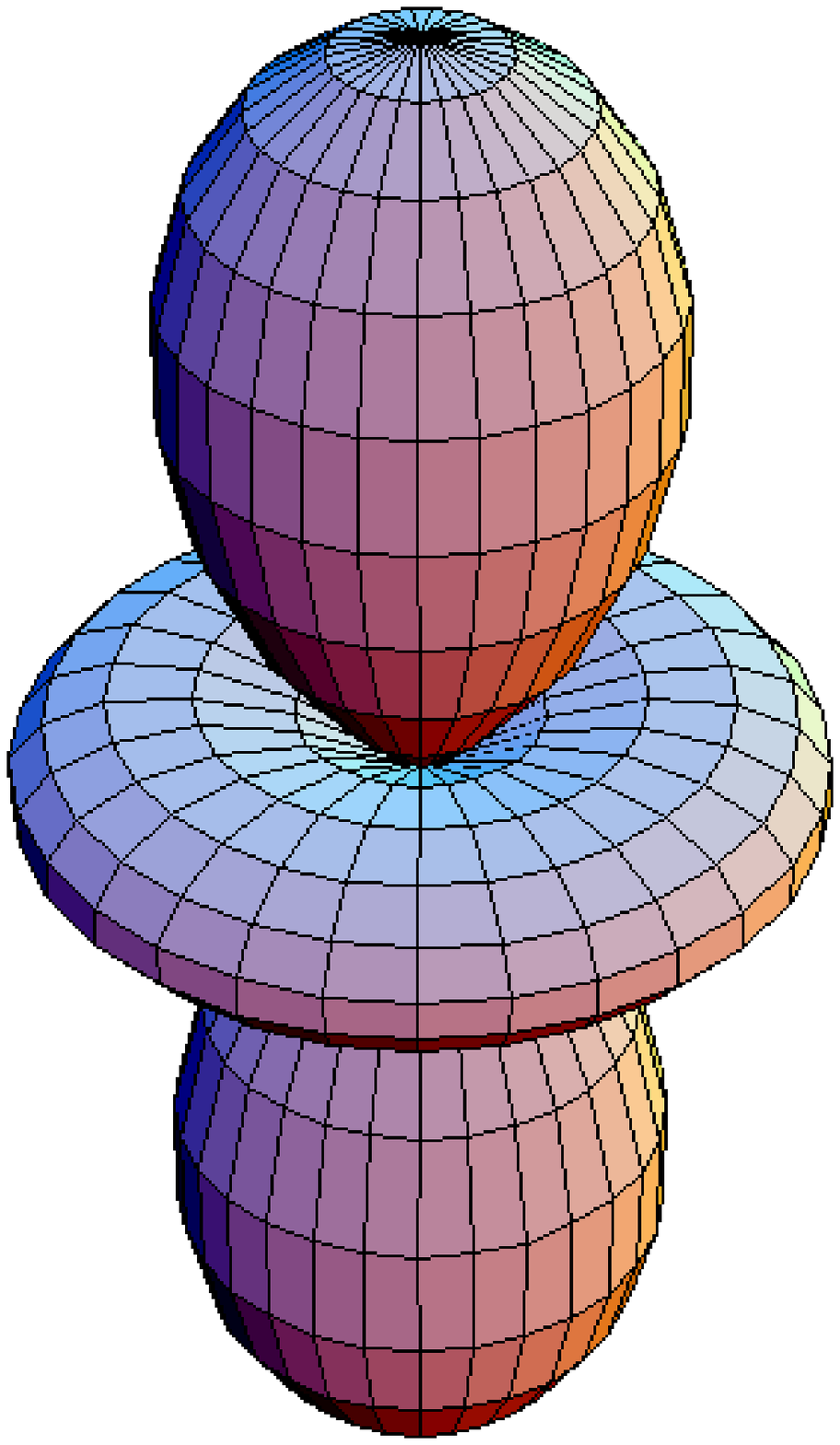}
\includegraphics[width=7cm]{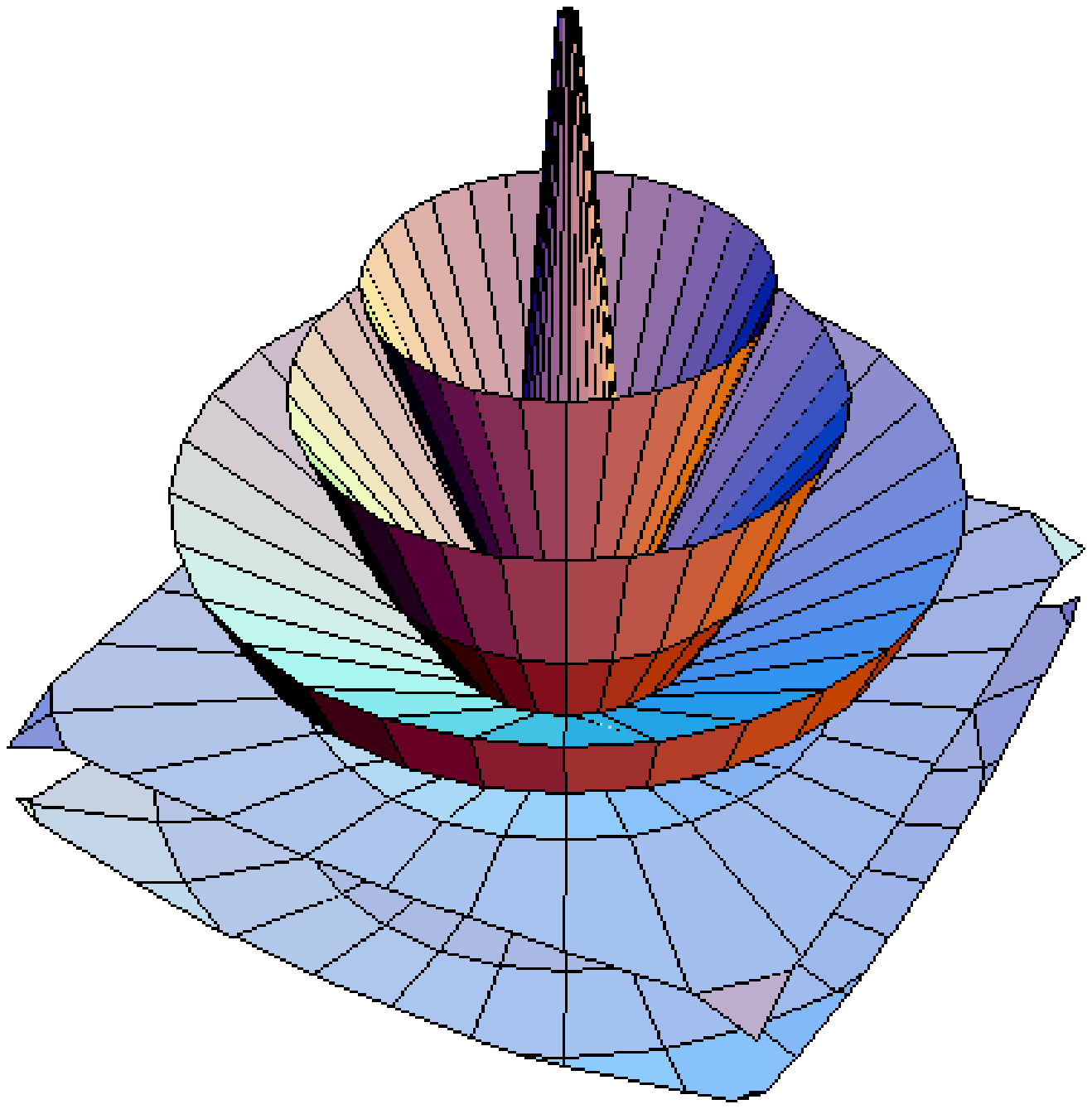}
\caption{$\mid Y^{0}_{2}(\theta,\varphi)\mid$ vs $\mid Z^{0}_{2}(\theta,\varphi)\mid$}
\end{figure}
\begin{figure}
\center
\includegraphics[width=7cm]{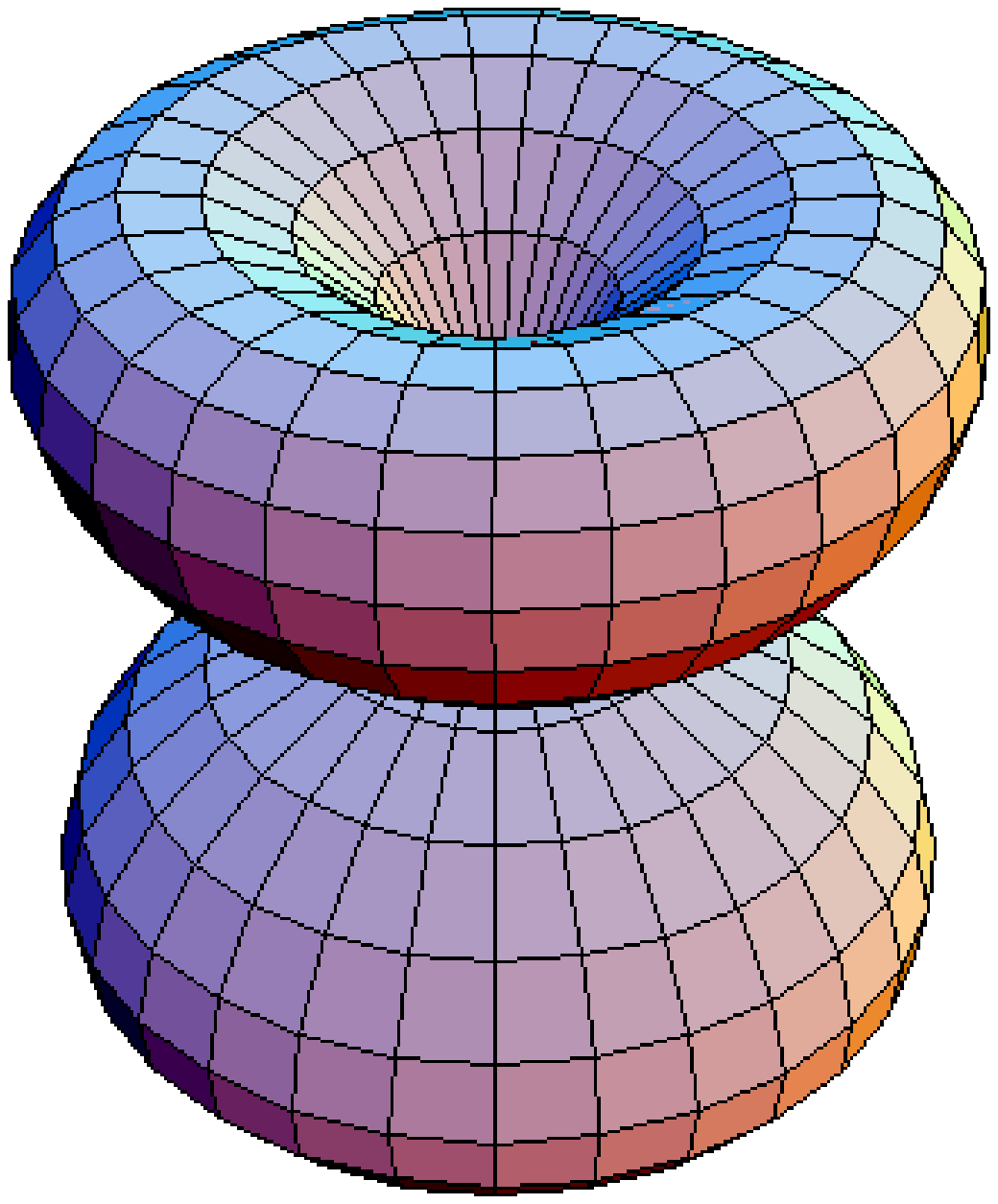}
\includegraphics[width=7cm]{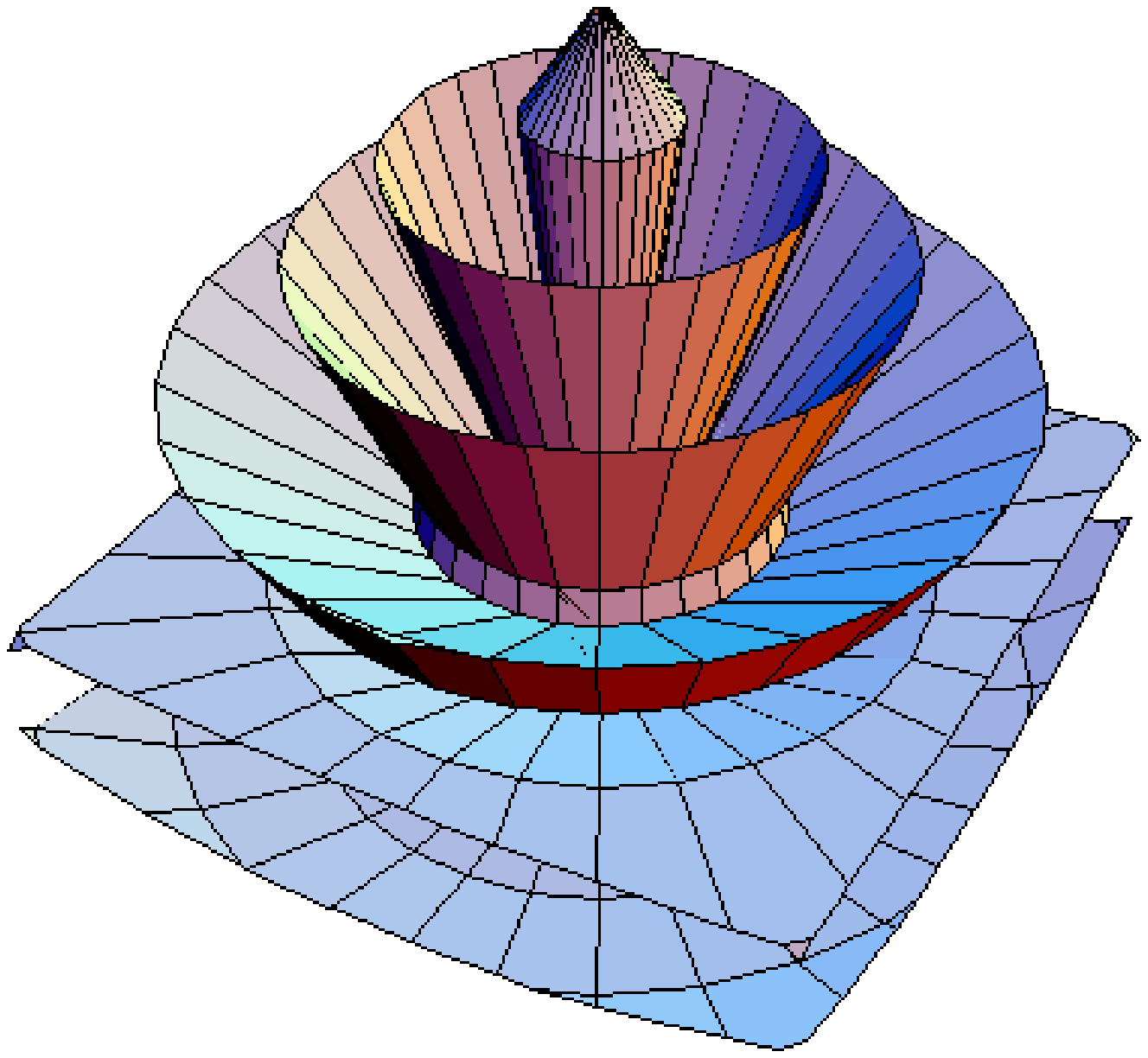}
\caption{$\mid Y^{\pm1}_{2}(\theta,\varphi)\mid$ vs $\mid Z^{\pm1}_{2}(\theta,\varphi)\mid$}
\end{figure}
\begin{figure}
\center
\includegraphics[width=7cm]{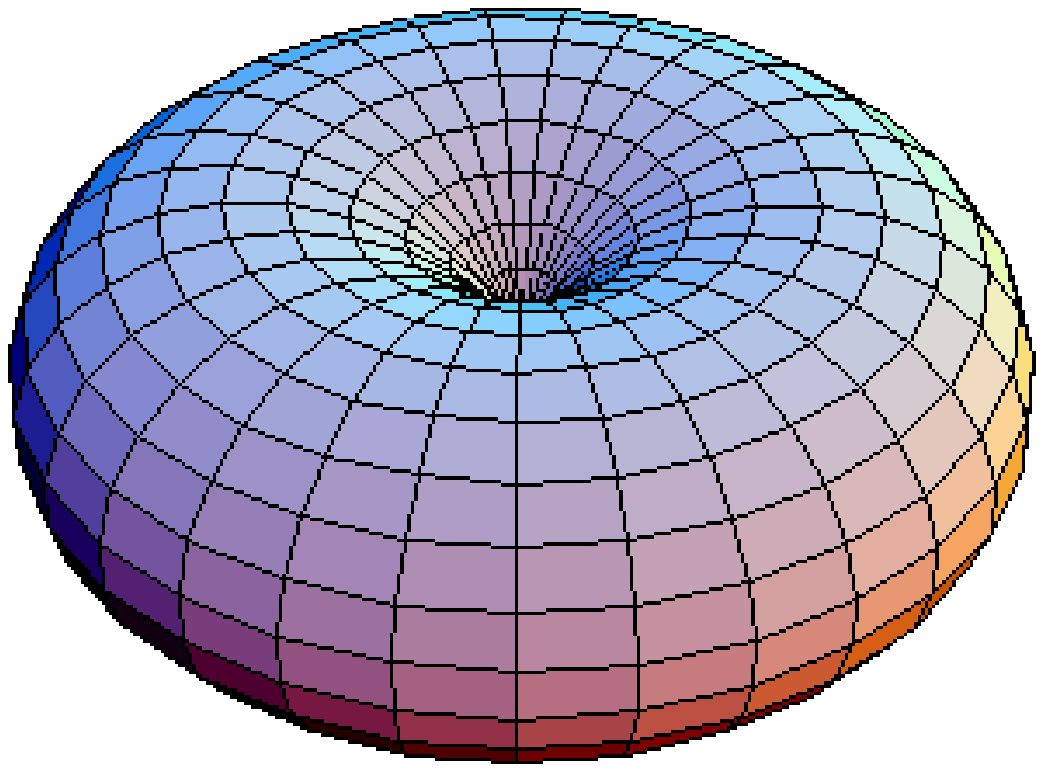}
\includegraphics[width=7cm]{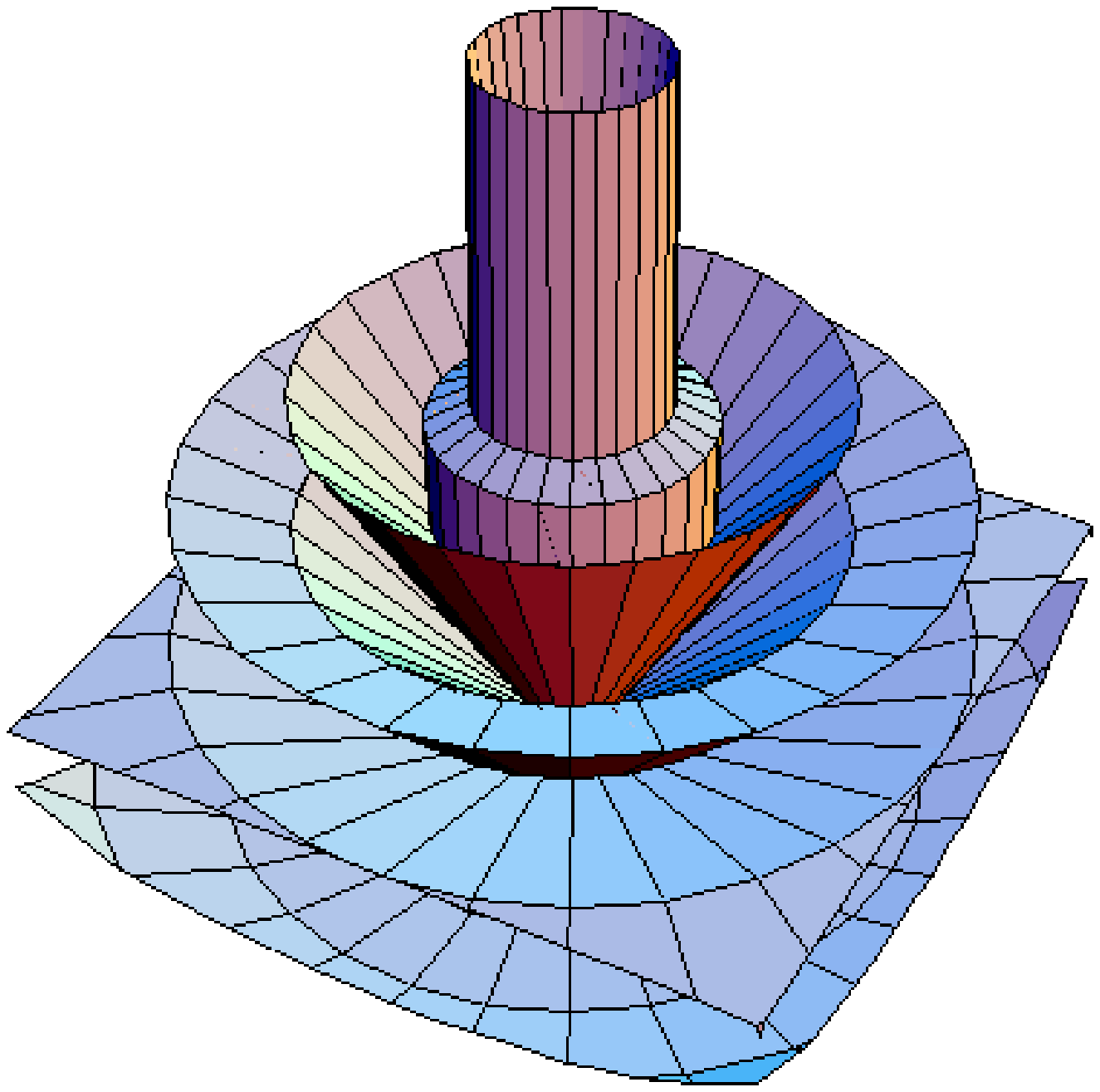}
\caption{$\mid Y^{\pm2}_{2}(\theta,\varphi)\mid$ vs $\mid Z^{\pm2}_{2}(\theta,\varphi)\mid$}
\end{figure}
\begin{figure}[!t]
\center
\includegraphics[width=7cm]{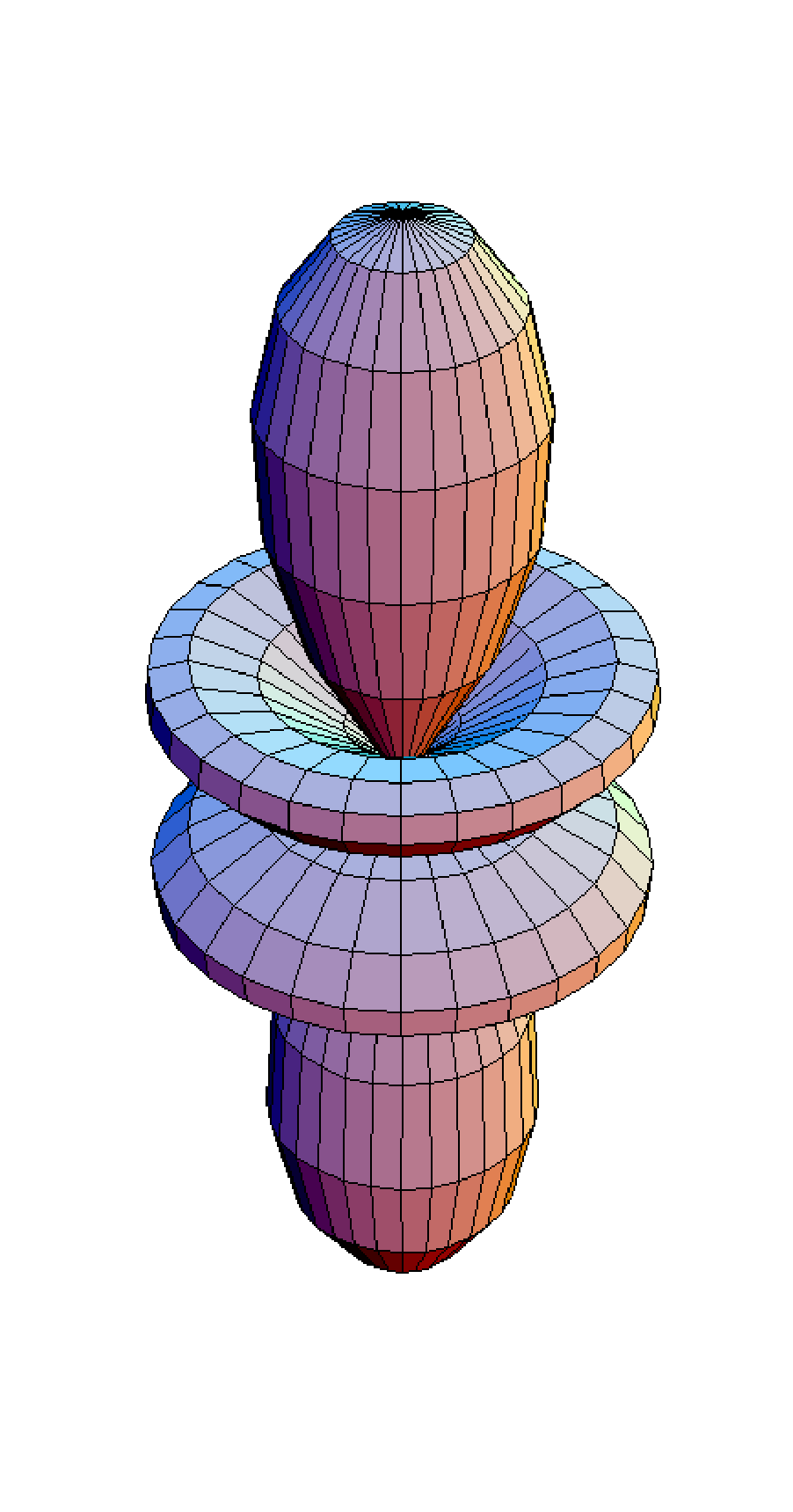}
\includegraphics[width=7cm]{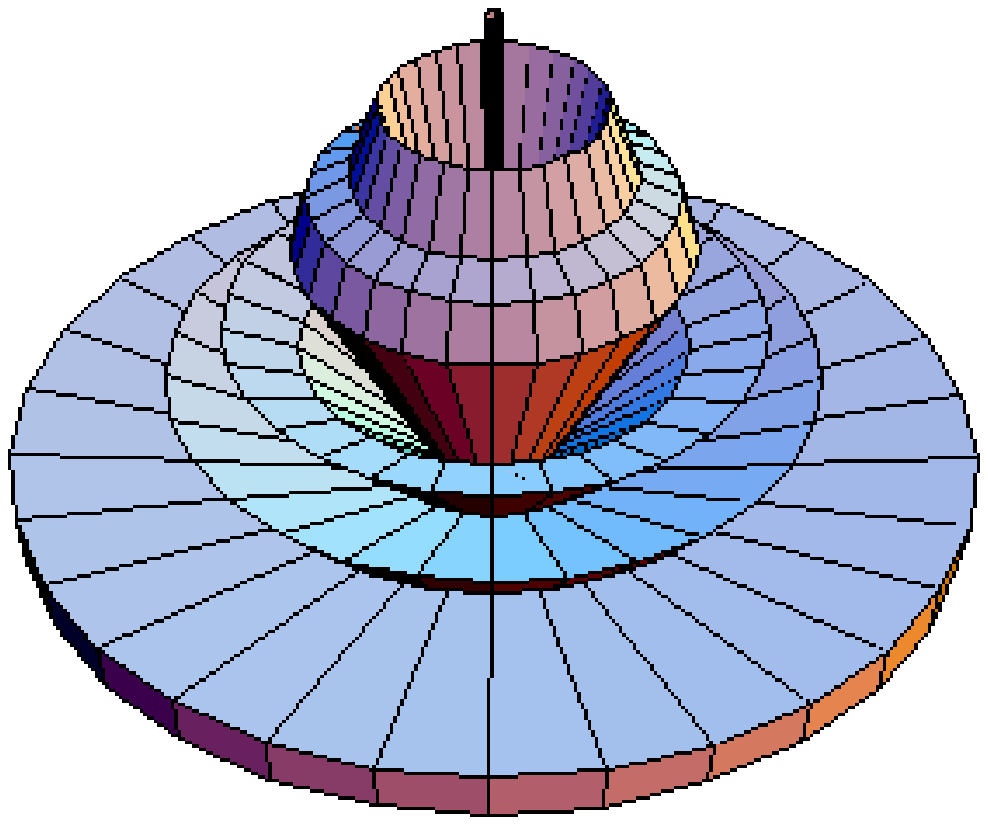}
\caption{$\mid Y^{0}_{3}(\theta,\varphi)\mid$ vs $\mid Z^{0}_{3}(\theta,\varphi)\mid$}
\end{figure}
\begin{figure}[!h]
\center
\includegraphics[width=7cm]{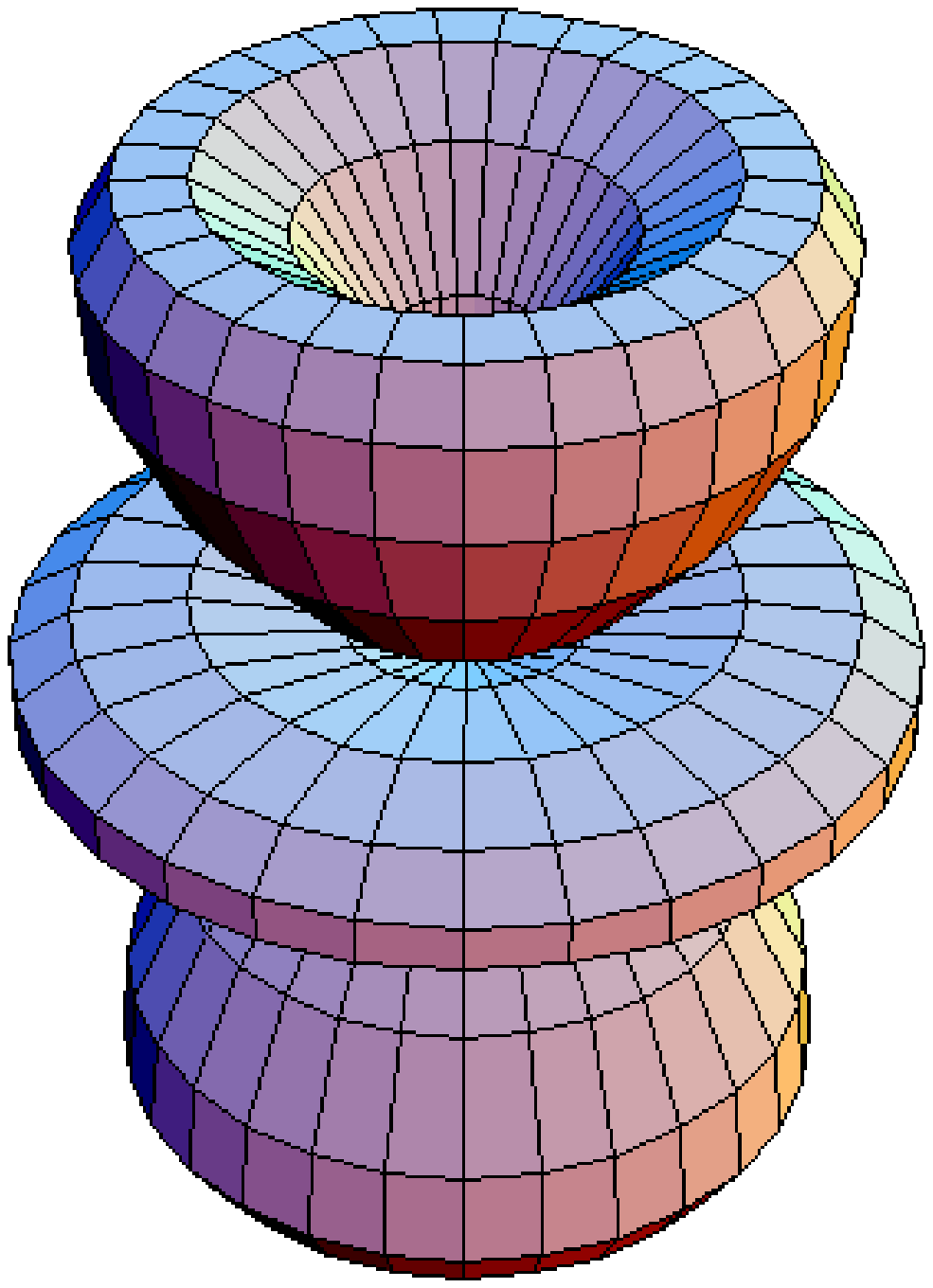}
\includegraphics[width=7cm]{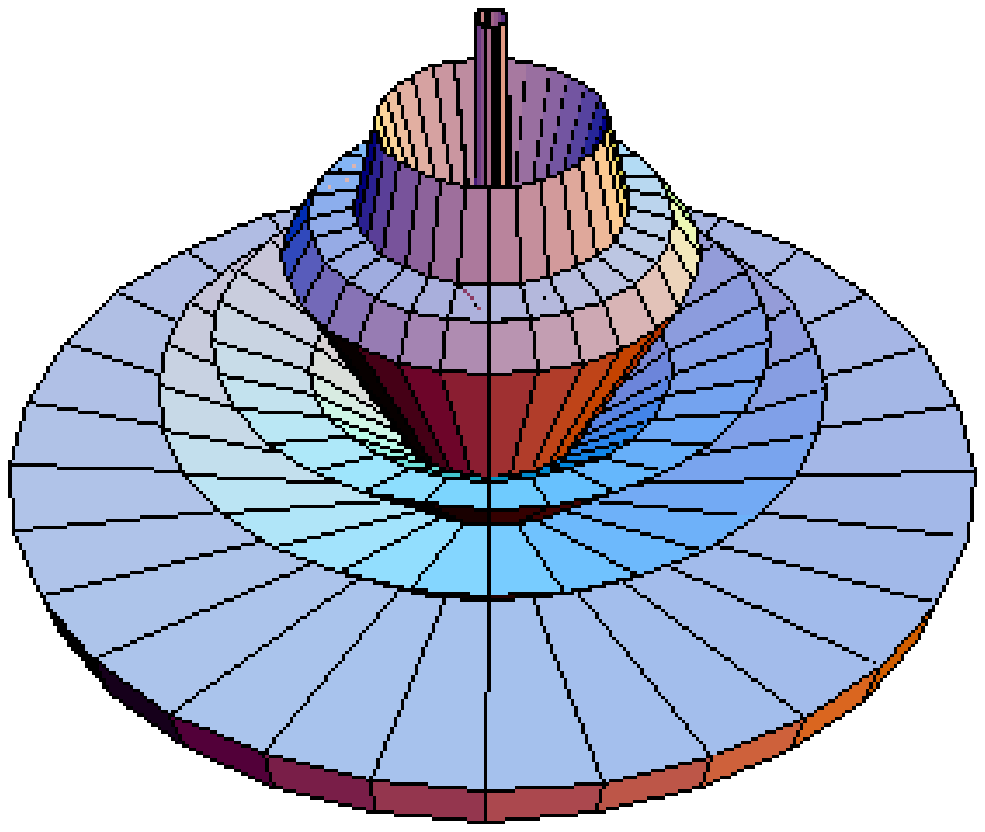}
\caption{$\mid Y^{\pm1}_{3}(\theta,\varphi)\mid$ vs $\mid Z^{\pm1}_{3}(\theta,\varphi)\mid$}
\end{figure}
\begin{figure}
\center
\includegraphics[width=7cm]{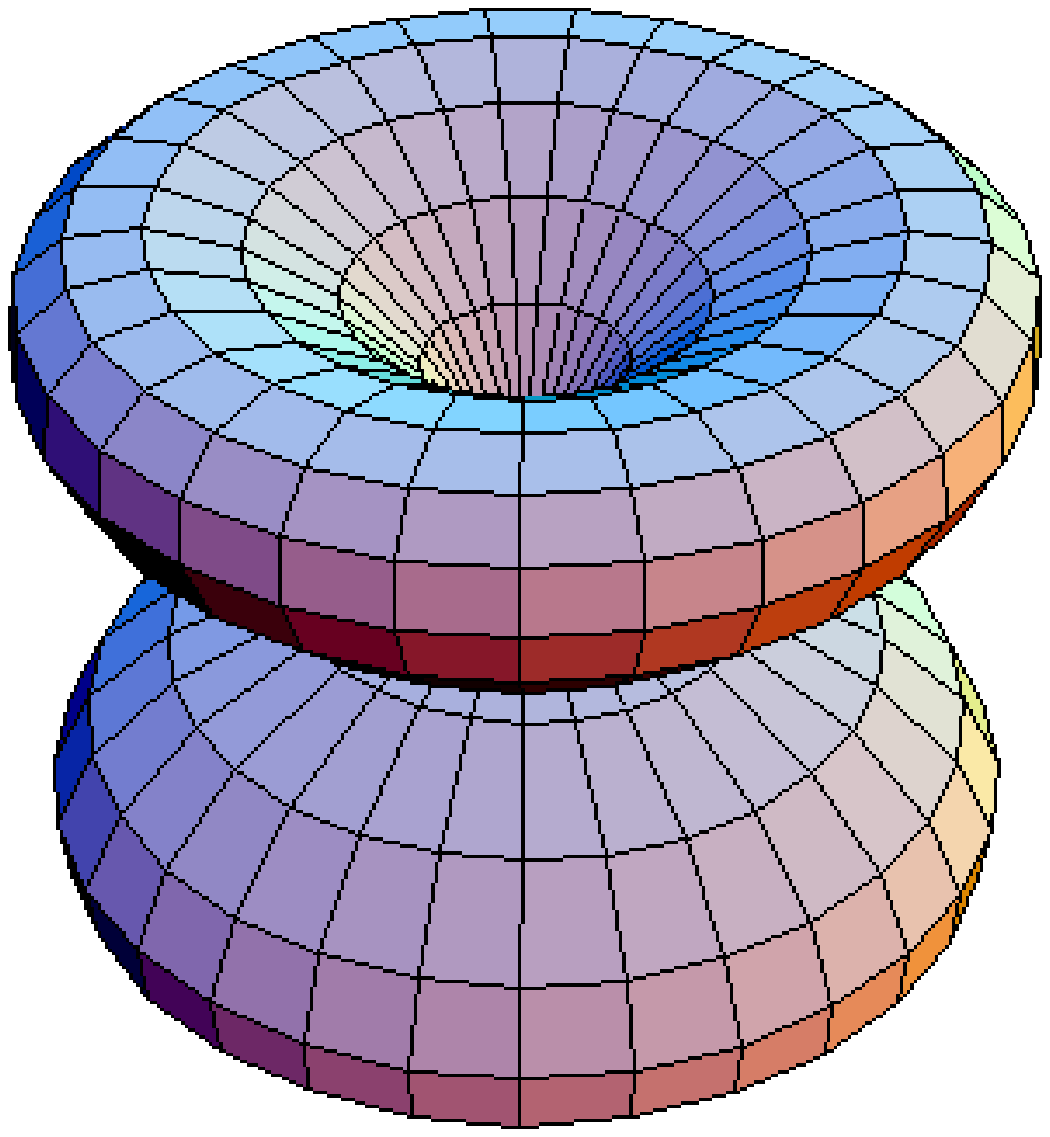}
\includegraphics[width=7cm]{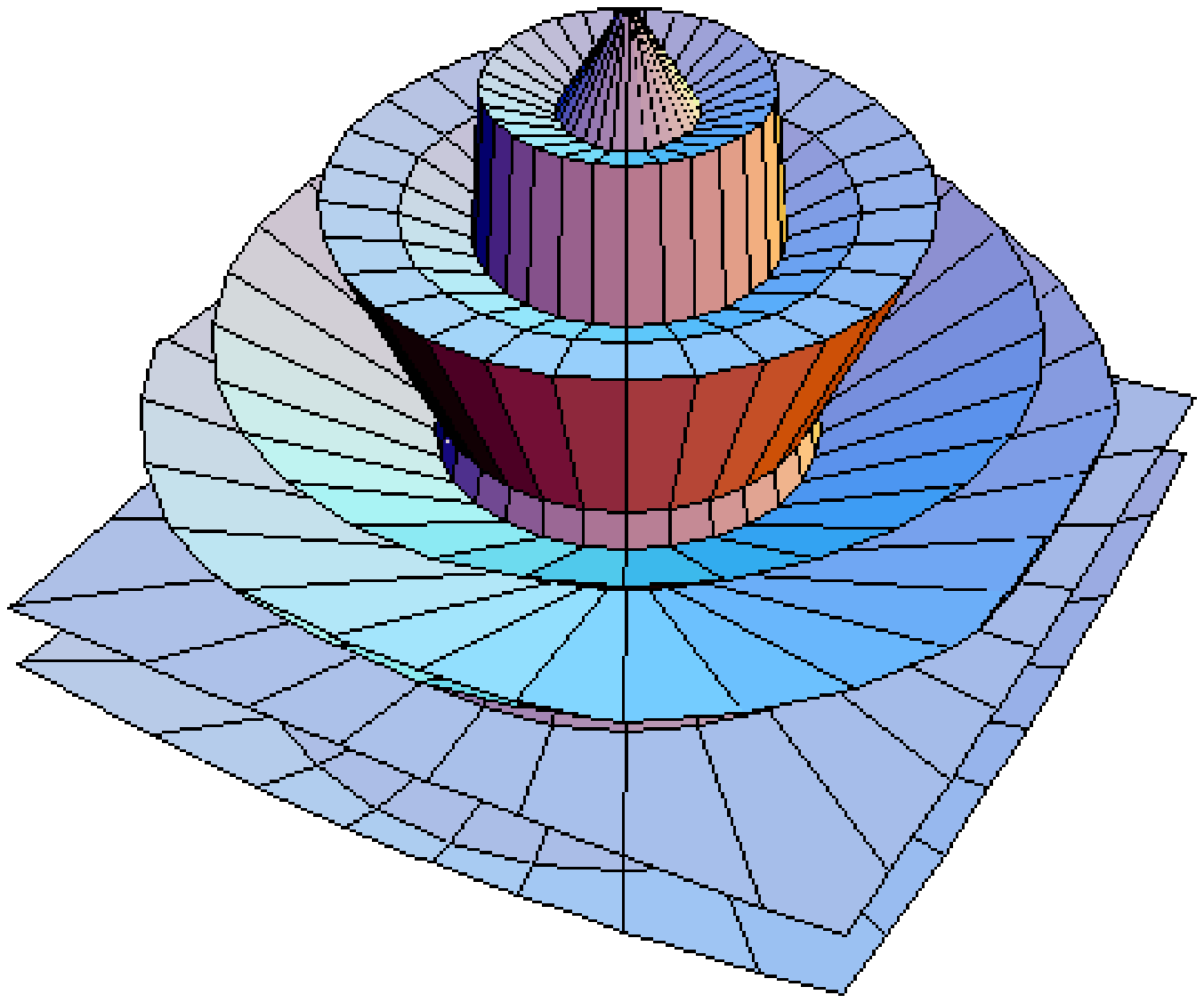}
\caption{$\mid Y^{\pm2}_{3}(\theta,\varphi)\mid$ vs $\mid Z^{\pm2}_{3}(\theta,\varphi)\mid$}
\end{figure}
\begin{figure}[!ht]
\center
\includegraphics[width=7cm]{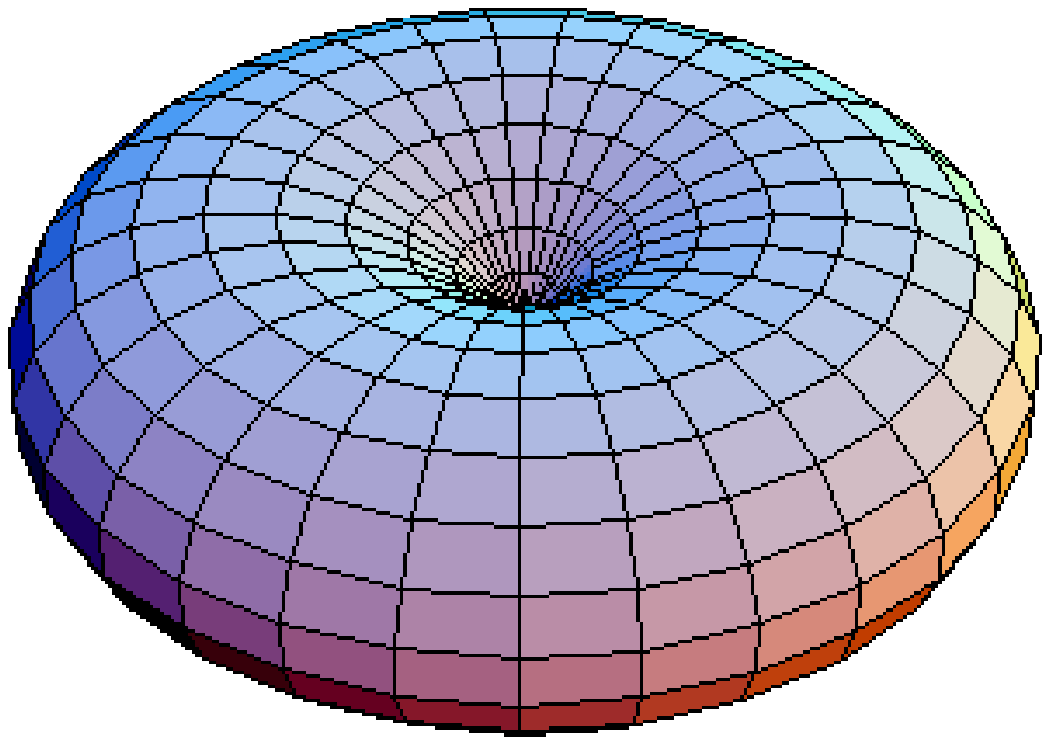}
\includegraphics[width=7cm]{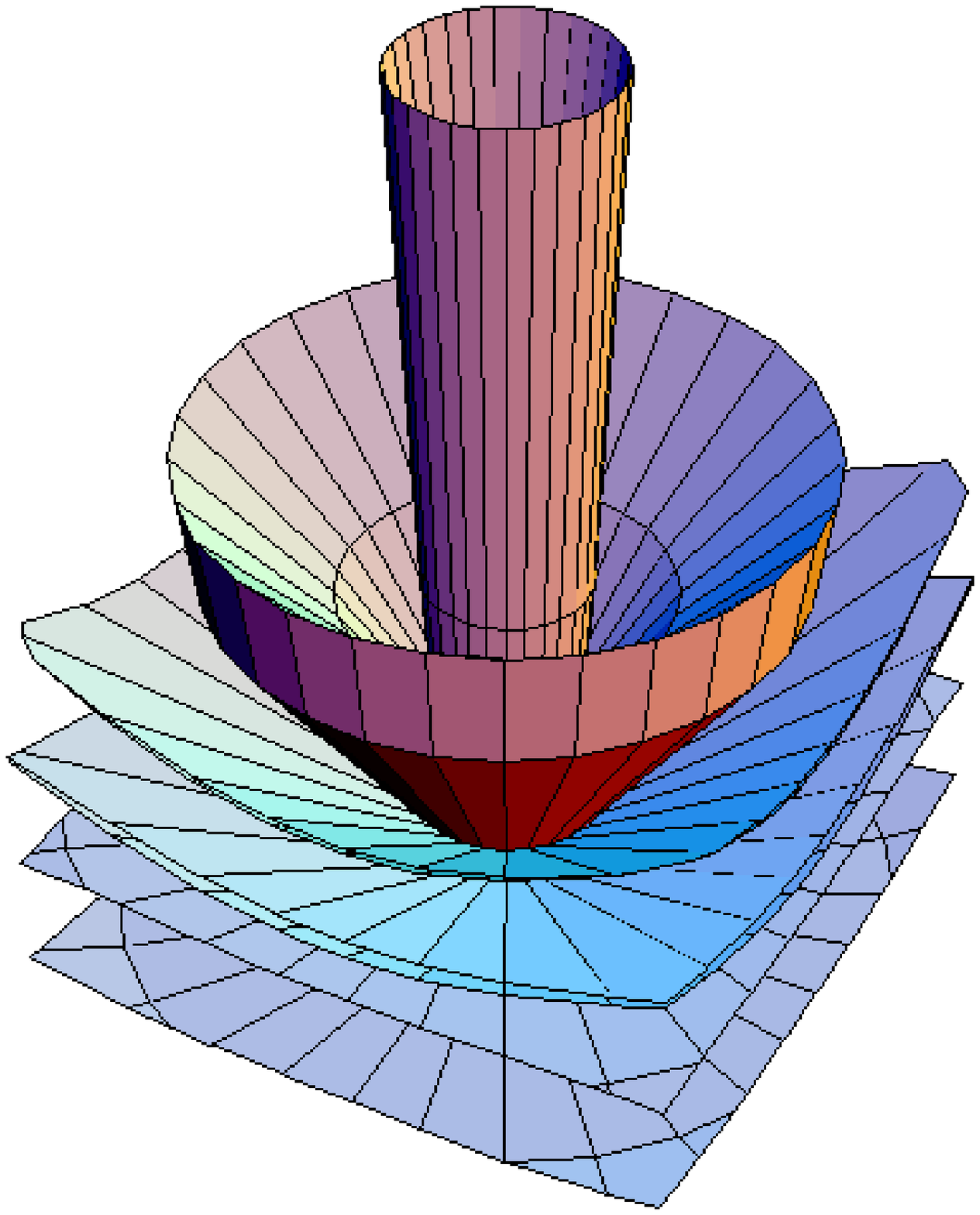}
\caption{$\mid Y^{\pm3}_{3}(\theta,\varphi)\mid$ vs $\mid Z^{\pm3}_{3}(\theta,\varphi)\mid$}
\end{figure}
\begin{figure}[!hb]
\center
\includegraphics[width=7cm]{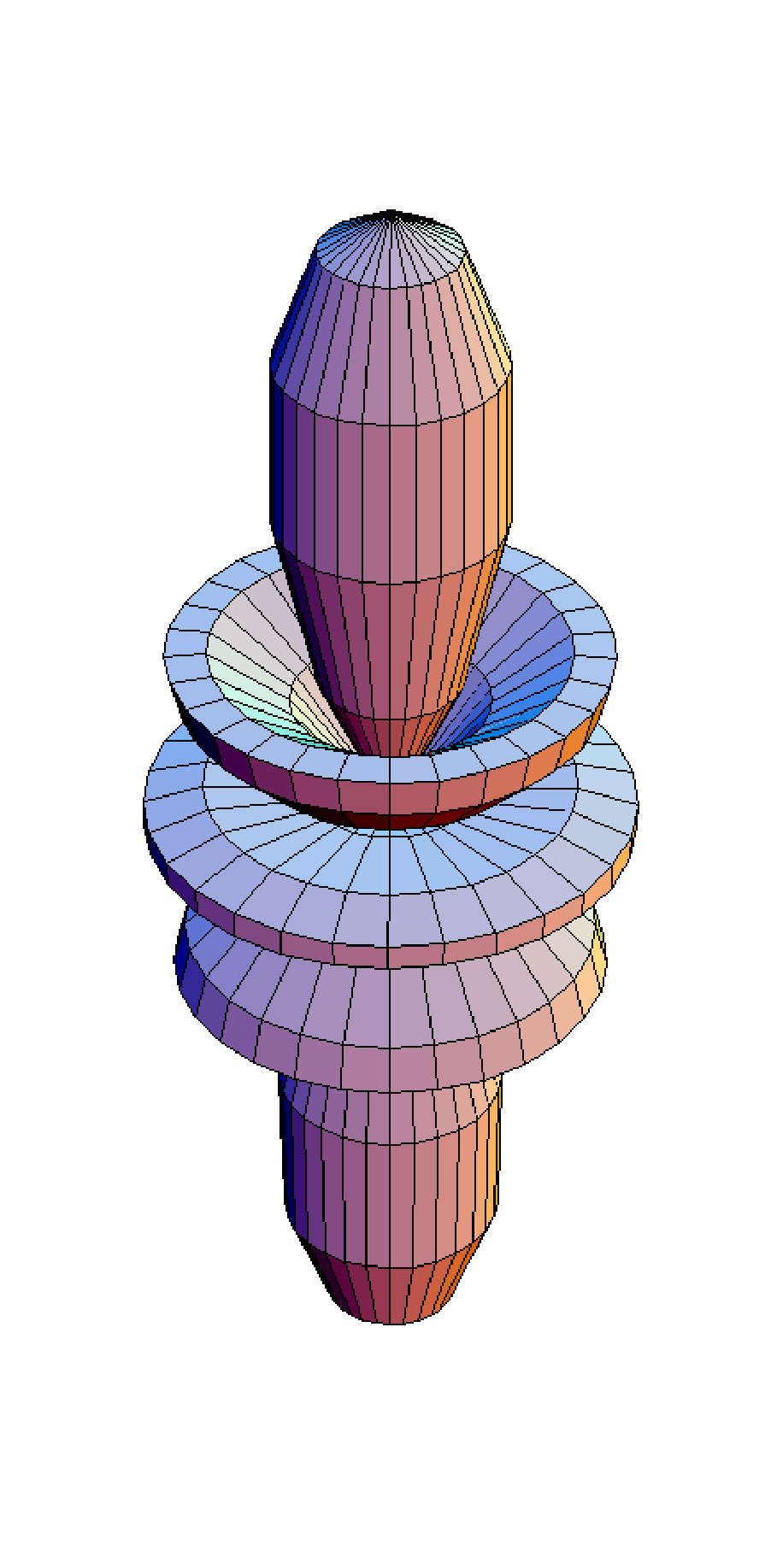}
\includegraphics[width=7cm]{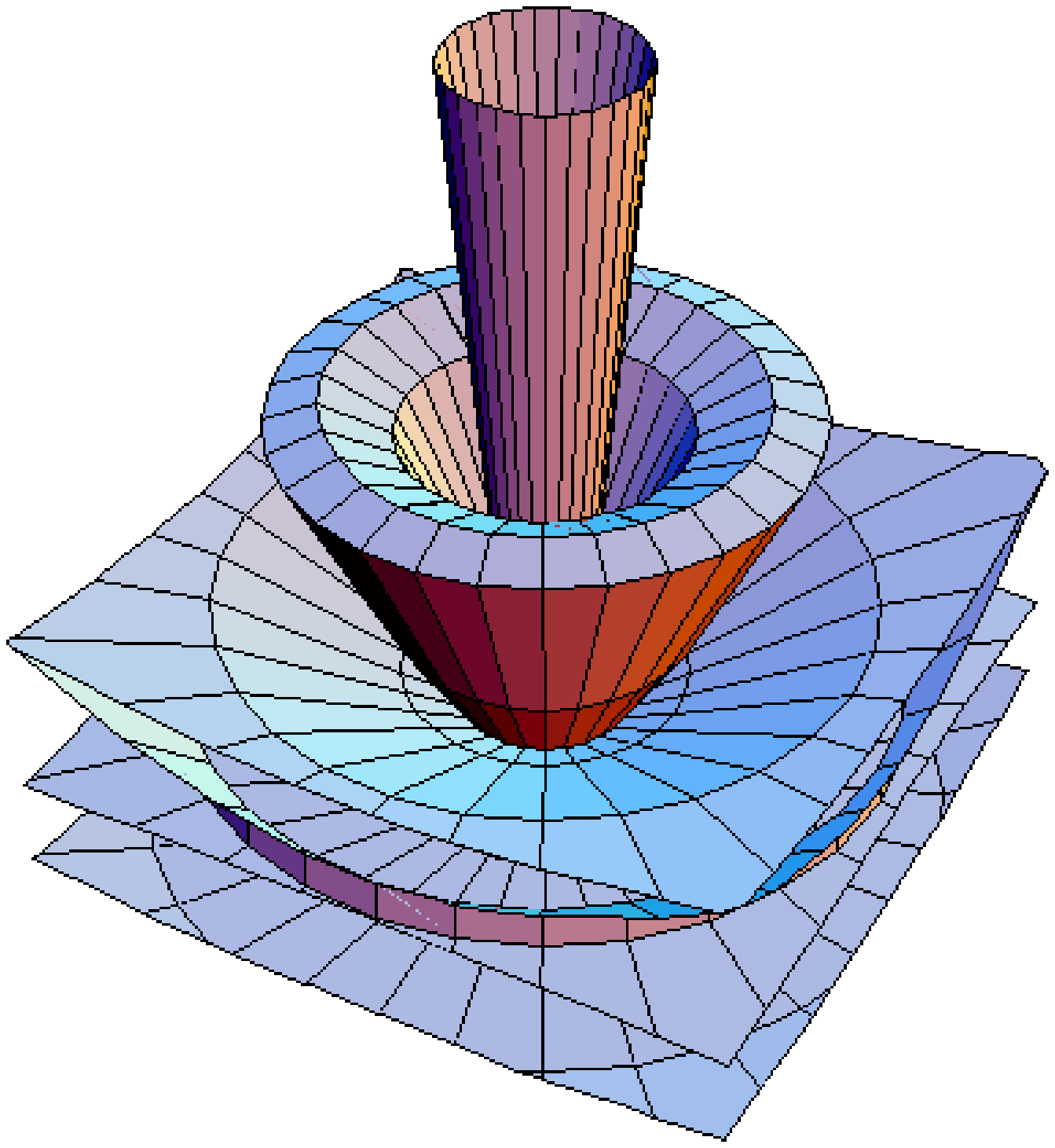}
\caption{$\mid Y^{0}_{4}(\theta,\varphi)\mid$ vs $\mid Z^{0}_{4}(\theta,\varphi)\mid$}
\end{figure}
\begin{figure}
\center
\includegraphics[width=7cm]{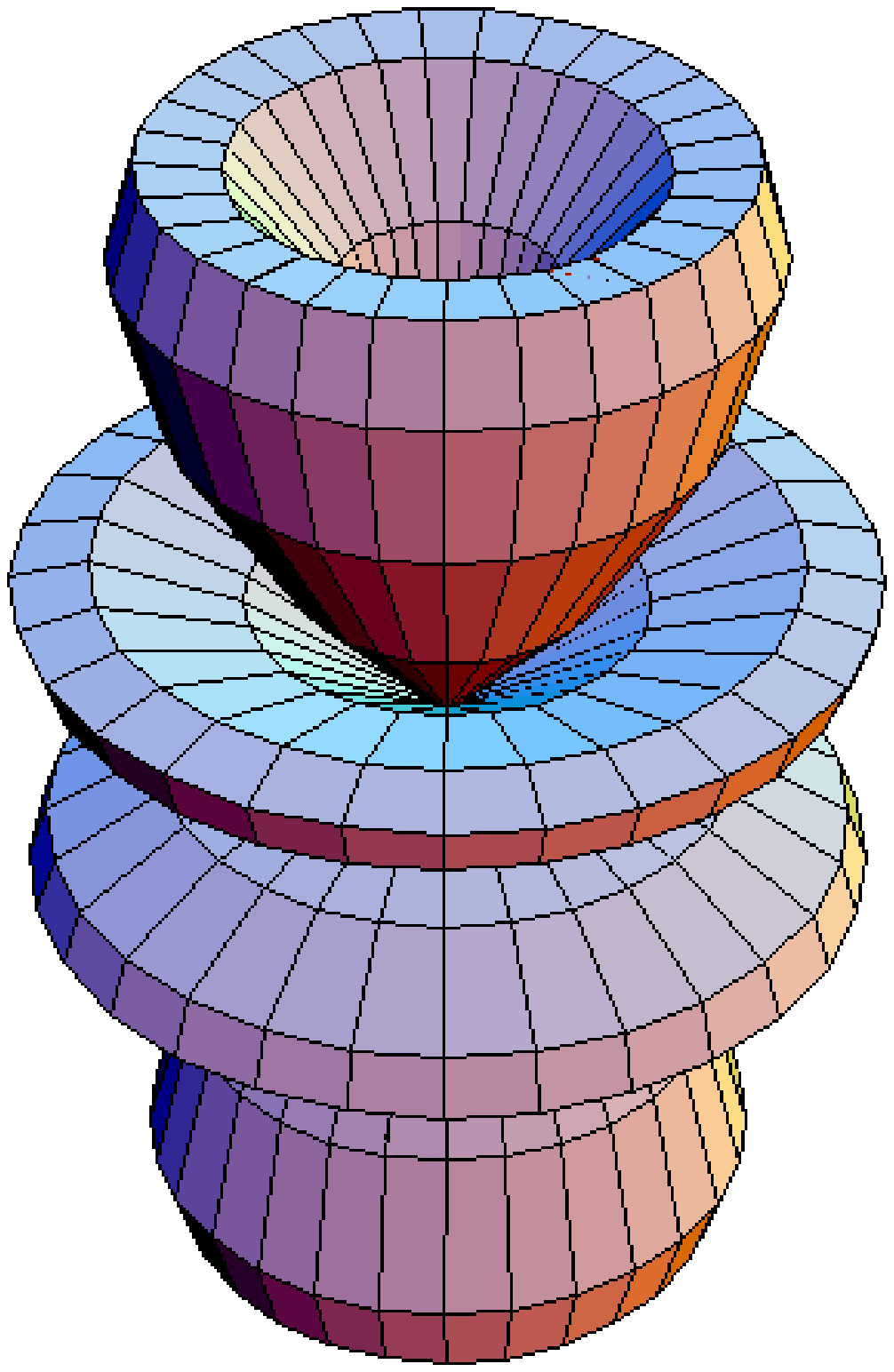}
\includegraphics[width=7cm]{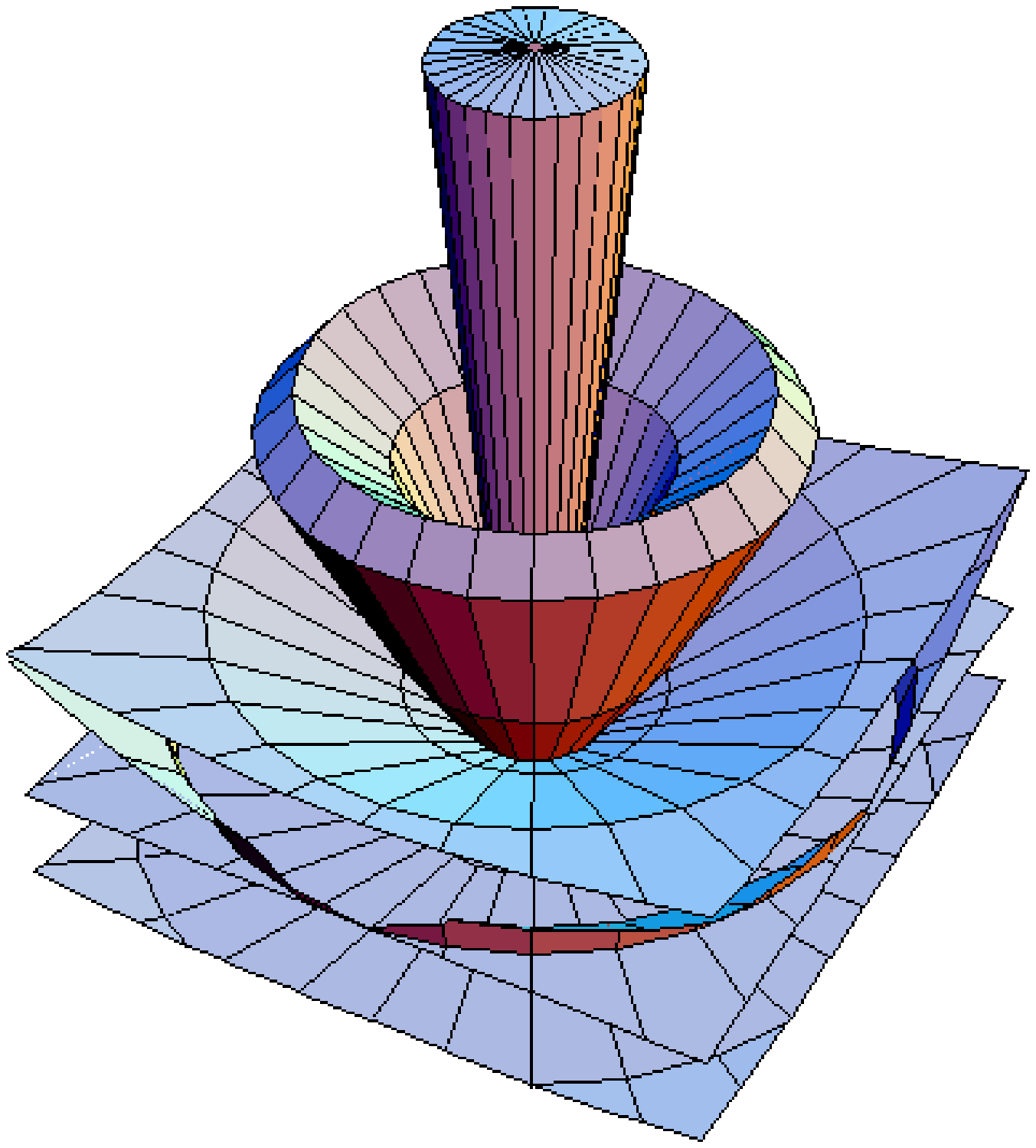}
\caption{$\mid Y^{\pm1}_{4}(\theta,\varphi)\mid$ vs $\mid Z^{\pm1}_{4}(\theta,\varphi)\mid$}
\end{figure}
\begin{figure}
\center
\includegraphics[width=7cm]{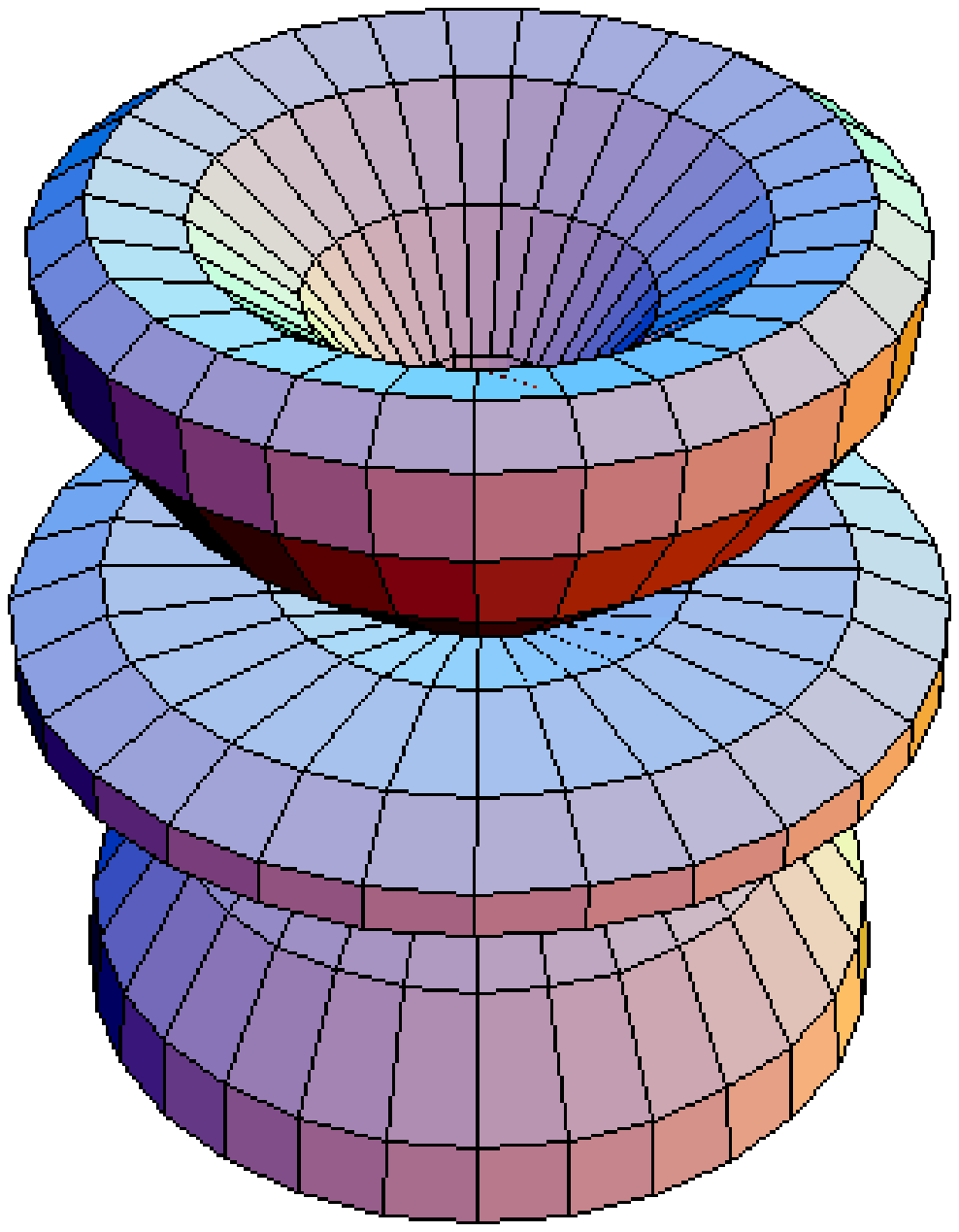}
\includegraphics[width=7cm]{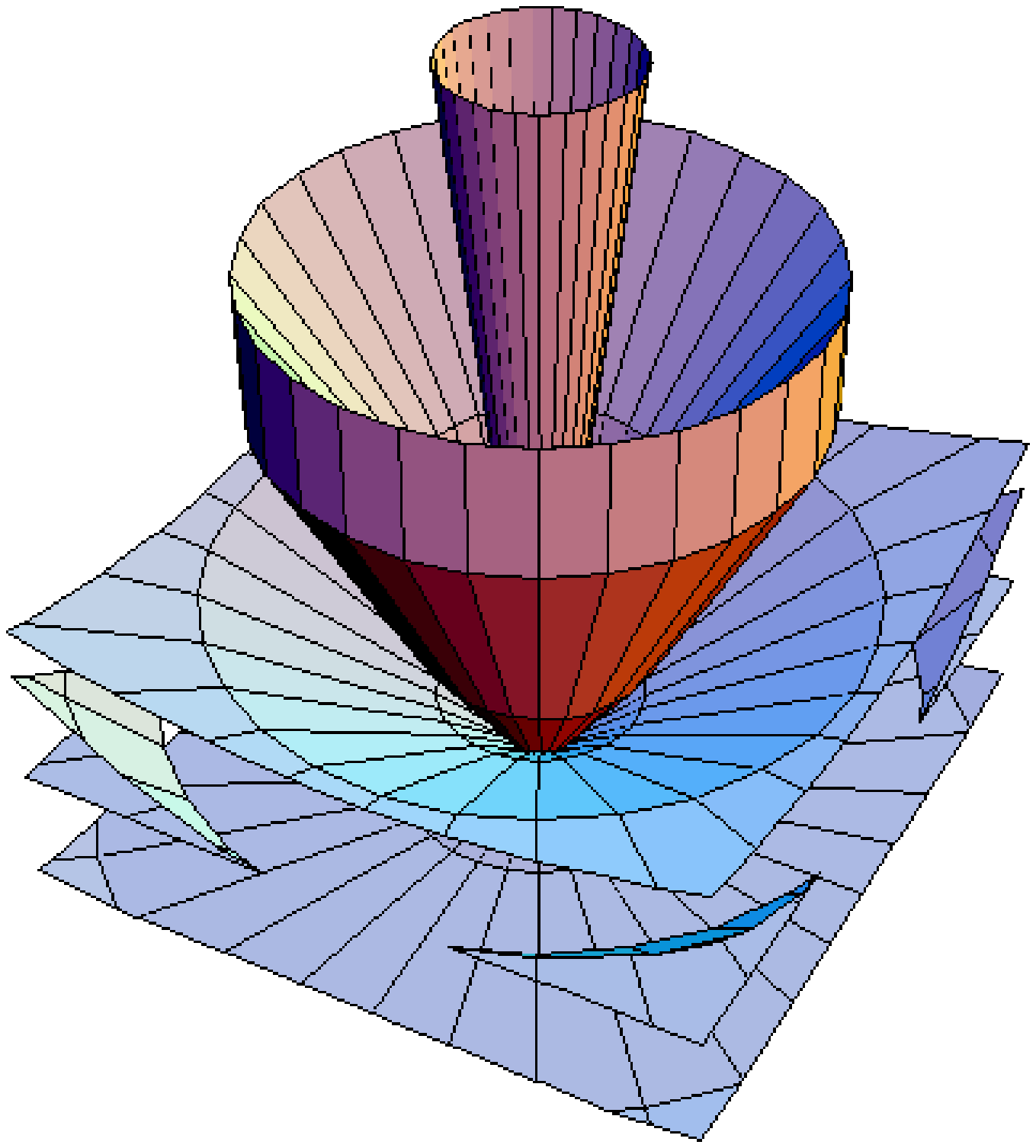}
\caption{$\mid Y^{\pm2}_{4}(\theta,\varphi)\mid$ vs $\mid Z^{\pm2}_{4}(\theta,\varphi)\mid$}
\end{figure}
\begin{figure}
\center
\includegraphics[width=7cm]{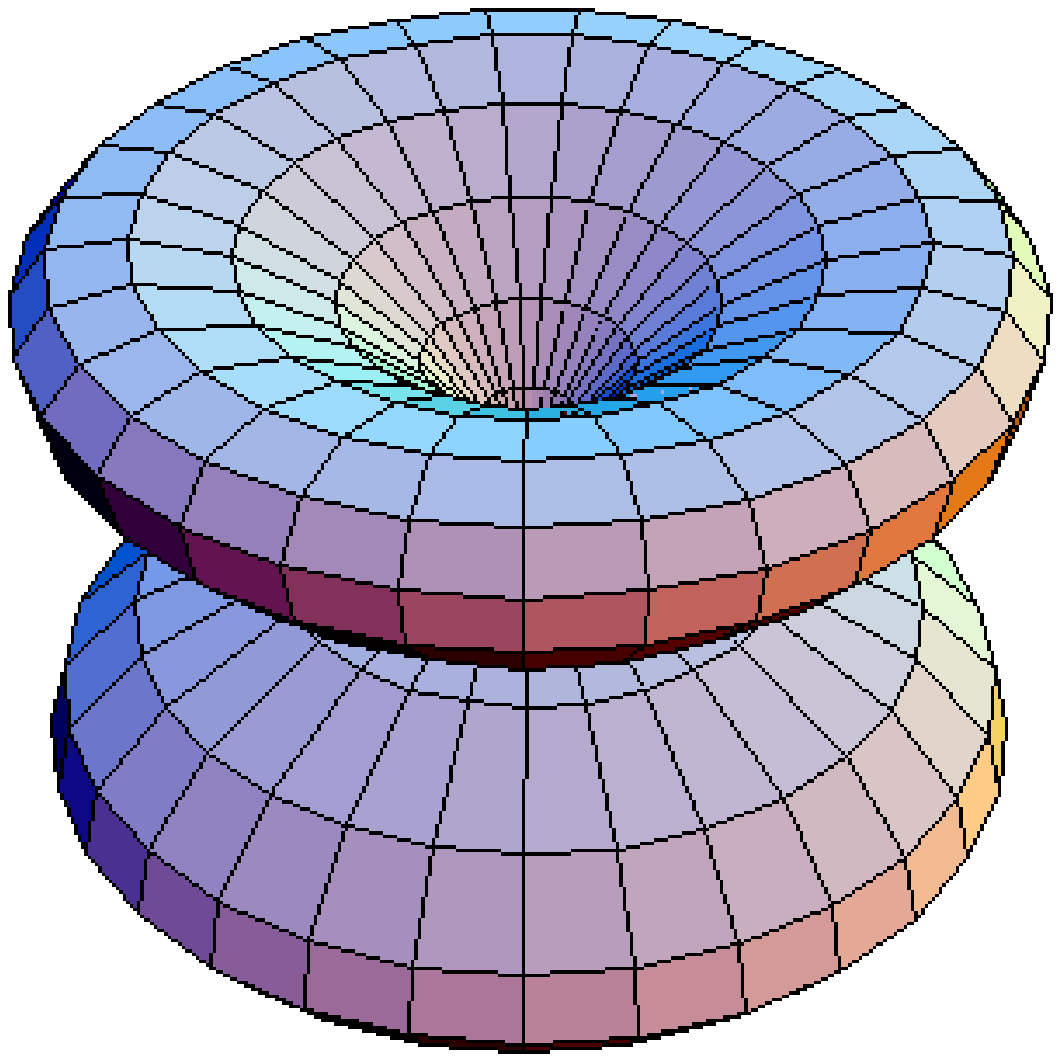}
\includegraphics[width=7cm]{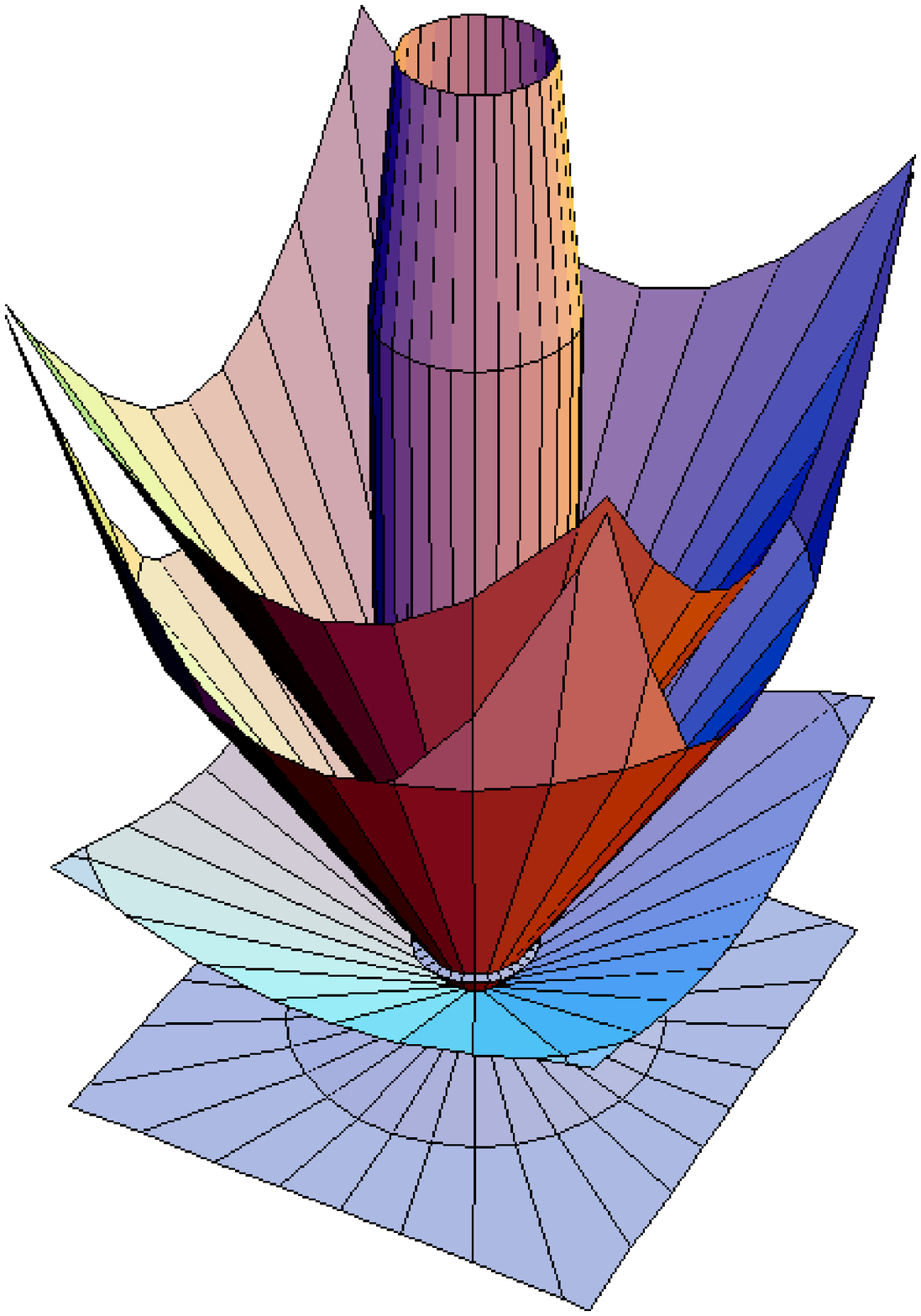}
\caption{$\mid Y^{\pm3}_{4}(\theta,\varphi)\mid$ vs $\mid Z^{\pm3}_{4}(\theta,\varphi)\mid$}
\end{figure}
\begin{figure}
\center
\includegraphics[width=7cm]{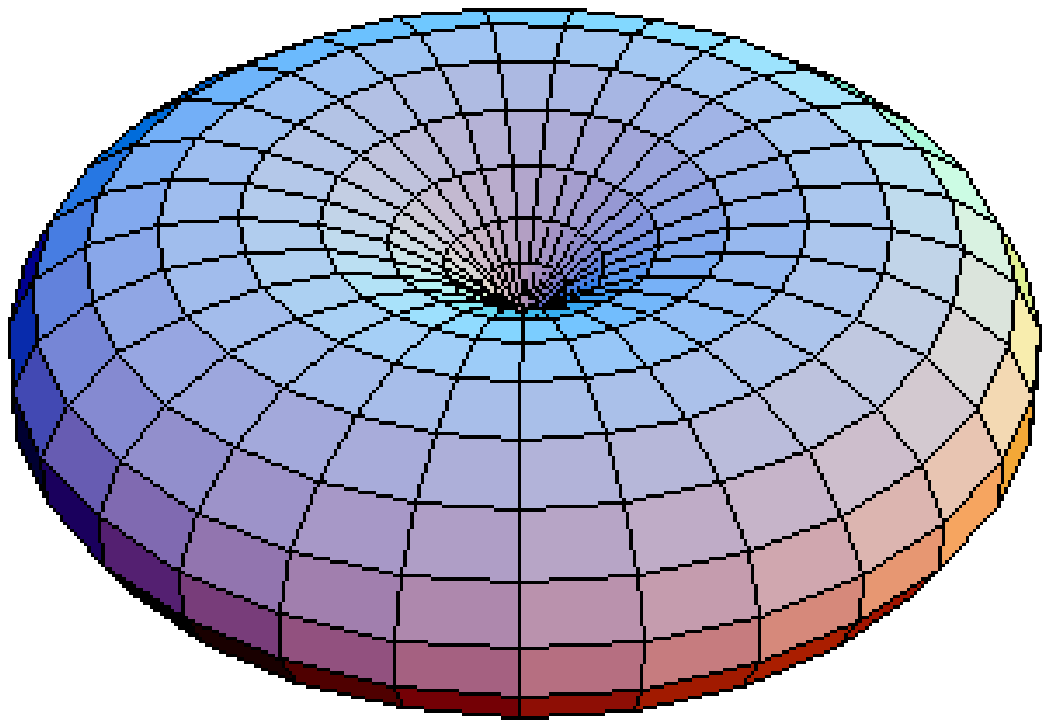}
\includegraphics[width=7cm]{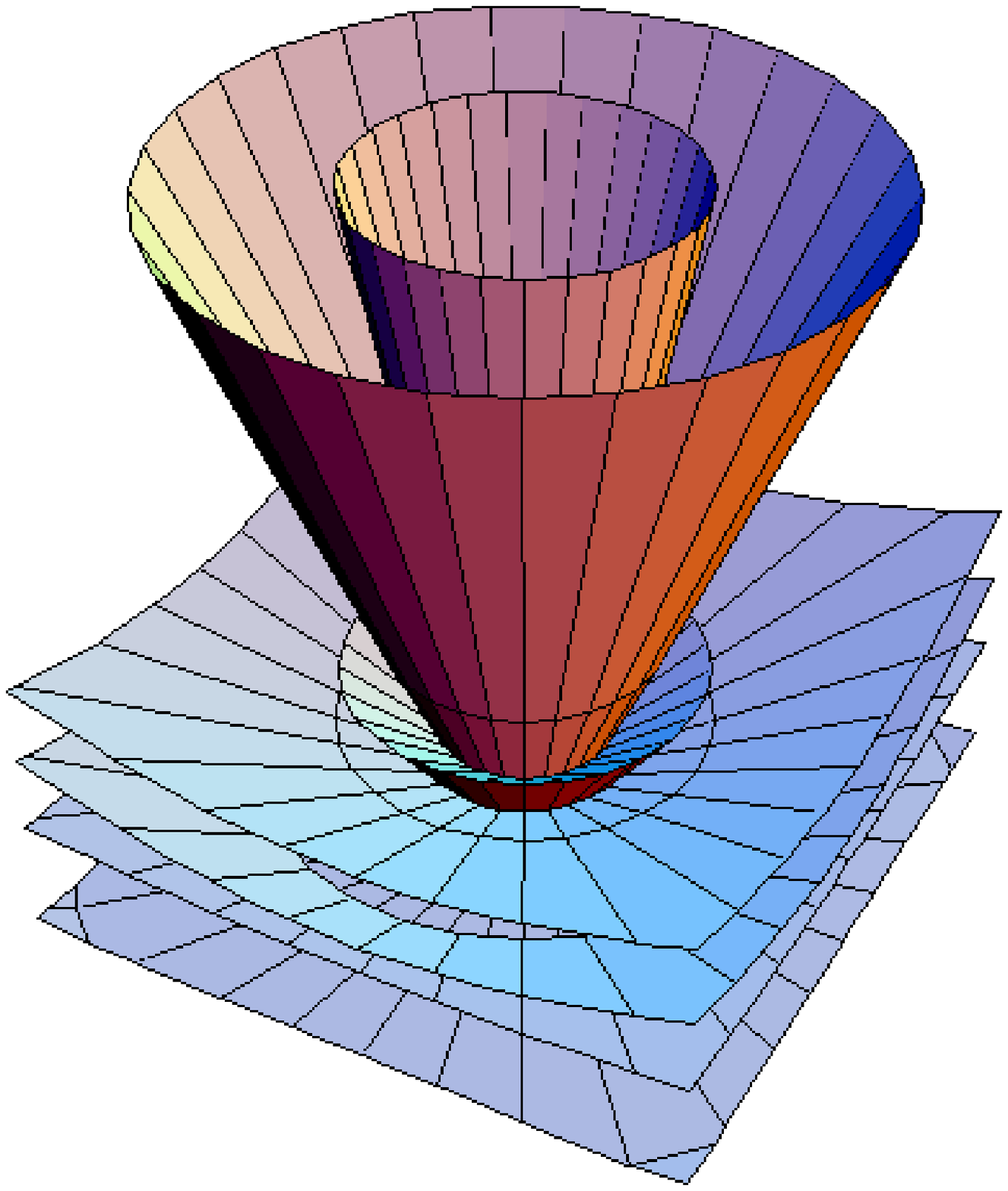}
\caption{$\mid Y^{\pm4}_{4}(\theta,\varphi)\mid$ vs $\mid Z^{\pm4}_{4}(\theta,\varphi)\mid$}
\end{figure}
\begin{figure}
\center
\includegraphics[width=7cm]{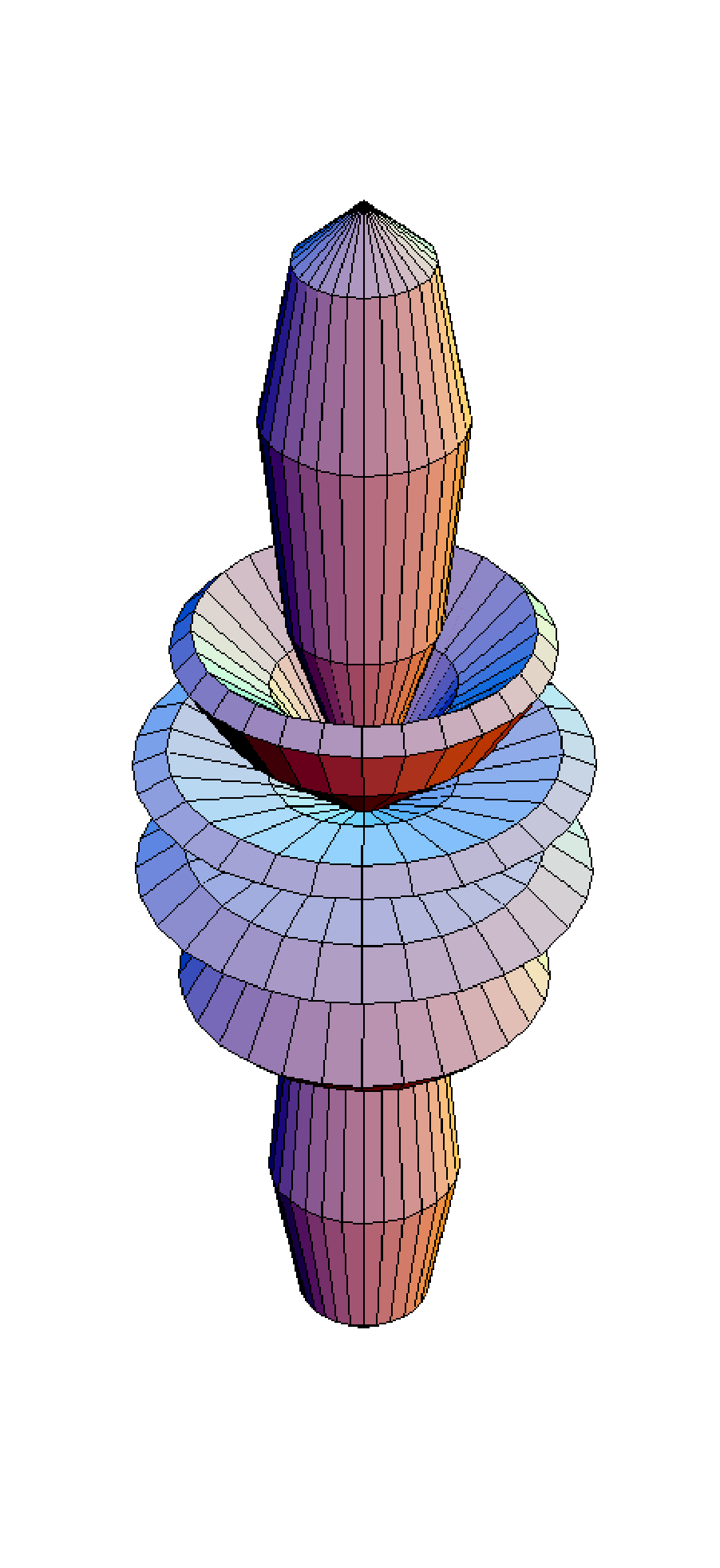}
\includegraphics[width=7cm]{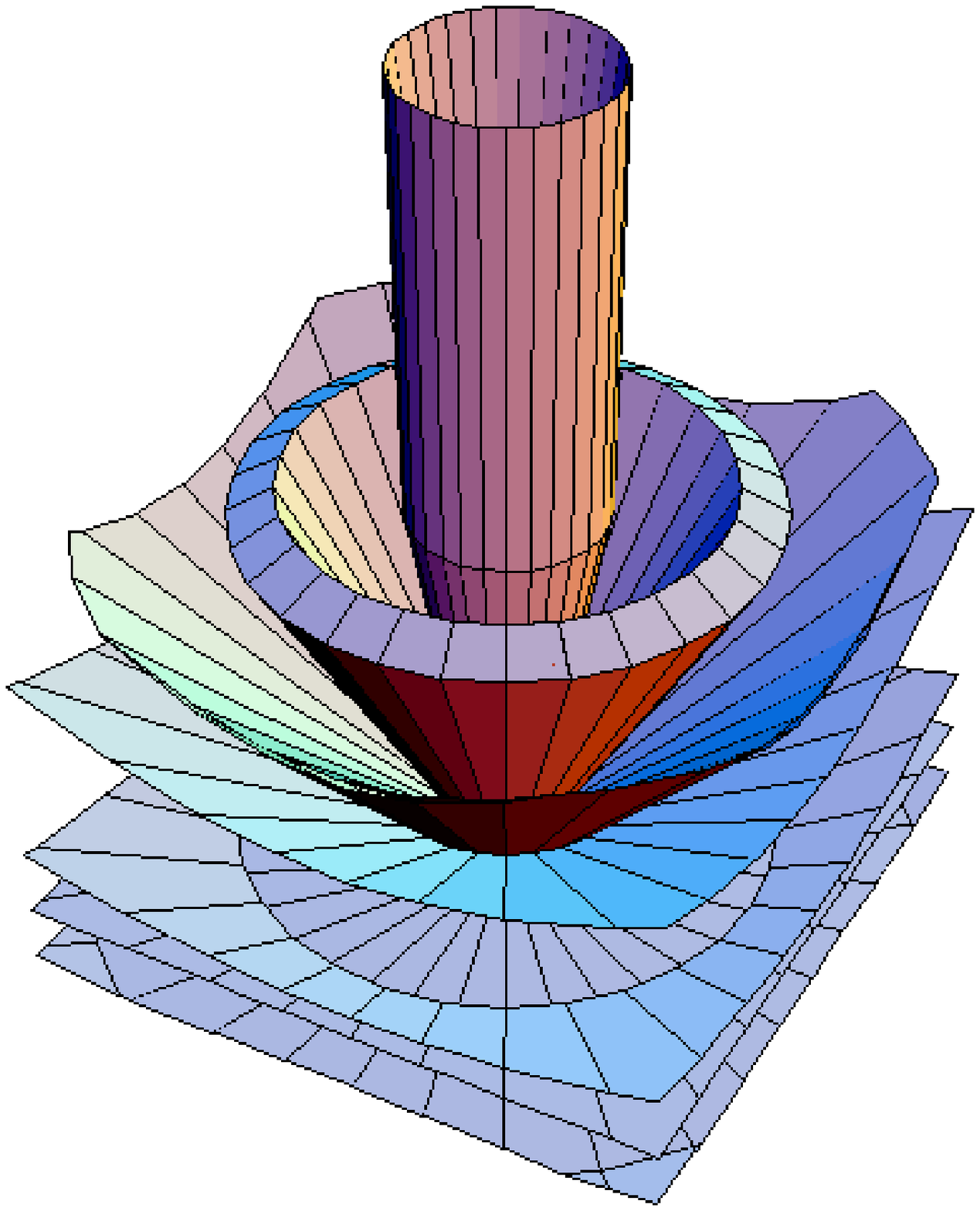}
\caption{$\mid Y^{0}_{5}(\theta,\varphi)\mid$ vs $\mid Z^{0}_{5}(\theta,\varphi)\mid$}
\end{figure}
\begin{figure}
\center
\includegraphics[width=7cm]{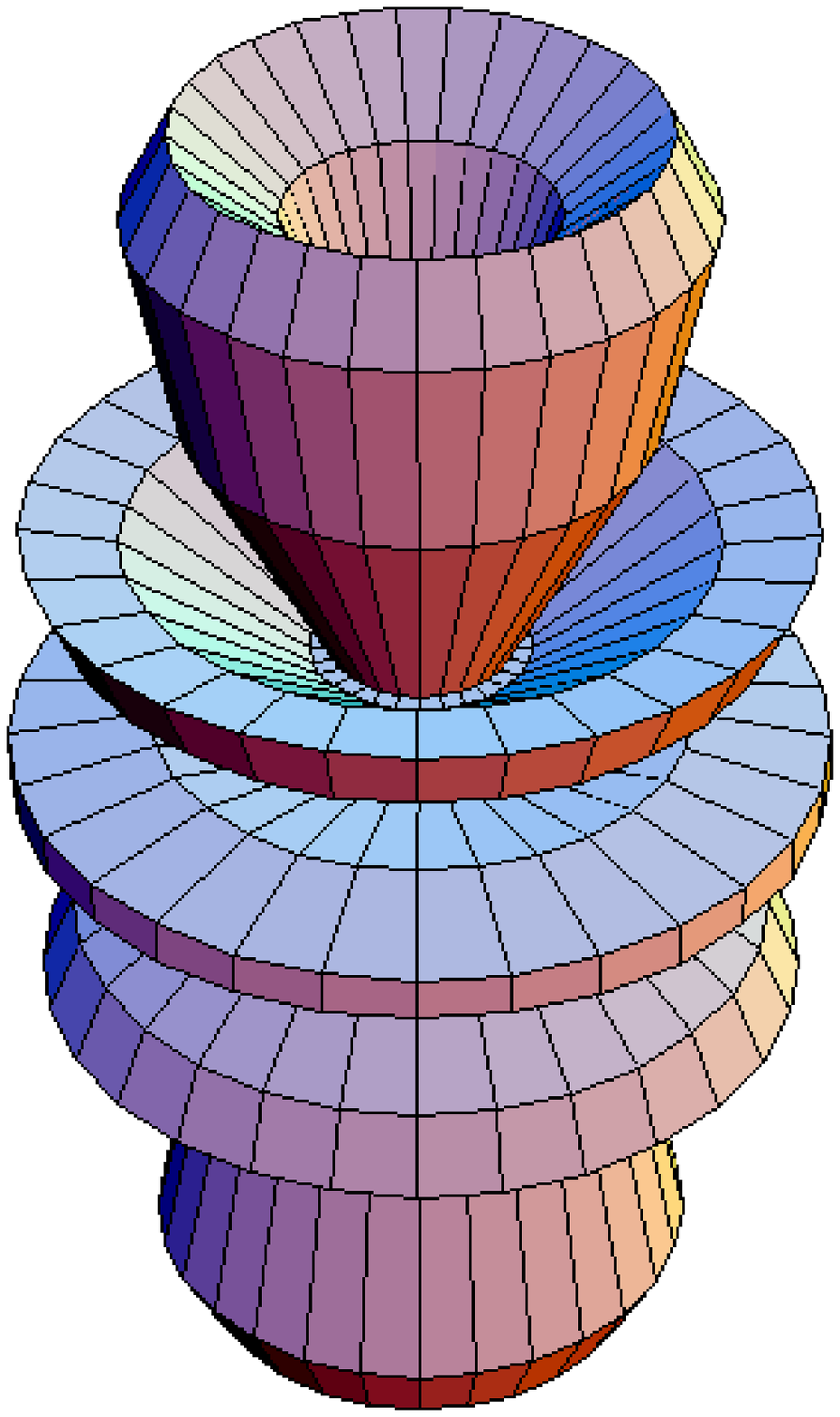}
\includegraphics[width=7cm]{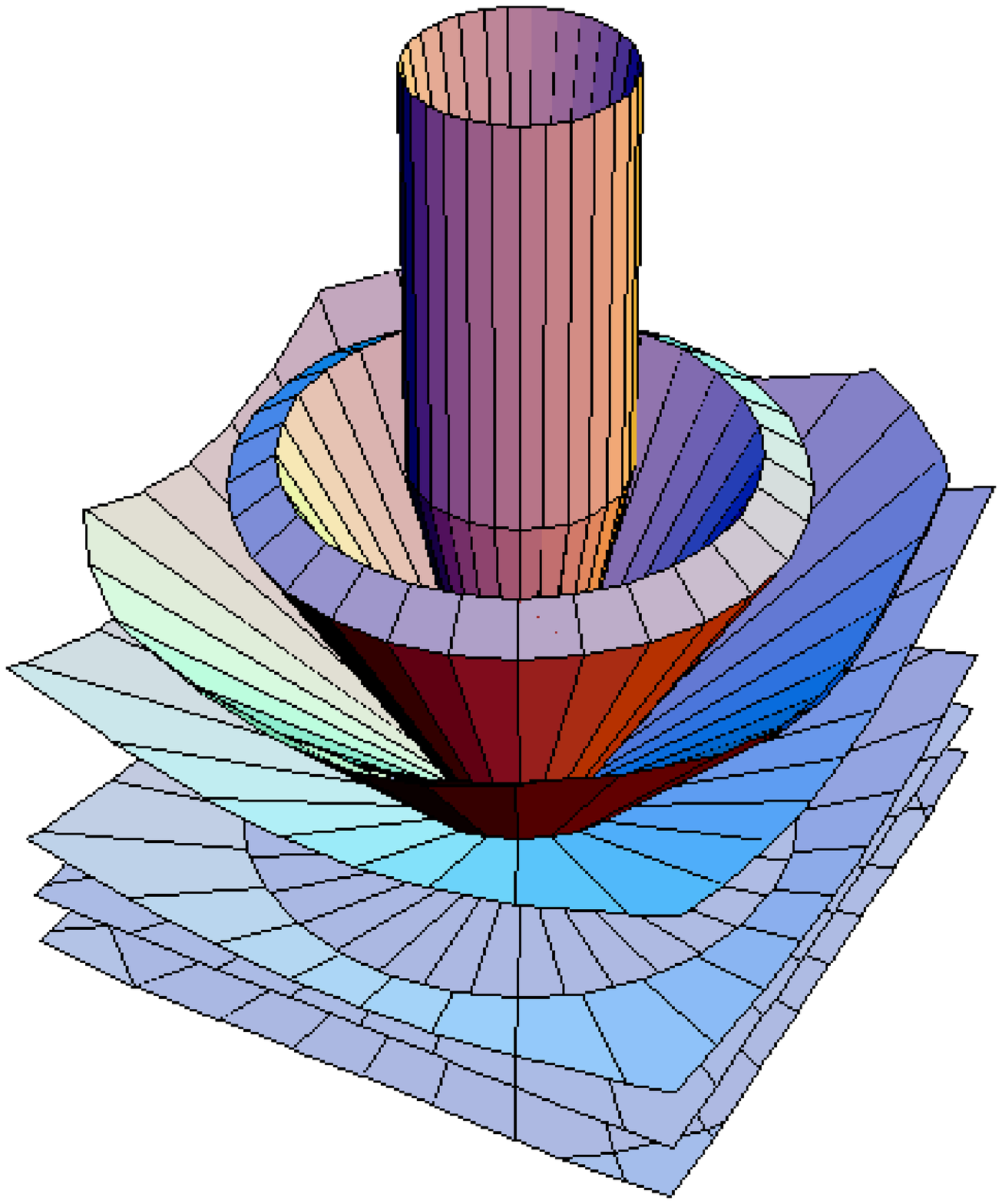}
\caption{$\mid Y^{\pm1}_{5}(\theta,\varphi)\mid$ vs $\mid Z^{\pm1}_{5}(\theta,\varphi)\mid$}
\end{figure}
\clearpage
\begin{figure}
\center
\includegraphics[width=7cm]{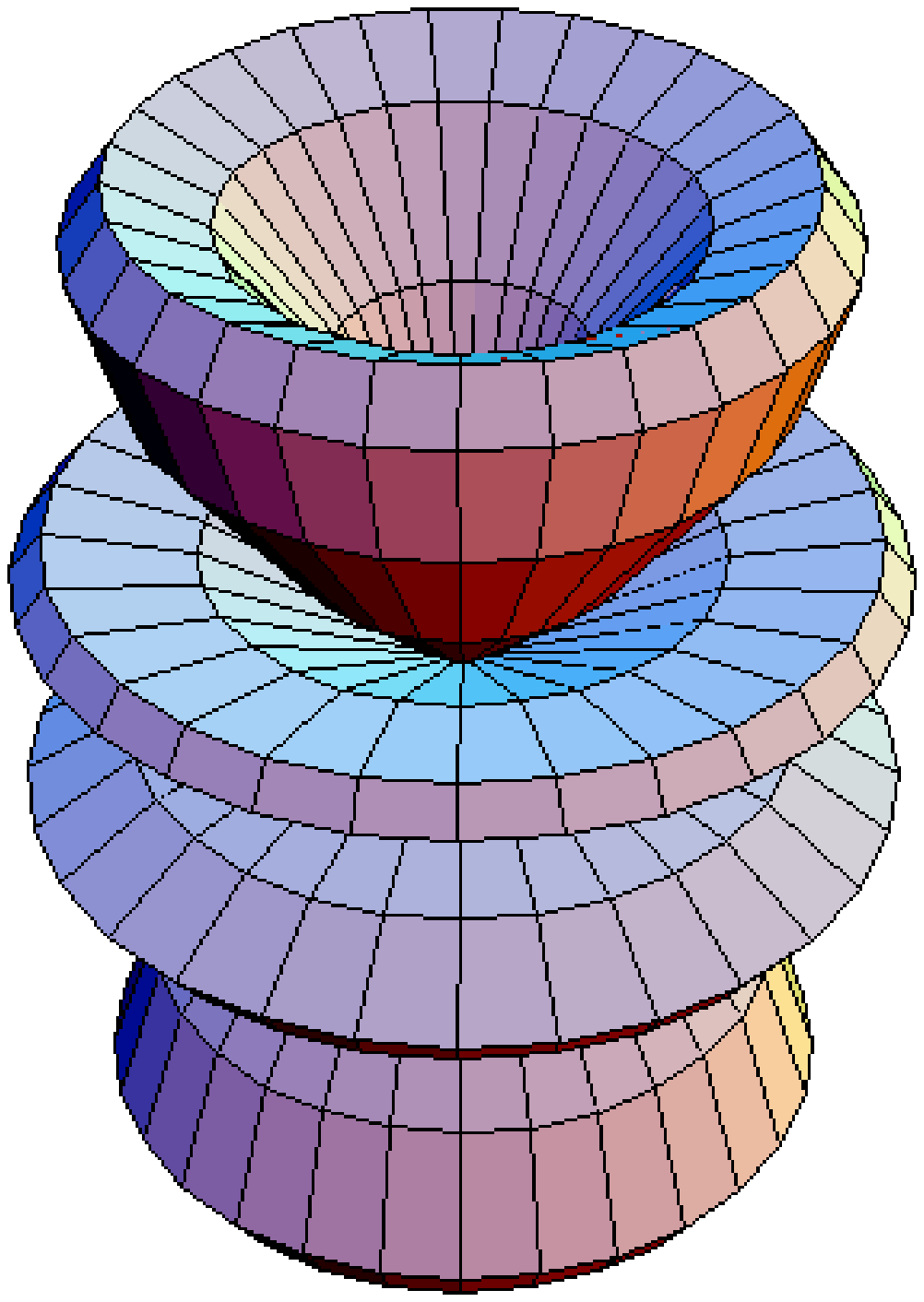}
\includegraphics[width=7cm]{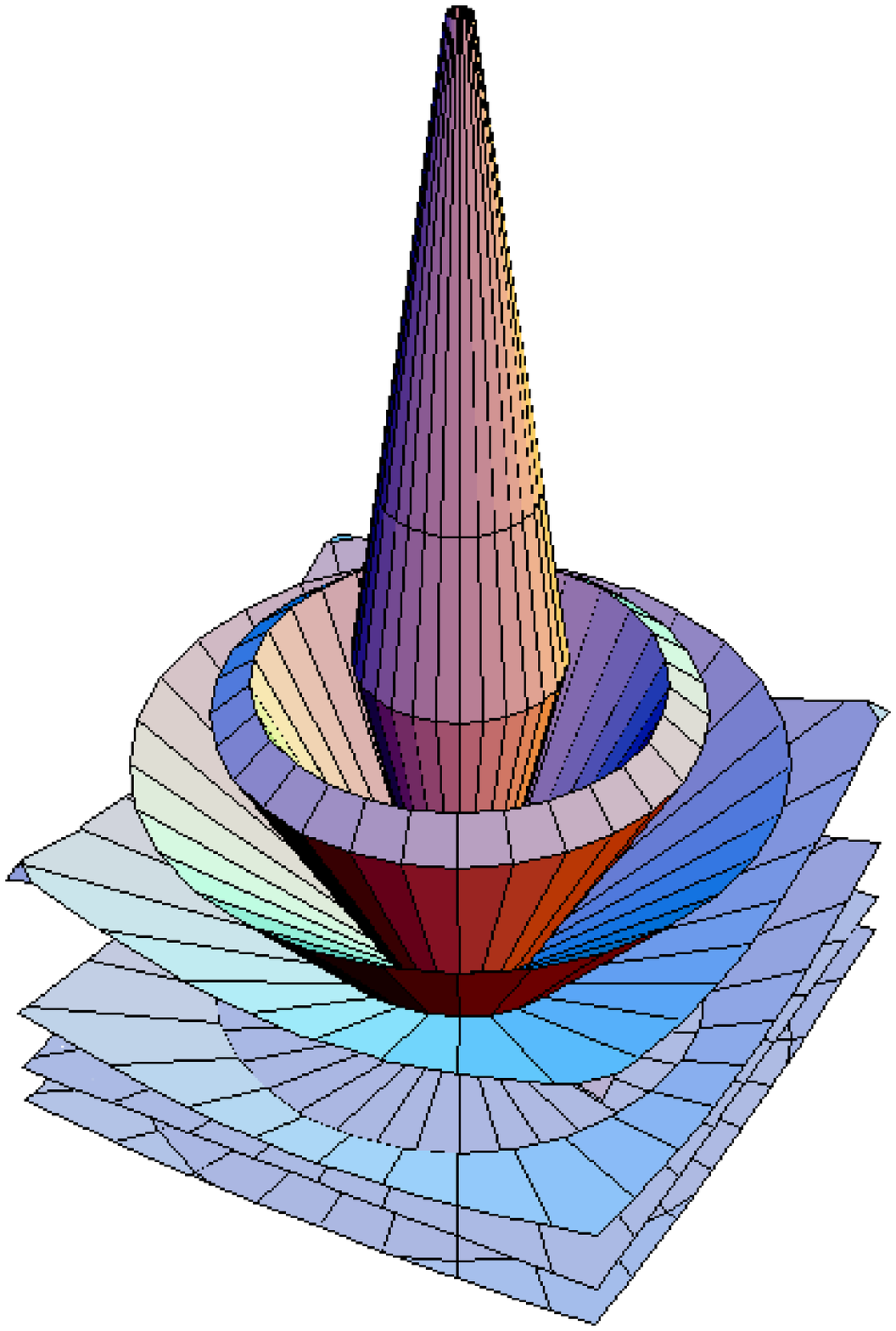}
\caption{$\mid Y^{\pm2}_{5}(\theta,\varphi)\mid$ vs $\mid Z^{\pm2}_{5}(\theta,\varphi)\mid$}
\end{figure}
\begin{figure}
\center
\includegraphics[width=7cm]{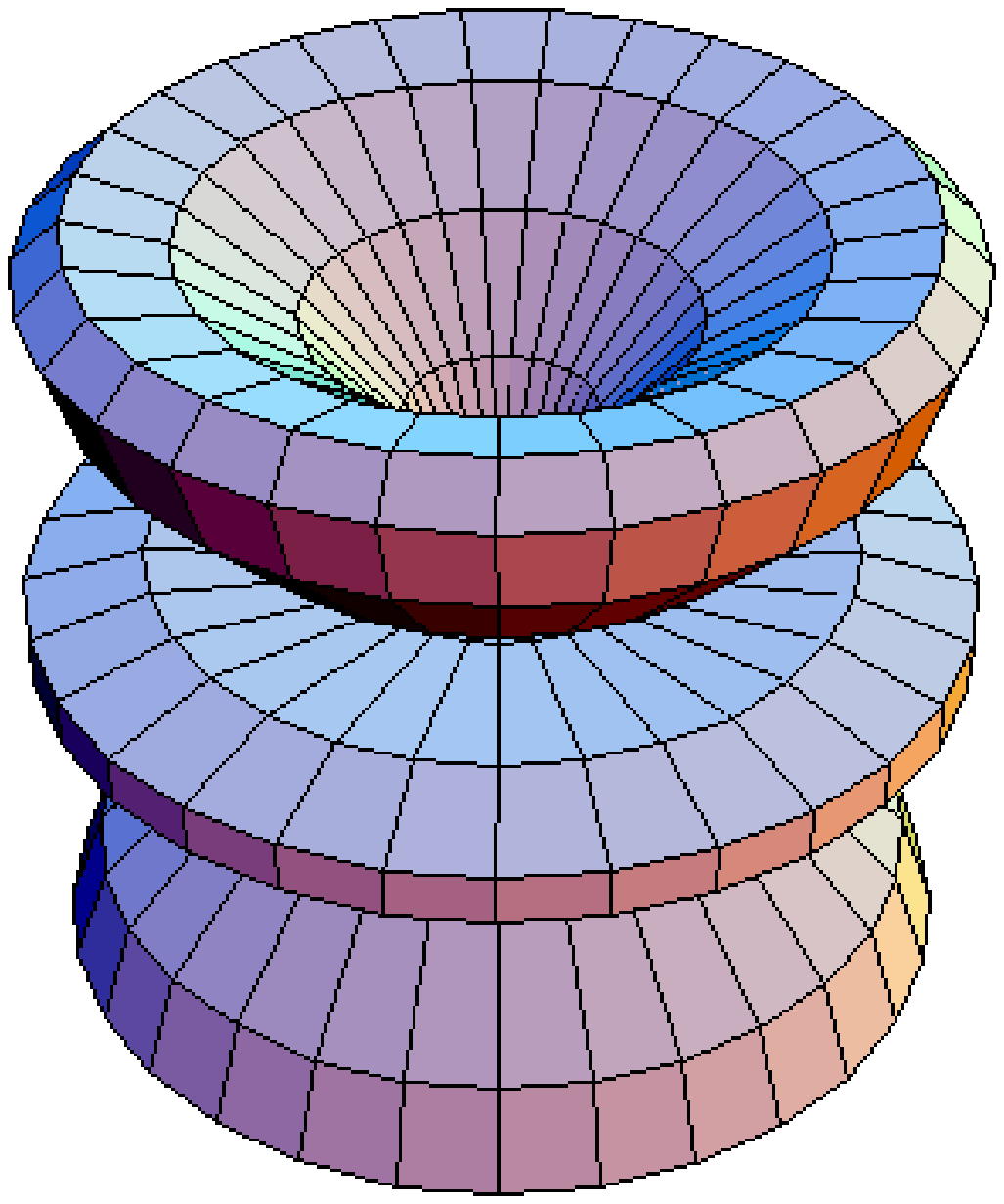}
\includegraphics[width=7cm]{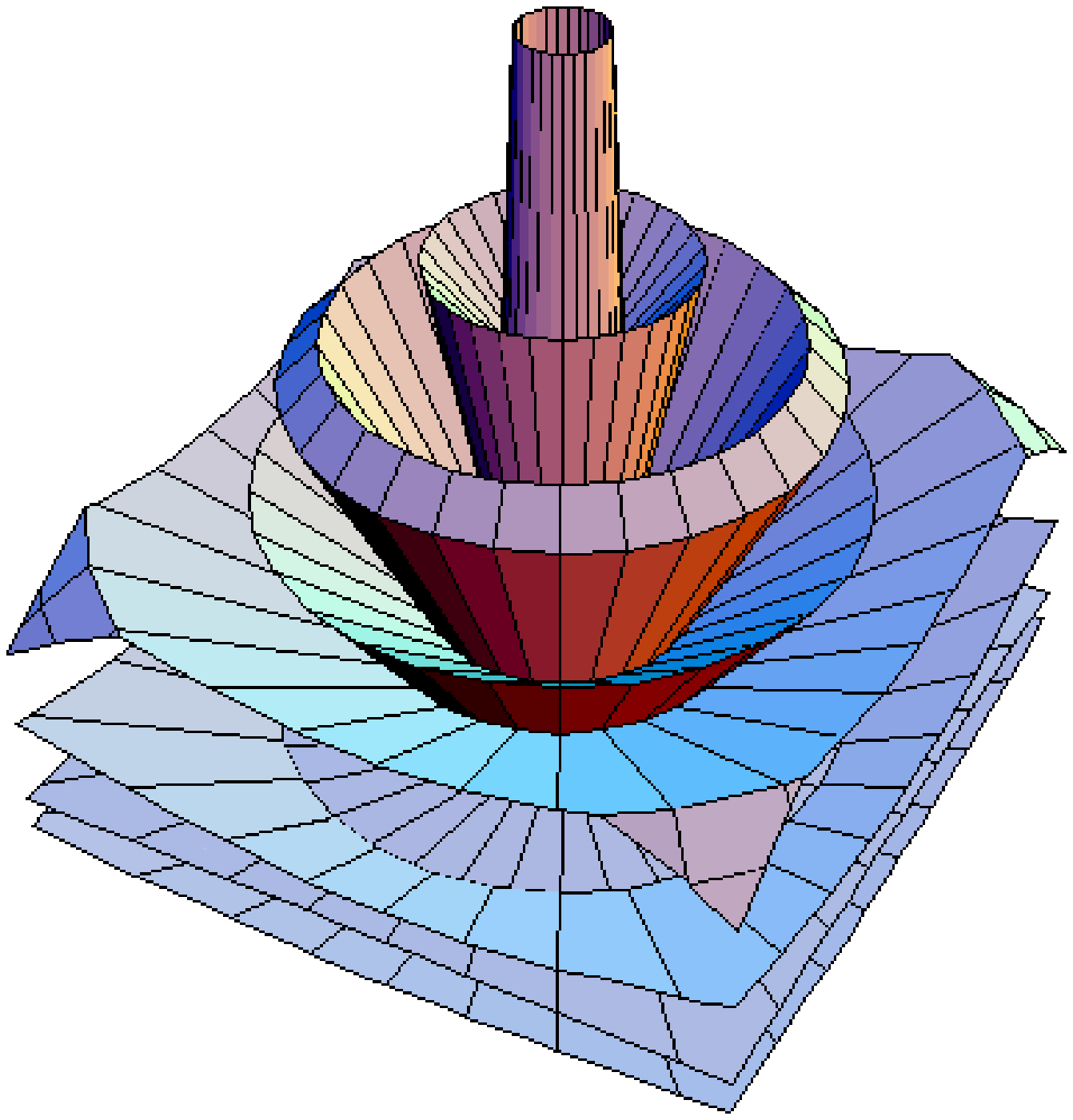}
\caption{$\mid Y^{\pm3}_{5}(\theta,\varphi)\mid$ vs $\mid Z^{\pm3}_{5}(\theta,\varphi)\mid$}
\end{figure}
\begin{figure}
\center
\includegraphics[width=7cm]{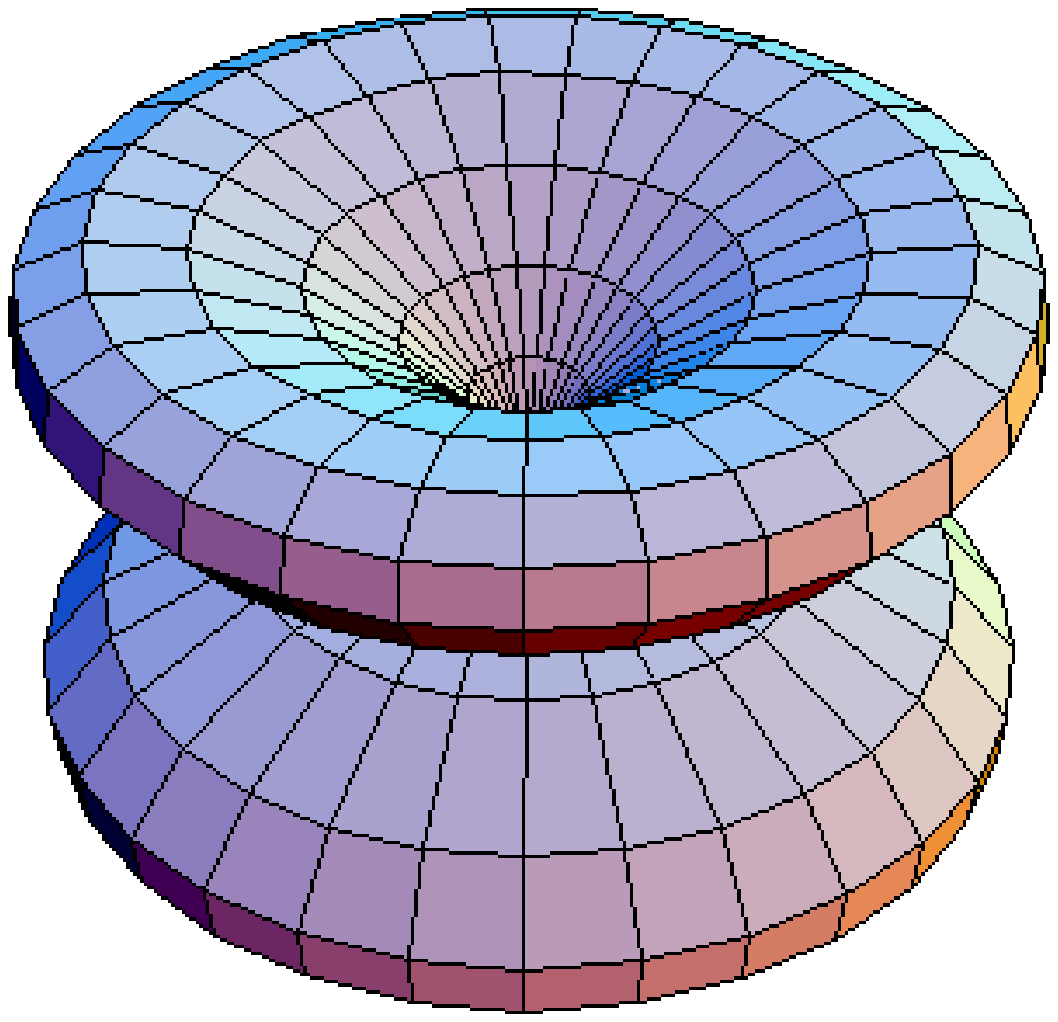}
\includegraphics[width=7cm]{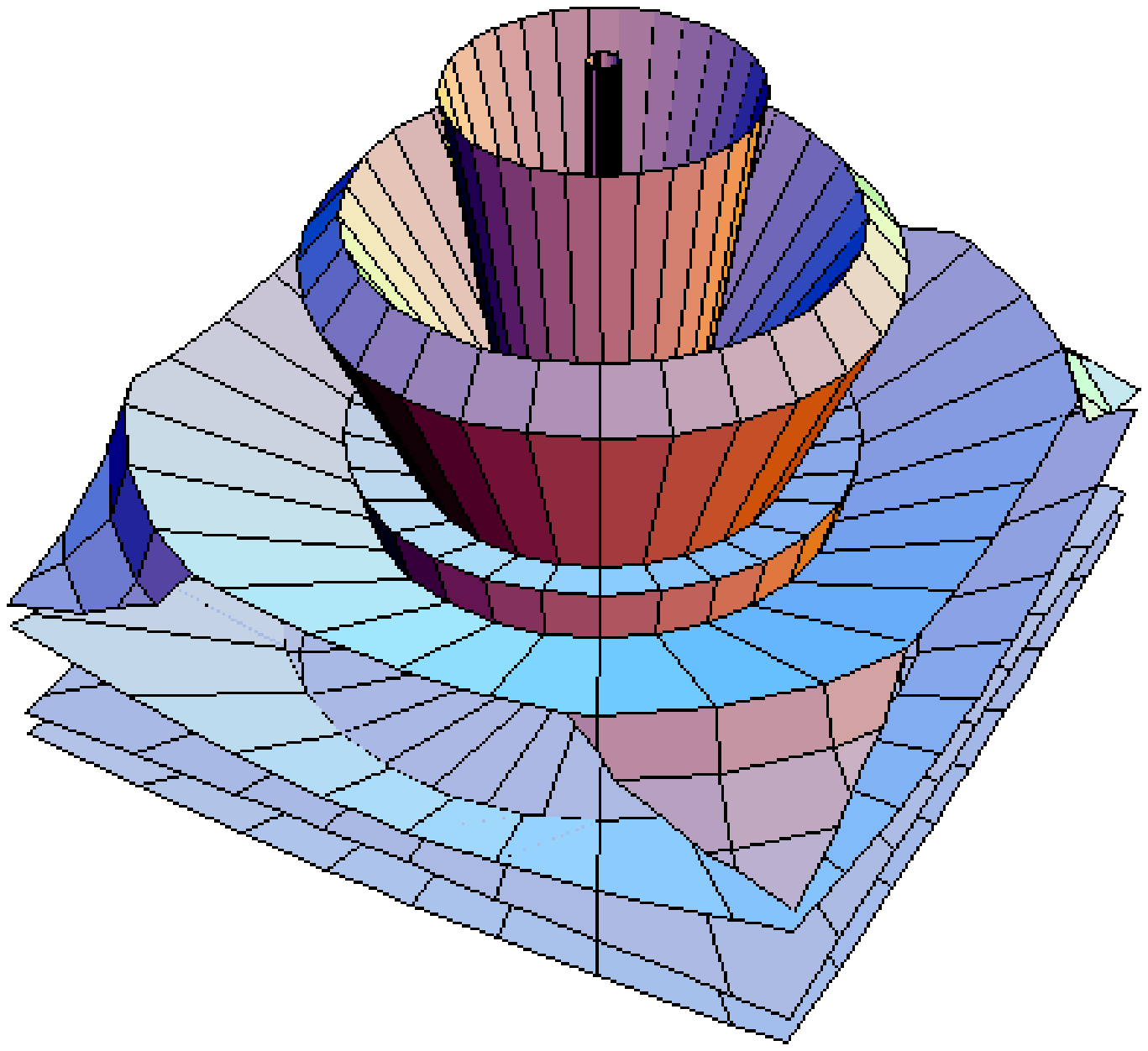}
\caption{$\mid Y^{\pm4}_{5}(\theta,\varphi)\mid$ vs $\mid Z^{\pm4}_{5}(\theta,\varphi)\mid$}
\end{figure}
\begin{figure}
\center
\includegraphics[width=7cm]{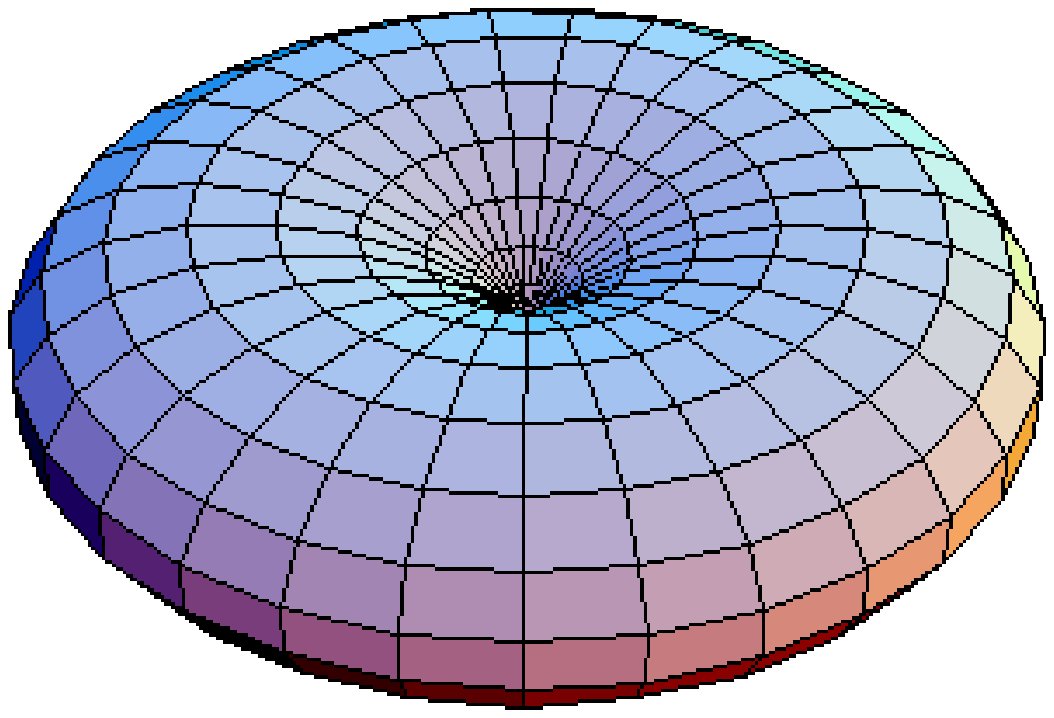}
\includegraphics[width=7cm]{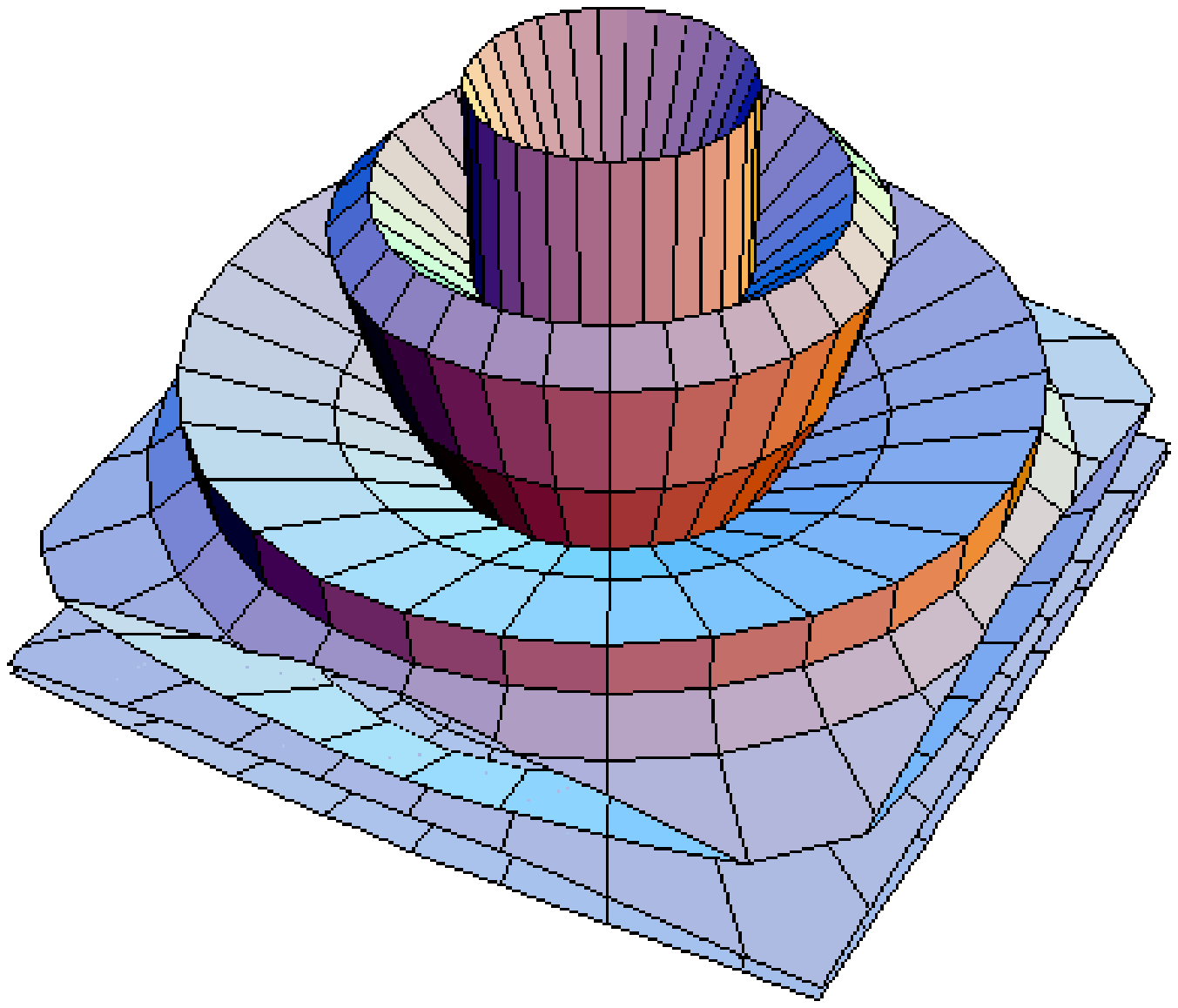}
\caption{$\mid Y^{\pm5}_{5}(\theta,\varphi)\mid$ vs $\mid Z^{\pm5}_{5}(\theta,\varphi)\mid$}
\end{figure}

\begin{figure}
\center
\includegraphics[width=7cm]{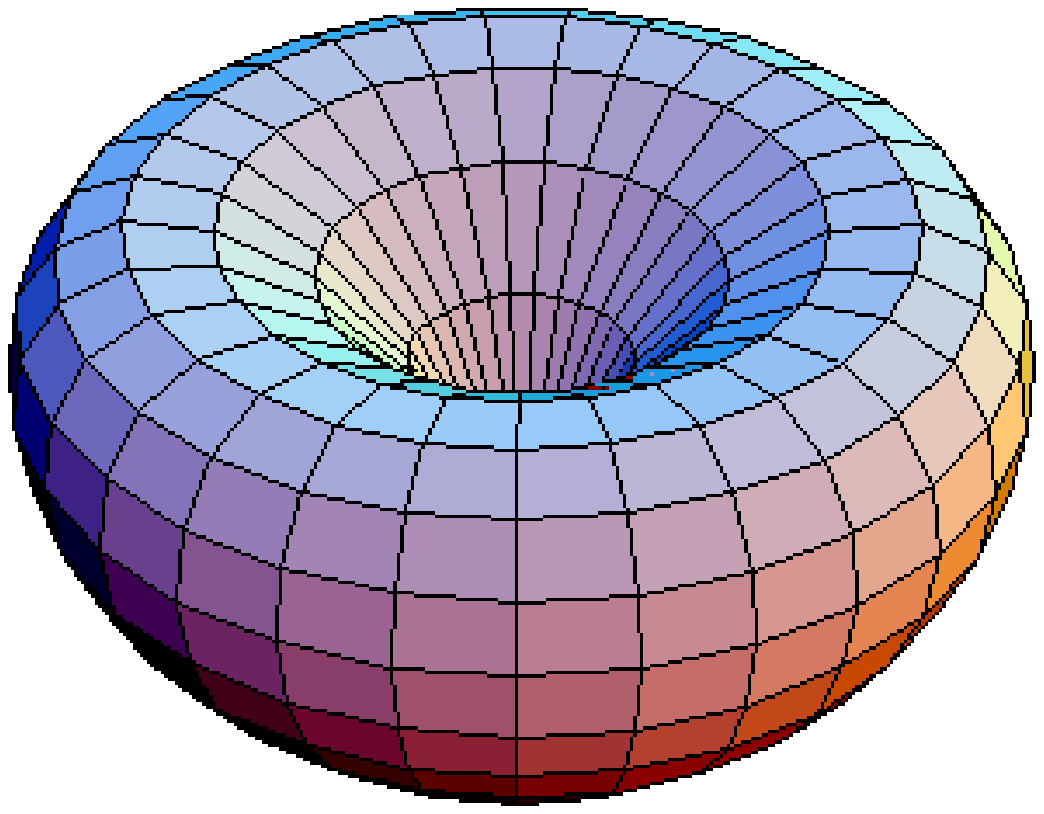}
\includegraphics[width=7cm]{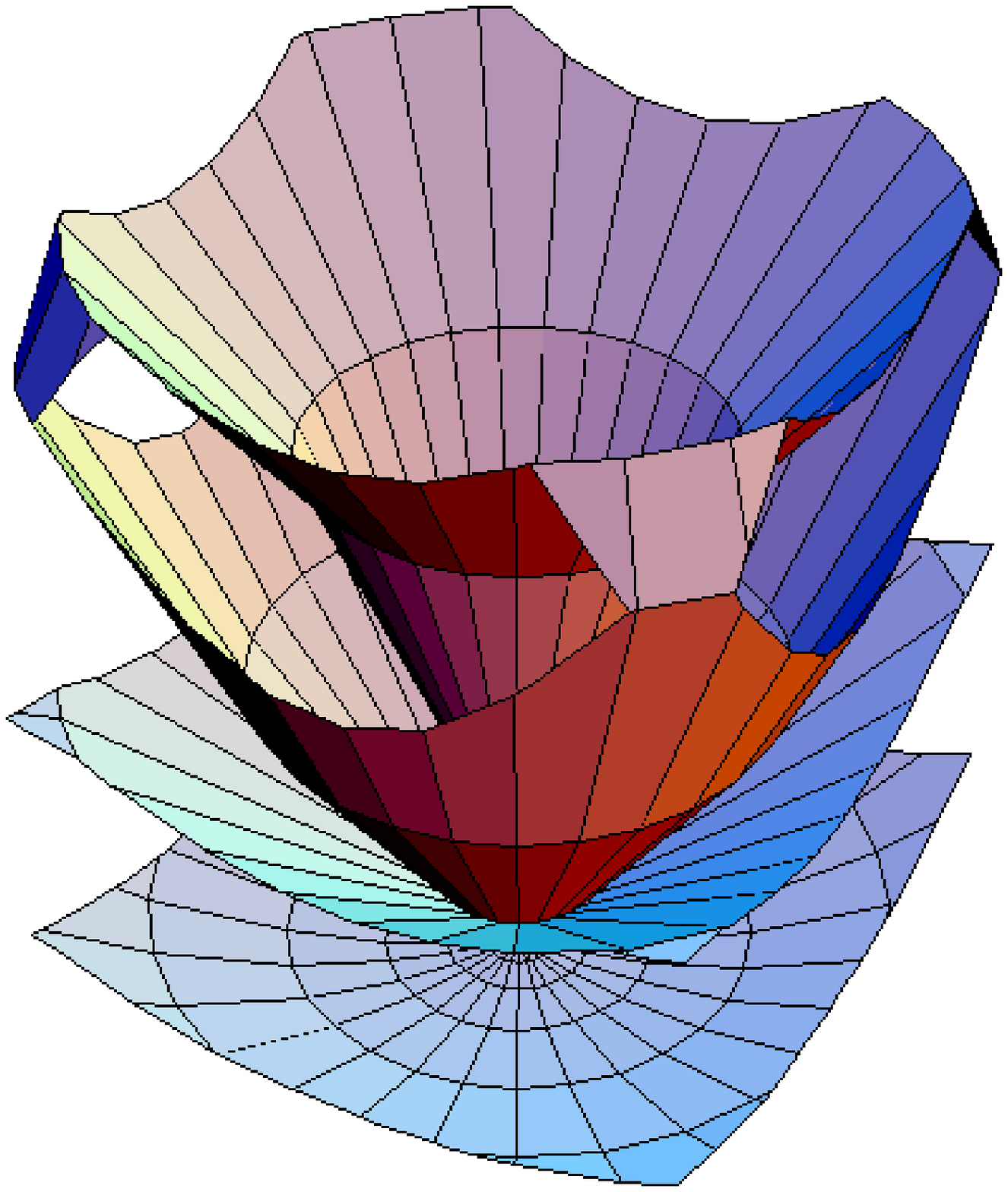}
\caption{$\mid Z_{1}^{2 }(\theta,\varphi)\mid$, and 
$\mid Z_{2}^{-5}(\theta,\varphi)\mid$ as examples for non-spherical
angular functions that don't have a spherical counterpart.}
\end{figure}


\listoffigures
\listoftables

\end{document}